\documentclass[10pt,conference,letterpaper]{IEEEtran}
\usepackage{times,amsmath,epsfig}

\usepackage{amsthm}
\usepackage{balance}  
\usepackage{paralist}
\usepackage{subfig}
\usepackage{graphicx}
\usepackage{siunitx}
\usepackage{nicefrac}
\usepackage{framed}
\usepackage{soul}
\usepackage{url}
\usepackage{amsmath,epsfig}
\usepackage{verbatim}
\usepackage{xspace}
\usepackage{xcolor,listings}
\usepackage{dsfont}
\usepackage[T1]{fontenc}
\usepackage[normalem]{ulem}
\usepackage{multirow}
\usepackage{amsmath}
\usepackage{amsthm}
\usepackage{mathtools}
\newcommand{\subparagraph}{}
\usepackage{titlesec}

\titlespacing*{\subsection}{0pt}{0.7\baselineskip}{0pt}

\newenvironment{paraenum}{\begin{inparaenum}[\itshape (i)\upshape]}{\end{inparaenum}}

\renewcommand{\sfdefault}{cmss}

\newcommand{\Eg}{{\it E.g.}, }
\newcommand{\eg}{{\it e.g.}, }

\newcommand{\ie}{{\it i.e.}, }

\newcommand\nth{\textsuperscript{th}\xspace}
\newcommand\Opt[1]{\operatorname{Opt}(#1)}
\newcommand\romCost[1]{\operatorname{romCost}(#1)}
\newcommand\migCost[1]{\operatorname{migCost}(#1)}
\newcommand\comCost[1]{\operatorname{comCost}(#1)}
\newcommand\rcvCost[1]{\operatorname{rcvCost}(#1)}
\newcommand\rmax{r_{max}\xspace}
\newcommand\cmax{c_{max}\xspace}

\newtheorem{problem}{Problem}
\newtheorem{theorem}{Theorem}
\newtheorem{definition}{Definition}
\newtheorem{example}{Example}
\newtheorem{observation}{Observation}

\newcommand{\tabname}[1]{\ensuremath{\textsf{#1}}} 
\newcommand{\attname}[1]{\mbox{{\small #1}}} 

\renewcommand{\O}[1]{$\mathcal{O}(#1)$}

\newcommand{\posmapping}{\ensuremath{\mathds{M}}}
\newcommand{\datamodel}{\ensuremath{\mathds{P}}}
\newcommand\size[1]{\operatorname{size}(#1)}
\newcommand\cost[1]{\operatorname{cost}(#1)}
\newcommand{\set}[1]{\left\lbrace #1\right\rbrace}

\newcounter{tctr}
\stepcounter{tctr}

\newcommand{\anon}[2]{#2}

\newcommand{\system}{\anon{{\sc XLSpread}\xspace}{{\sc DataSpread}\xspace}}
\usepackage{color}
\newcommand{\cut}[1]{}
\newcommand{\longlong}[1]{}

\newcommand{\tr}[1]{#1}
\newcommand{\paper}[1]{}
\newcommand{\later}[1]{}

\newcommand{\code}[1]{\textsf{\small\mdseries #1}}

\newcommand{\agp}[1]{\textcolor{red}{Aditya: #1}}
\newcommand{\msb}[1]{\textcolor{red}{\hl{Mangesh: #1}}}

\newcommand{\squishlist}{
   \begin{list}{$\bullet$}
    { \setlength{\itemsep}{0pt}
      \setlength{\parsep}{2pt}
      \setlength{\topsep}{2pt}
      \setlength{\partopsep}{0pt}      
      \setlength{\leftmargin}{12pt}
    }
}
\newcommand{\squishend}{\end{list}}

\newcommand{\squishenum}{
	\begin{enumerate}
		{ \setlength{\itemsep}{-50pt}
			\setlength{\parsep}{0pt}
			\setlength{\topsep}{0pt}
			\setlength{\partopsep} \migCost{0pt}
		}
	}
	\newcommand{\squishenumend}{\end{enumerate}}
\newcommand{\stitle}[1]{\vspace{0.2em}\noindent\textbf{#1}}

\newcommand{\emtitle}[1]{\vspace{0.1em}\noindent\emph{#1}}

\newcommand{\ta}[2]{\vspace{-5pt}\begin{framed}\small \vspace{-7pt}\noindent\textit{\underline{Takeaway \thetctr\stepcounter{tctr} (#1):} #2}\vspace{-7pt}\end{framed}\vspace{-5pt}}

\newcommand{\boxy}[1]{\vspace{-5pt}\begin{framed}\vspace{-7pt}\noindent#1\vspace{-7pt}\end{framed}\vspace{-5pt}}
\newcommand{\boxyta}[1]{\vspace{-5pt}\begin{framed}\small \vspace{-7pt}\noindent{\em \ul{Takeaway:} #1}\vspace{-7pt}\end{framed}\vspace{-5pt}}

\paper{\renewcommand{\baselinestretch}{0.98}}
\tr{\renewcommand{\baselinestretch}{0.985}}

\makeatletter
\def\@copyrightspace{\relax}
\makeatother

\title{Towards a Holistic Integration of Spreadsheets with Databases:
A Scalable Storage Engine for Presentational Data Management
\tr{\\\Large [Technical Report]}}

\author{
{Mangesh Bendre, Vipul Venkataraman, Xinyan Zhou, Kevin Chang, Aditya Parameswaran}%
\vspace{1.6mm}\\
\fontsize{10}{10}\selectfont\itshape
Department of Computer Science, University of Illinois at Urbana-Champaign (UIUC), USA\\
\fontsize{9}{9}\selectfont\ttfamily\upshape
\{bendre1\,|\,vvnktrm2\,|\,xzhou14\,|\,kcchang\,|\,adityagp\}@illinois.edu}

\begin{document}
\maketitle
\begin{abstract}
Spreadsheet software is the tool of choice for interactive ad-hoc data management,
with adoption by billions of users. 
However, spreadsheets are not scalable, unlike database systems.
On the other hand, database systems, while highly scalable, do not support interactivity as
a first-class primitive. 
We are developing \system\anon{\footnote{Name anonymized for double-blind purposes.}}{}, 
to holistically integrate spreadsheets as a front-end interface with
databases as a back-end datastore, providing
scalability to spreadsheets, and interactivity to databases,
an integration we term {\em presentational data management (PDM)}.
In this paper, we make the first step towards this vision: developing a
storage engine for PDM, studying how to flexibly represent 
spreadsheet data within a database and
how to support and maintain access by position. 
We first conduct an extensive survey of spreadsheet use
to motivate our functional requirements for a storage engine for PDM.
We develop a natural set of mechanisms for flexibly representing spreadsheet data
and demonstrate that identifying the optimal representation
is {\sc NP-Hard}; however, we develop an efficient approach to
identify the optimal representation from an important and intuitive subclass of representations.
We extend our mechanisms with positional access mechanisms that don't suffer
from cascading update issues, leading to constant time access and modification performance. 
We evaluate these 
representations on a workload of typical spreadsheets and spreadsheet operations, providing
up to 50\% reduction in storage, and up to 50\% reduction in formula evaluation time. 
\end{abstract}

\section{Introduction} \label{sec:introduction}

We are witnessing an increasing availability of data
across a spectrum of domains, necessitating interactive
ad-hoc management of this data: a business owner may want to manage
customer data and invoices, a scientist experimental measurements,
and a fitness enthusiast heart rate and activity traces.
However, while there are two major software paradigms
for supporting interactive ad-hoc data management---spreadsheets
and databases---neither of them fulfill the desired requirements, as we
illustrate using two real use-cases below:

\begin{example}[Using Spreadsheets for Genomic Data Analysis]
During the course of genomic data analysis, biologists,
such as our collaborators at the KnowEnG center at Mayo Clinic, generate 
data describing genomic variants as VCF (variant cell format) files,
akin to CSV files\footnote{\url{www.internationalgenome.org/data}; \url{www.ncbi.nlm.nih.gov/projects/SNP/}}.
These VCF files are large, 
with tens of millions of rows and hundreds of columns, plus a raw
size of many gigabytes. 
Unfortunately, many biologists, 
like scientists in many other domains,
are adept at using spreadsheet software, but are not comfortable enough with programming to use databases.
To interactively explore or browse their VCF data, they struggle to load such files into spreadsheet software: 
\eg Microsoft Excel limits uploaded datasets to 1M rows and Google Sheets to 2M cells.
And even when one can load large datasets, these tools become sluggish and unresponsive.
In fact, many biologists are unable to explore the datasets they themselves create, instead sending them
to bioinformatics collaborators to analyze.
\longlong{; and several special tools like VCFtools!\cite{Danecek:2011:VCF:2012215.2012217} have been developed for browsing and manipulating such data in essentially ``relational'' ways.}
\end{example}

\begin{example}[Using Databases for Customer Management]
The owner of a small retail startup
in Champaign, Illinois created a MySQL database for managing customers and sales, 
organized in a schema comprising 15 tables. 
There are several actions that he and his staff would like to routinely perform, such as
insert (customers), modify (due dates of invoices),
filter (overdue invoices), 
join (invoices and payments), and aggregate (the total amounts).
To perform these operations without requiring SQL, he has to employ a programmer to develop database applications.
Instead, he wants to interactively manipulate data for ad-hoc operations, but no such tools exist.  
\end{example}

\begin{figure}[t]	
    \centering
	\includegraphics[width=0.38\textwidth,clip]{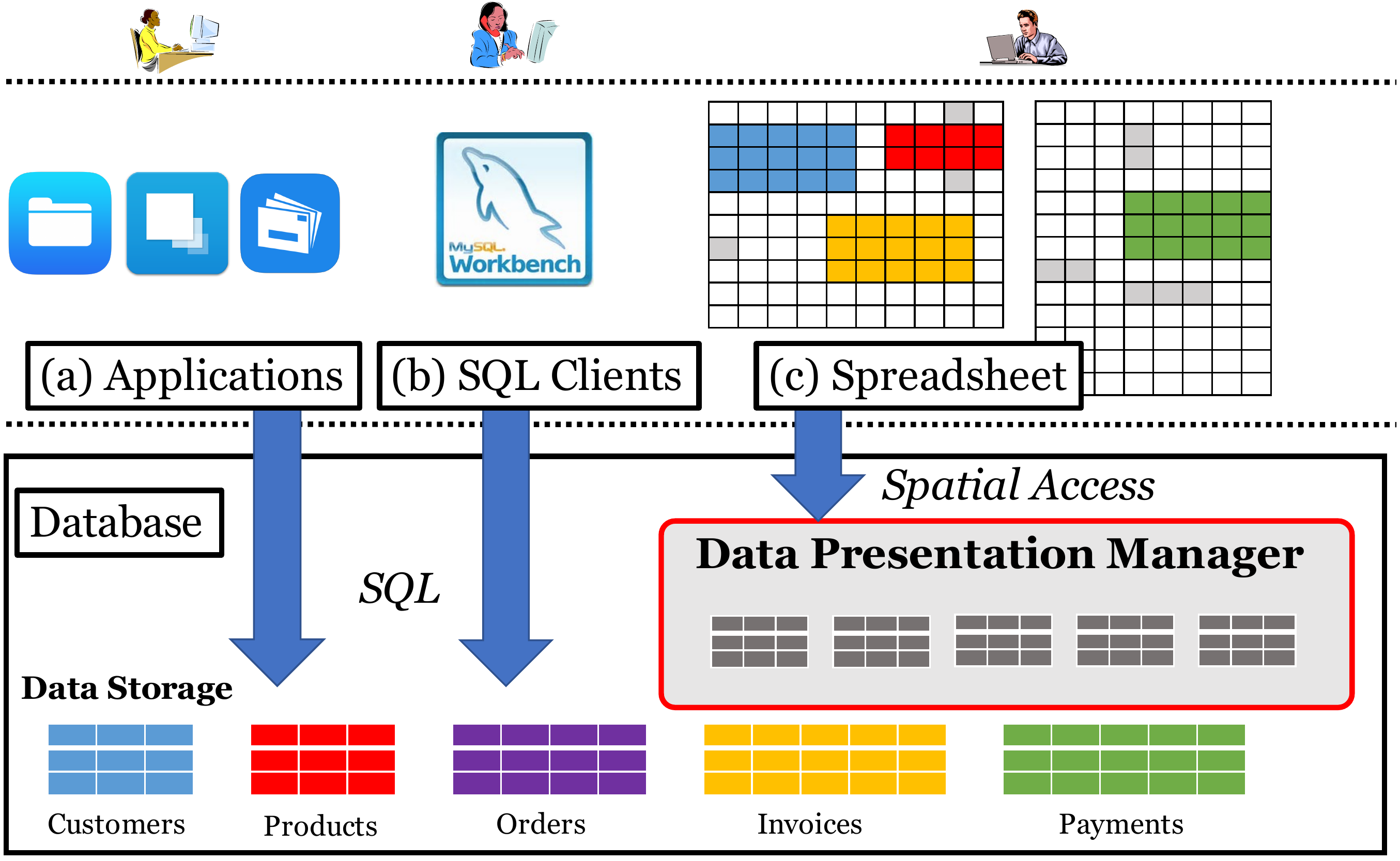}
	\vspace{-3pt}
	\caption{Connecting spreadsheets and databases for {\em Presentational Data Management}.}
	\label{fig:interfacemodel}
\vspace{-22pt}
	\end{figure}

\noindent As these common use cases demonstrate,
we are critically lacking a solution 
for interactive ad-hoc management
of data.
On the one hand, spreadsheet software, while being
heralded as a prime example of a direct manipulation~\cite{shneiderman_direct_1983} tool, 
lacks {\em scalability}, due to its inability to operate
on datasets that go beyond main memory capabilities,
and {\em expressiveness}, since its formulae only operate on 
one cell at a time, necessitating complex means (\eg \code{VLOOKUP}) to orchestrate simple operations like joins. 
On the other hand, while databases provide both scalability and expressiveness, 
they lack support for direct manipulation vital for interactive ad-hoc data management.
Thus, users access databases either via pre-programmed
database applications (Figure~\ref{fig:interfacemodel}a),
or SQL clients (Figure~\ref{fig:interfacemodel}b), 
which only support operations on entire relations at a time,
as opposed to directly interacting with data
for ad-hoc updates and analysis.
To this end, there has been a number of papers on making databases
usable, \eg~\cite{jagadish_making_2007,nandi_querying_2013,li2014constructing,idreos2013dbtouch,abouzied2012dataplay}, 
but this research has not witnessed widespread adoption.

To address this, 
we are building a system, \system, with {\em spreadsheets as a front-end interface,
and databases as a back-end datastore}, with dual objectives:
%
\begin{paraenum}  
\item  allowing users to manipulate data from databases on a spreadsheet interface, without relying on pre-programmed applications or SQL clients---thereby enabling interactive ad-hoc data management for a database, while 
%
\item operating on datasets not limited by main memory---thereby
addressing the key limitation of present-day spreadsheets.
\end{paraenum}
We call this new research direction of holistically integrating spreadsheets and databases  
{\bf \em presentational data management} (PDM).
\longlong{-- where the spreadsheet modality 
enables a database with ad-hoc querying and manipulation while presenting data for browsing.}
In PDM, using a system like \system, a user can view, analyze, and manipulate
data in a presentational (\ie spatial) interface (Figure~\ref{fig:interfacemodel}c), in addition to 
standard approaches (Figure~\ref{fig:interfacemodel}a,b). 
They can import large tables (\eg VCF files) from/to the interface or database,
and arrange them on the interface.
They can also operate at various granularities, embodying the principles of \emph{direct manipulation}~\cite{shneiderman_direct_1983}---from
cells (like a spreadsheet) to tables (like a database)---adding computation in the form
of formulae or queries as part of the interface, in addition to data.
This can result in various forms of arrangement of data, ranging from 
structured tables, reports, and forms, to ad-hoc presentations of data embedded with computation. 
They can also refer to data by tables or attributes (as in a database)
or position (as in a spreadsheet).
While we primarily focus on spreadsheets, similar considerations apply 
for other presentational interfaces for interactive ad-hoc data management.


While developing \system is a multi-year vision, we have already 
made significant
headway, with a functional prototype (see
\code{http://dataspread.github.io}). 
In this paper, we focus on
the following fundamental question---{\em how do we develop a storage manager to support 
presentational data management (PDM)?} 

\stitle{Requirements for a PDM Storage Engine.} 
In developing a storage engine for PDM,  we conduct a 
survey and user study (Section~\ref{sec:empirical_study}) to characterize
two key functional requirements for such a storage engine to support the direct manipulation of data in 
a presentational interface (such as a spreadsheet):

\longlong{Towards developing a storage engine for PDM, we identify
two requirements-- which are fundamental for direct manipulation of data on a spreadsheet or any interactive ad-hoc data management interface.
As today's databases lack such support, all the SQL clients--despite their "graphical interfaces"-- failed to deliver natural, direct manipulation.}

\vspace{2pt}
\noindent {\em (i) Presentational Awareness.} A storage engine
for PDM must be aware of the layout of data within the spreadsheet interface
and be flexible enough to adapt to various ad-hoc modalities
users might choose to lay out and manage data (and queries) on spreadsheets, ranging from
fully structured tables, to data scattered across the spreadsheet,
along with formulae.

\vspace{2pt}
\noindent {\em (ii) Presentational Access.} A storage engine for
PDM must support access of a range of data by position: for example,
users may scroll to a certain region of the spreadsheet, or a formula
may access a range of cells; this access must be supported as a first-class primitive.





\stitle{Challenges in Supporting PDM.}
In supporting these functional requirements, 
our first set of challenges emerge
in how we can {\bf \em flexibly represent presentational
information within a database}.
A user may manage several
table-like regions within a spreadsheet, 
interspersed with empty rows or columns,
along with formulae.
One option is to store the spreadsheet as a single relation,
with tuples as spreadsheet rows, and 
attributes as spreadsheet columns---this can be very wasteful due to sparsity.
Another option is to store
the filled-in cells 
as key-value pairs: [(row \#, column \#), value];
this can be effective for sparse spreadsheets,
but is wasteful for dense spreadsheets with well-defined
tables.
One can imagine hybrid representation schemes
using both ``dense'' and ``sparse'' schemes, as
well as those that take access patterns, \eg via formulae, into account.
Unfortunately, we show that it is {\sc NP-Hard}
{\em to identify the optimal representation}.

Our second set of challenges emerge in 
{\bf \em supporting and maintaining presentational access}. 
Say we use a single relation to record information
about a sheet, with one tuple for each spreadsheet row,
and one attribute for each spreadsheet column;
with an additional attribute that records the spreadsheet row number. 
Now, inserting a single row in the spreadsheet
can lead to an expensive {\em cascading update}
of the row numbers of all subsequent rows; thus, we must
develop techniques that allow us to avoid this issue.
Moreover, we need {\em positional indexes} that can access
a range of rows at a time, say, when a user scrolls to a certain
region of the spreadsheet. 
While one could use a traditional index (\eg a \mbox{B+ tree})
on the attribute corresponding to row number, the cascading update
makes it hard to maintain such an index across edit operations. 

\stitle{Our Contributions.}
In this paper, we address the aforementioned challenges
in developing a scalable storage manager for PDM. 
Our contributions are the following:

 

\vspace{-4pt}
{\flushleft{\textbf{1. Understanding Present-day Solutions for PDM}}}. 
We perform an empirical study of four spreadsheet datasets
plus a user survey to understand how spreadsheets are presently used for
data manipulation and analysis (Section \ref{sec:empirical_study}).

\vspace{-4pt}
{\flushleft{\textbf{2. Abstracting the Functional Requirements}}}.
Based on our study, we define our {\em conceptual data model}, 
as well as the operations necessary for PDM\paper{, and describe our prototype, 
drilling into the storage engine} 
(Section \ref{sec:interface_model}).

\vspace{-4pt}
{\flushleft{\textbf{3. Primitive Representation Schemes for PDM}}}.
We propose four {\em primitive data models}
that implement the conceptual data model, and demonstrate
that they represent ``optimal extreme choices'' (Section~\ref{sec:primitive_datamodels}).

\vspace{-4pt}
{\flushleft{\textbf{4. Near-Optimal Hybrid Representation Schemes for PDM}}}.
We develop a space of {\em hybrid data models}, utilizing these primitive data models,
and demonstrate that identifying the optimal
hybrid is {\sc NP-Hard} (Section~\ref{sec:hybrid_data_model});
we further develop multiple PTIME solutions that
provide near-optimality (Section~\ref{sec:optimal-recursive-decomposition}), 
plus greedy heuristics (Section~\ref{sec:greedy}),
and show that they can be incrementally maintained (Section~\ref{sec:extensions}).

\vspace{-4pt}
{\flushleft{\textbf{5. Presentational Access Schemes for PDM}}}.
We develop solutions to maintain positional information,
while reducing the impact of cascading updates (Section~\ref{sec:positional_mapping}).

\vspace{-4pt}
\tr{
{\flushleft{\textbf{6. Prototype of \system}}}.
We have developed a fully functional prototype of \system, and describe
its functionalities that go beyond the storage manager (Section~\ref{sec:architecture}).

\vspace{-4pt}
{\flushleft{\textbf{7. Experimental Evaluation}}}.
We evaluate our data models and presentational access
schemes on a variety of real-world
and synthetic datasets, demonstrating that our storage engine
is scalable and efficient (Section~\ref{sec:experiments}). 
We also conduct a small qualitative evaluation to 
illustrate how \system can handle the use-cases described earlier.
}

\stitle{Related Work.} 
While there have been attempts at combining spreadsheets
and relational database functionality, ultimately, all of these
attempts fall short because they do not let spreadsheet users
perform ad-hoc data manipulation operations~\cite{witkowski_advanced_2005, witkowski_query_2005,liu_spreadsheet_2009};
there have also been efforts that enhance spreadsheets or databases without combining them \eg\cite{tyszkiewicz_spreadsheet_2010}. 
We describe this and other related work in detail in Section~\ref{sec:related}.





\begin{table*}[t]
\setlength\tabcolsep{2pt}
\vspace{-5pt}
    \centering
    \footnotesize
    \begin{tabular}{l||r||r|r|r||r|r||r|r||r|r}
     \multirow{3}{*}{\bf Dataset}   & \multirow{3}{*}{\bf Sheets}  & \multicolumn{3}{c||}{\bf{Formulae Distribution}} &  \multicolumn{2}{c||}{\bf{Density Distribution}} &  \multicolumn{2}{c||}{\bf{Tabular Regions}}  &  \multicolumn{2}{c}{\bf{Formula Access}} \\    
        &  & Sheets with & Sheets with  & \%formulae & Sheets with & Sheets with & \multirow{2}{*}{Tables} & \multirow{2}{*}{\%Coverage} & Cells Accessed & Tabular Regions \\
        &  & formulae &  $>20\%$ formulae & coverage & $<50\%$ density & $<20\%$ density & & & per formula & per formula \\
        \hline
    Internet  & 52,311 & 29.15\% & 20.26\%  & 1.30\%    & 22.53\% & 6.21\% & 67,374 & 66.03\% & 334.26 & 2.50\\
    ClueWeb09 & 26,148 & 42.21\% & 27.13\%  & 2.89\%    & 46.71\% & 23.8\% & 37,164 & 67.68\% & 147.99 & 1.92\\
    Enron     & 17,765 & 39.72\% & 30.42\%  & 3.35\%    & 50.06\% & 24.76\% & 9,733 & 60.98\% & 143.05 & 1.75\\
    Academic  &    636 & 91.35\% & 71.26\%  & 23.26\%   & 90.72\% & 60.53\% & 286   & 12.10\% &   3.03 & 1.54\\
    \end{tabular}
    \caption{Spreadsheet Datasets: Preliminary Statistics.}
    \label{tab:spreadsheet-datasets}
    \vspace{-14pt}
\end{table*}

\tr{
\begin{figure*}[t]
	\centering
	\subfloat{}{\includegraphics[width=0.23\textwidth,clip]{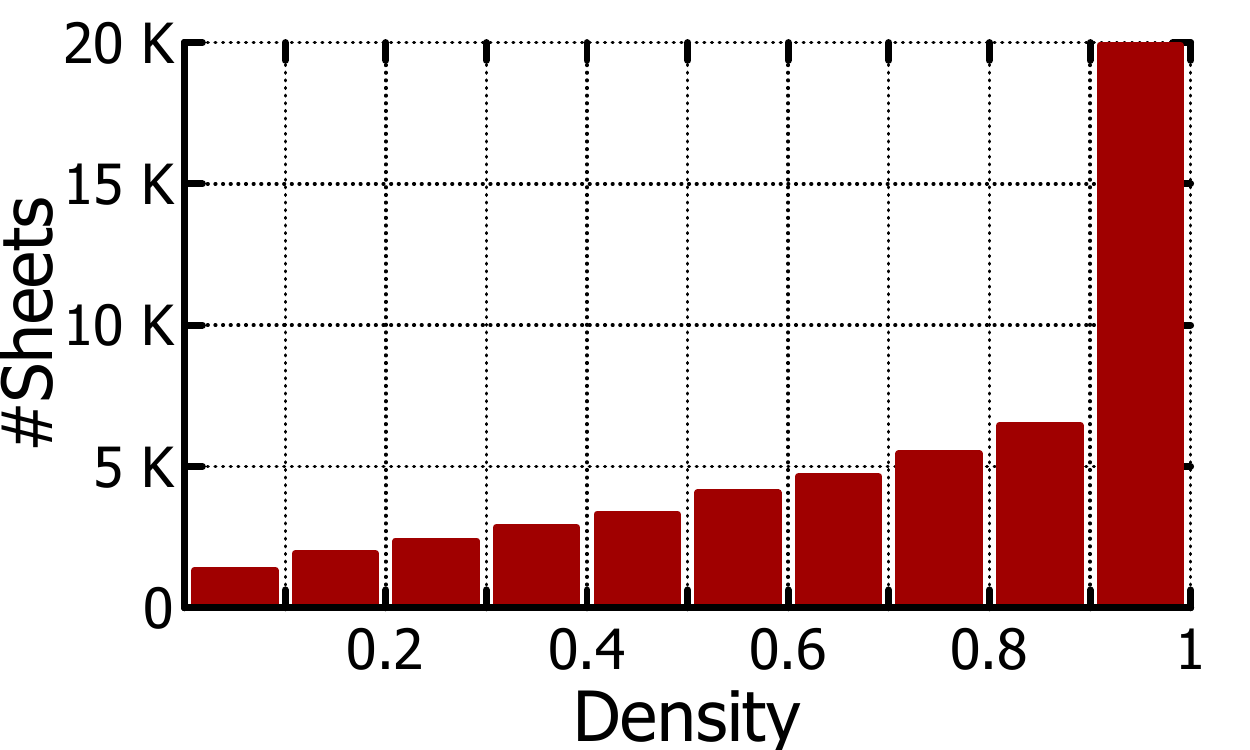}}
	\hspace{5pt}
	\subfloat{}{\includegraphics[width=0.23\textwidth,clip]{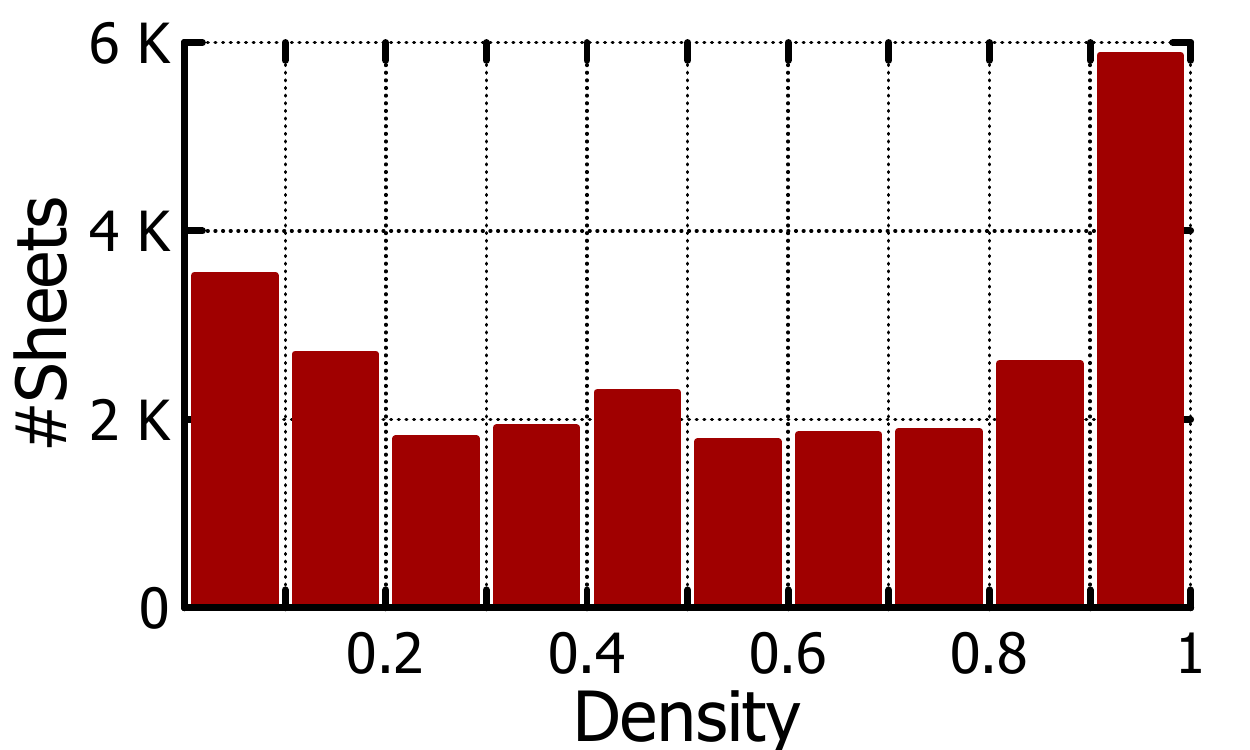}}
	\hspace{5pt}
	\subfloat{}{\includegraphics[width=0.23\textwidth,clip]{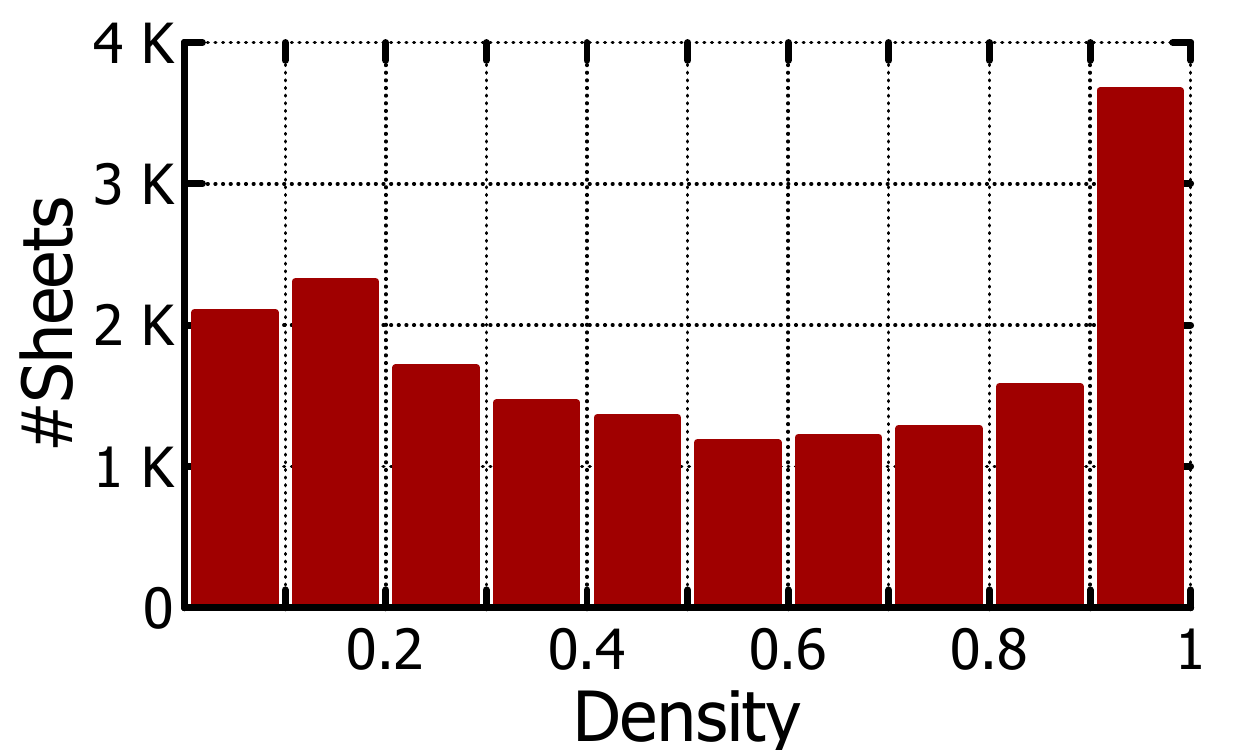}}
	\hspace{5pt}
	\subfloat{}{\includegraphics[width=0.23\textwidth,clip]{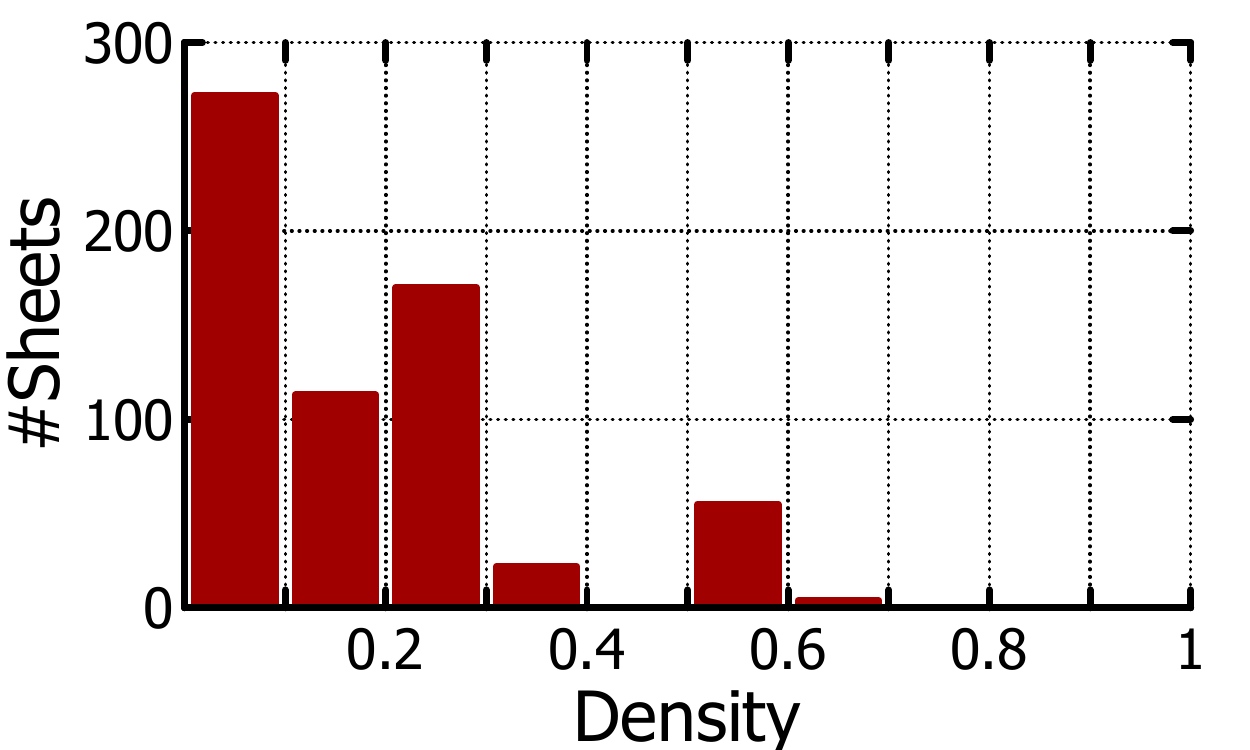}}
	\vspace{-4pt}
	\caption{Data Density --- (a) Internet. (b) ClueWeb09. (c) Enron. (d) Academic.}
	\label{fig:density}
	\vspace{-8pt}
\end{figure*}
}

\tr{
\begin{figure*}[t]
 	\centering
 	\subfloat{}{\includegraphics[width=0.23\textwidth,clip]{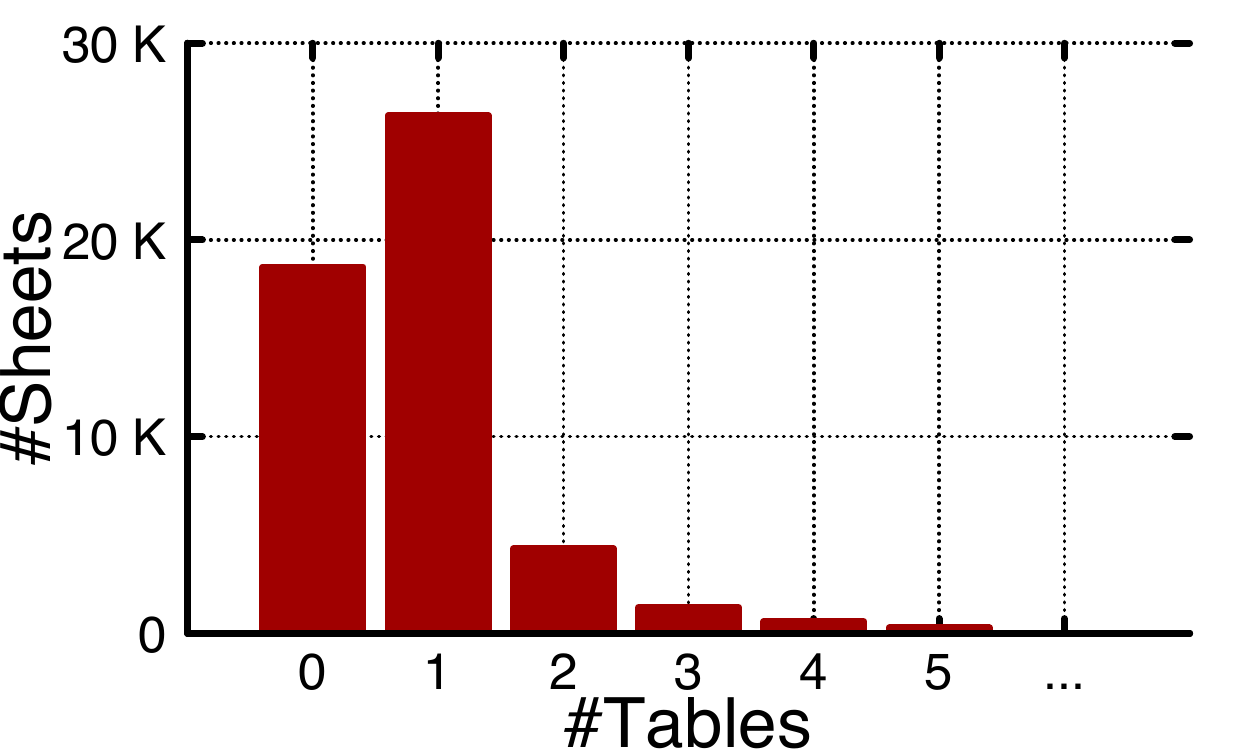}}
 	\hspace{5pt}
 	\subfloat{}{\includegraphics[width=0.23\textwidth,clip]{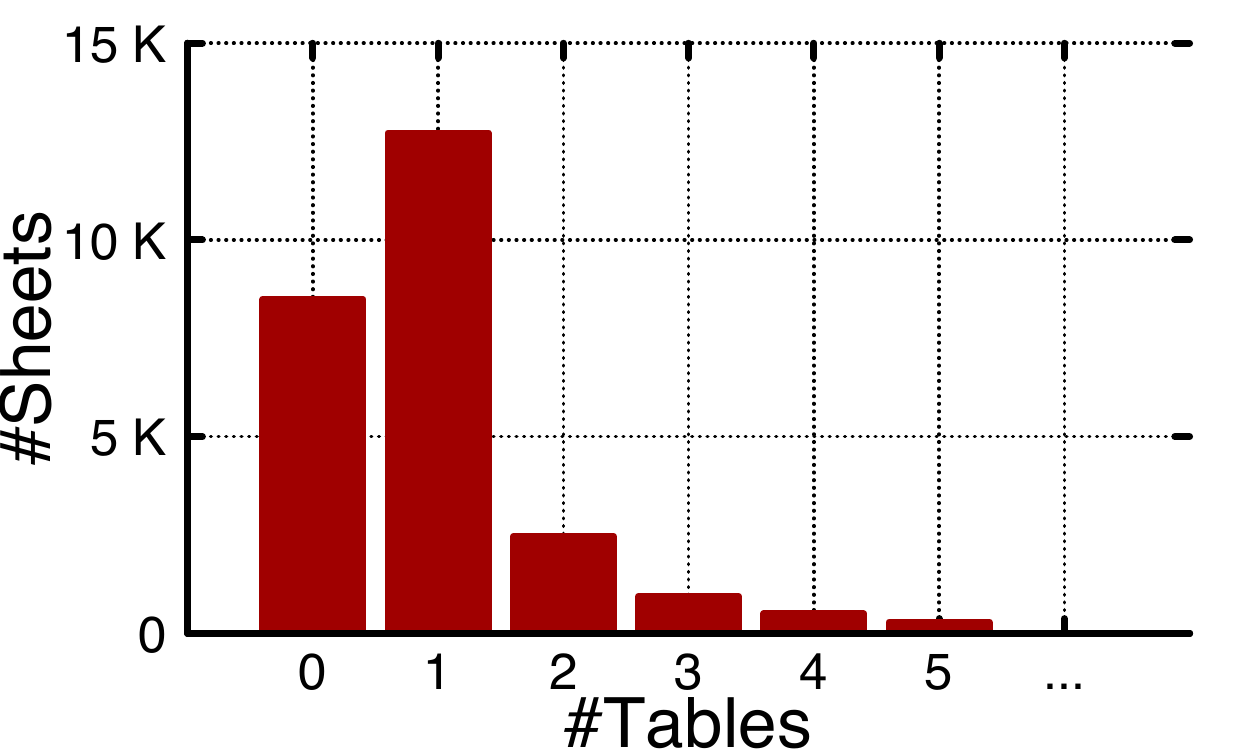}}
 	\hspace{5pt}
 	\subfloat{}{\includegraphics[width=0.23\textwidth,clip]{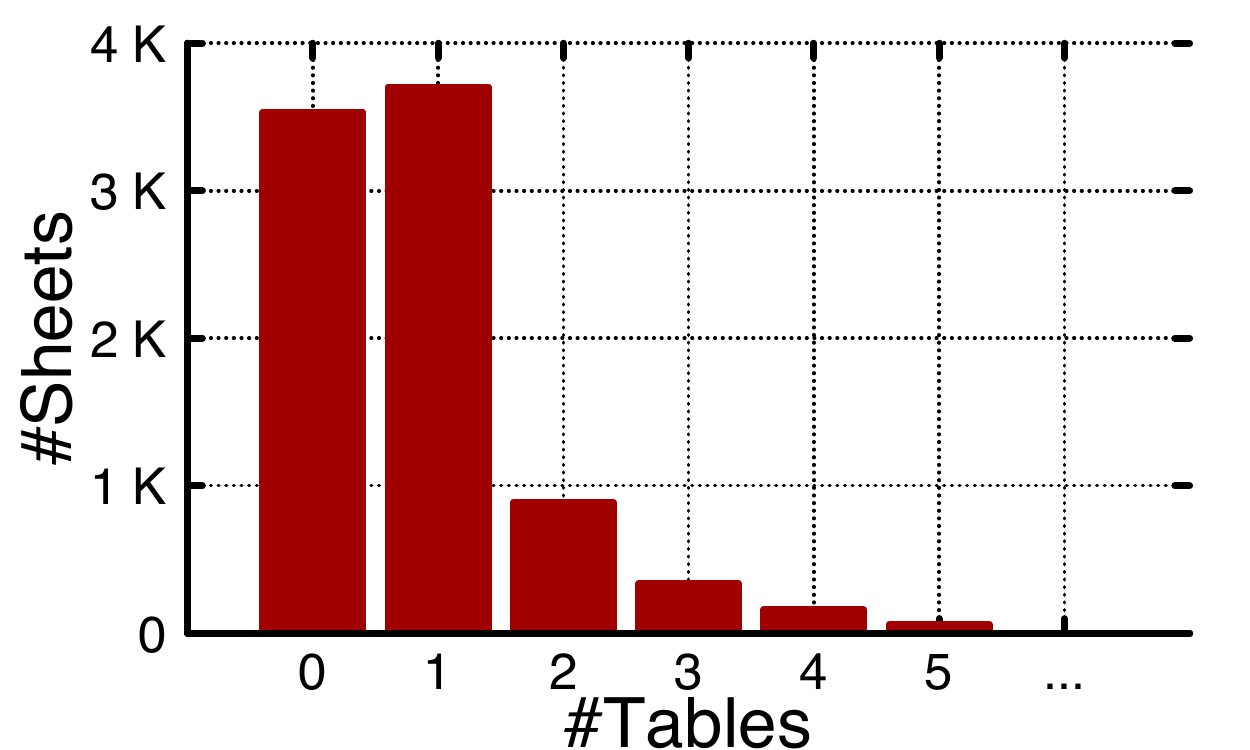}}
 	\hspace{5pt}
 	\subfloat{}{\includegraphics[width=0.23\textwidth,clip]{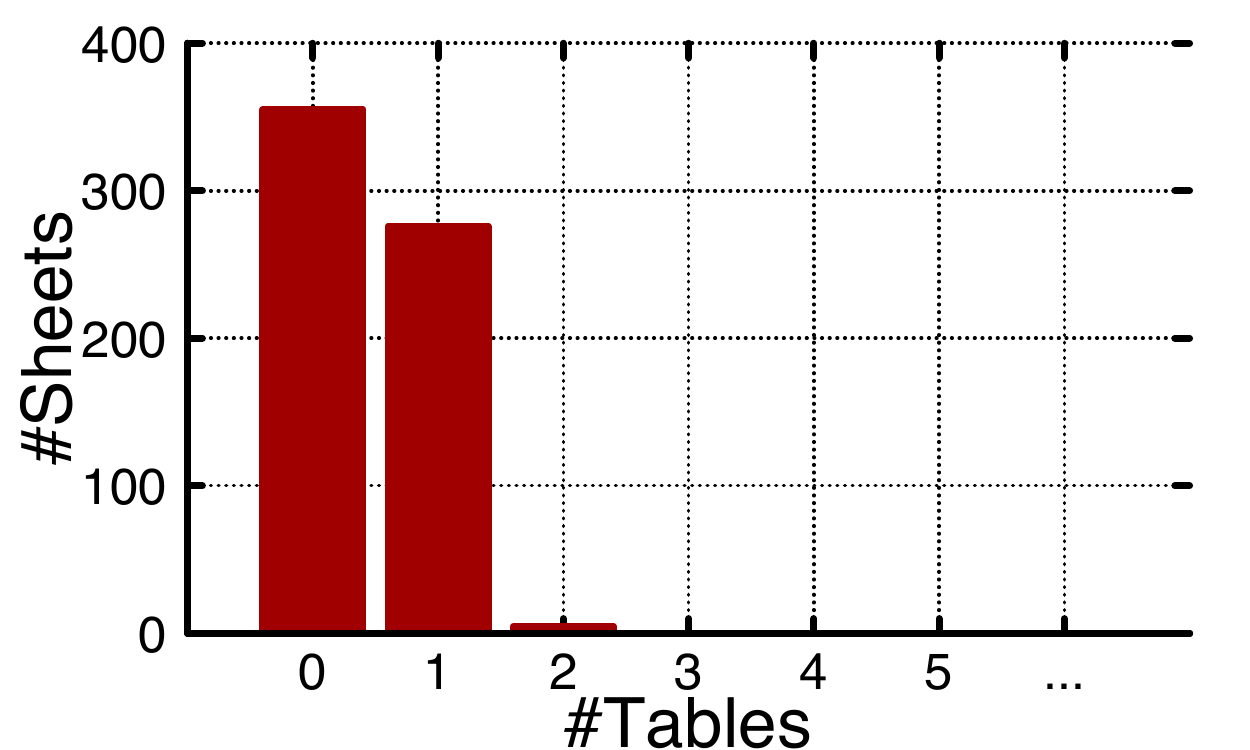}}
 	\caption{Tabular Region Distribution - (a) Internet. (b) ClueWeb09. (c) Enron. (d) Academic.}
 	\label{fig:table_dist}
 	\vspace{-8pt}
 \end{figure*}}

\tr{
\begin{figure*}[t]
	\centering
	\subfloat{}{\includegraphics[width=0.23\textwidth,clip]{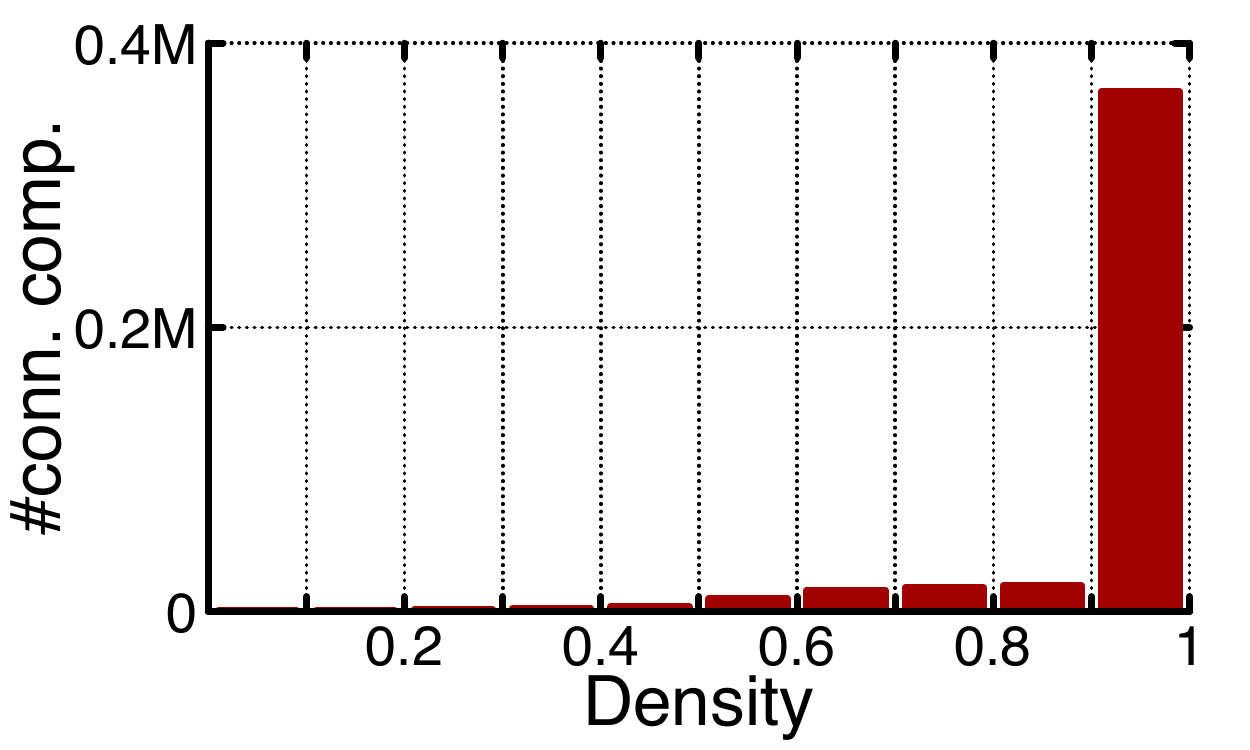}}
	\hspace{5pt}
	\subfloat{}{\includegraphics[width=0.23\textwidth,clip]{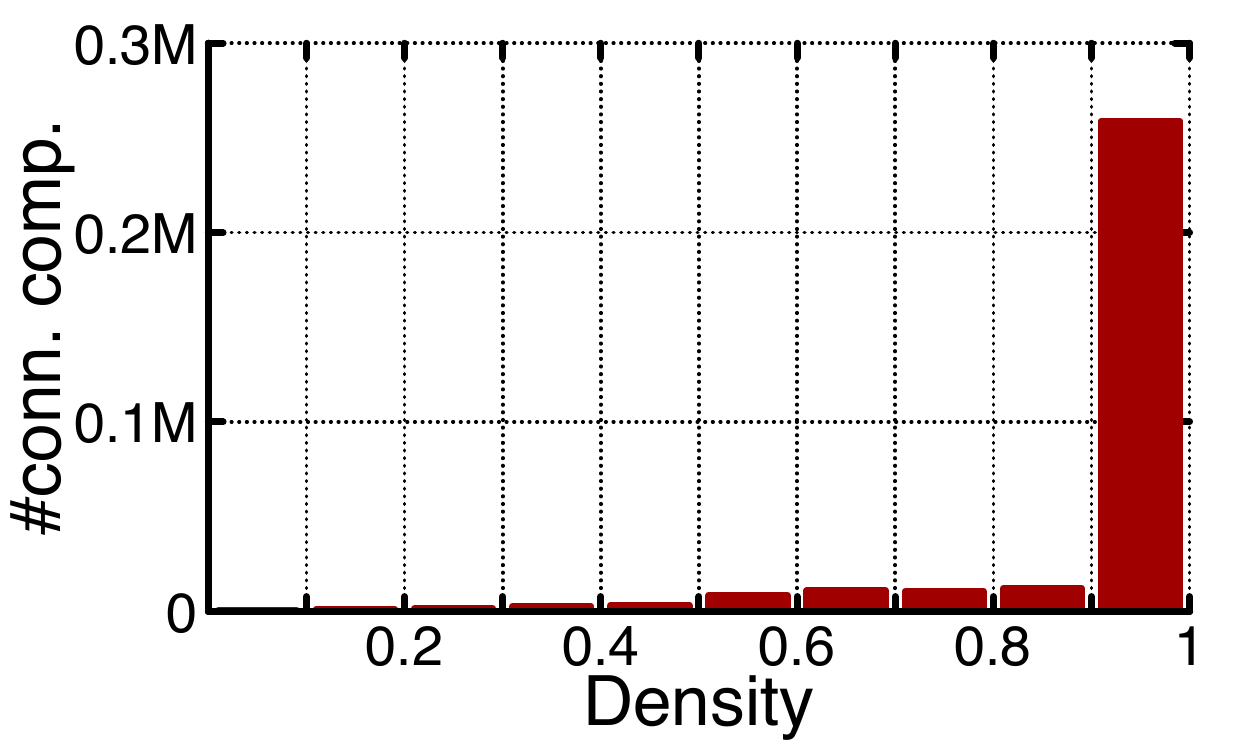}}
	\hspace{5pt}
	\subfloat{}{\includegraphics[width=0.23\textwidth,clip]{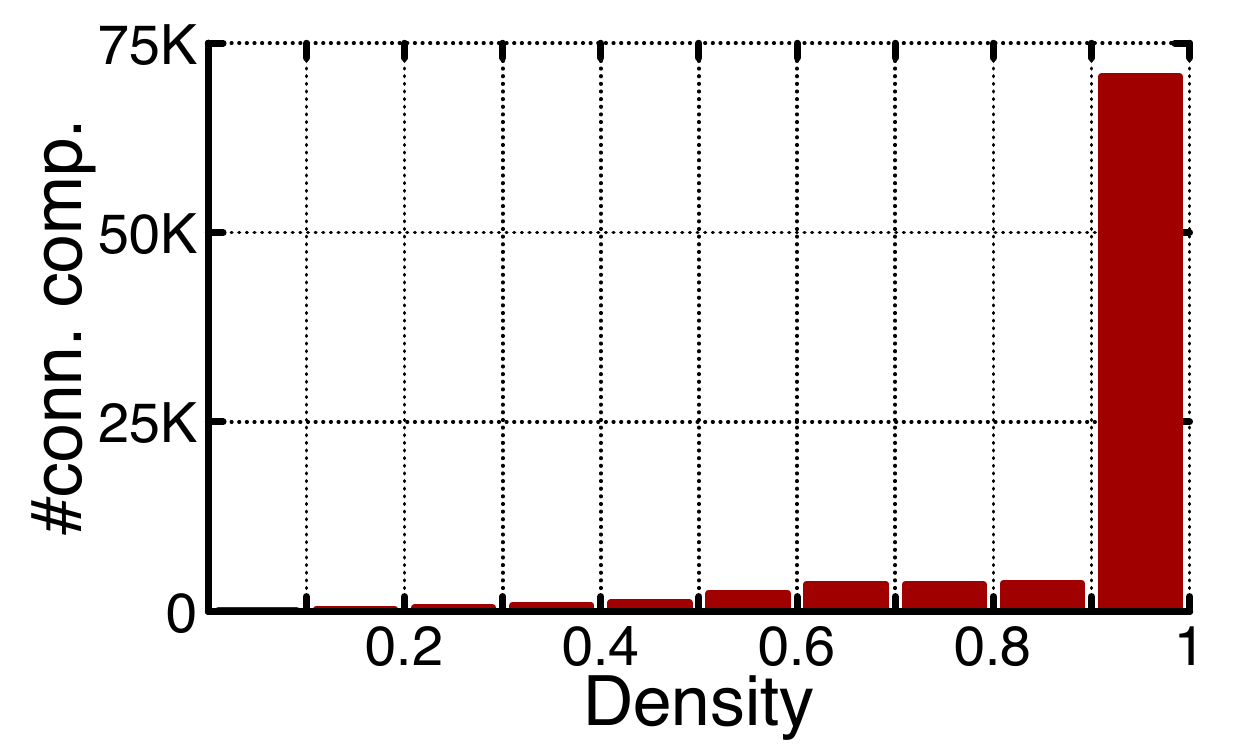}}
	\hspace{5pt}
	\subfloat{}{\includegraphics[width=0.23\textwidth,clip]{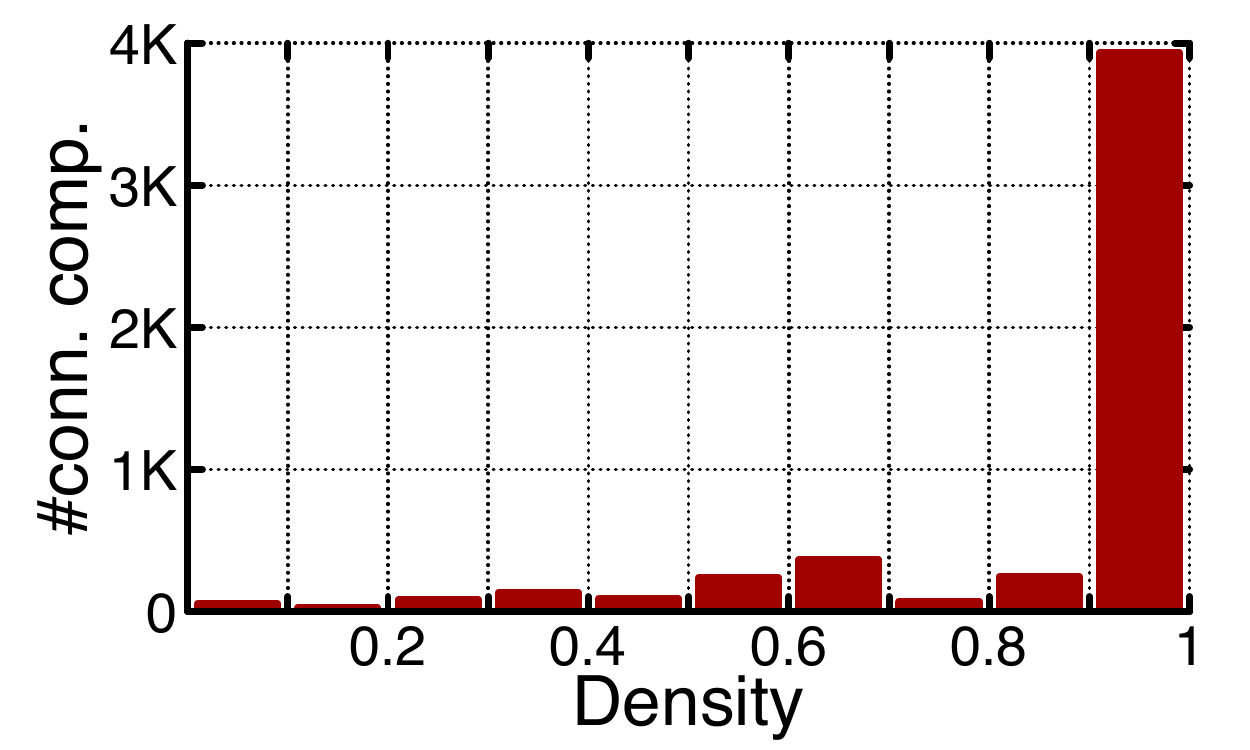}}
	\vspace{-4pt}
	\caption{Connected Component Data Density --- (a) Internet. (b) ClueWeb09. (c) Enron. (d) Academic.}
	\label{fig:ccdensity}
	\vspace{-8pt}
\end{figure*}
}

\cut{
\begin{figure*}[t]
	\centering
	\subfloat{}{\fbox{\includegraphics[width=0.22\textwidth,clip]{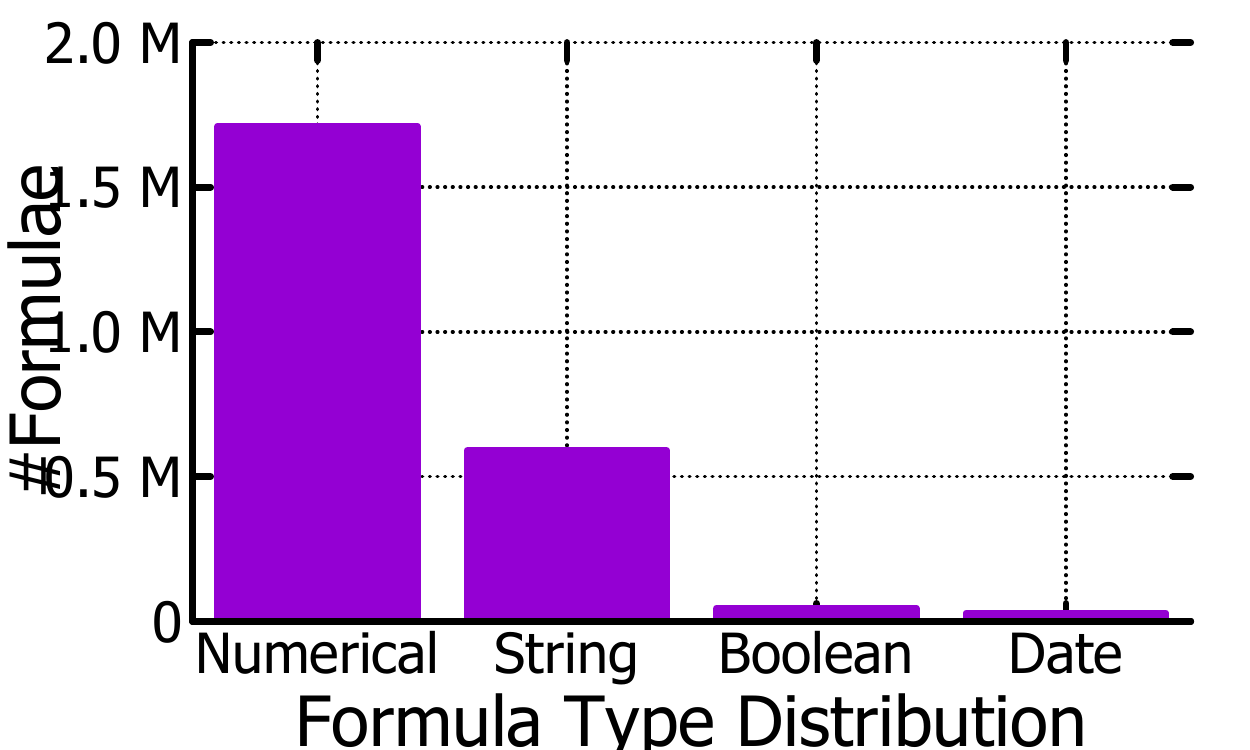}}}
	\hspace{5pt}
	\subfloat{}{\fbox{\includegraphics[width=0.22\textwidth,clip]{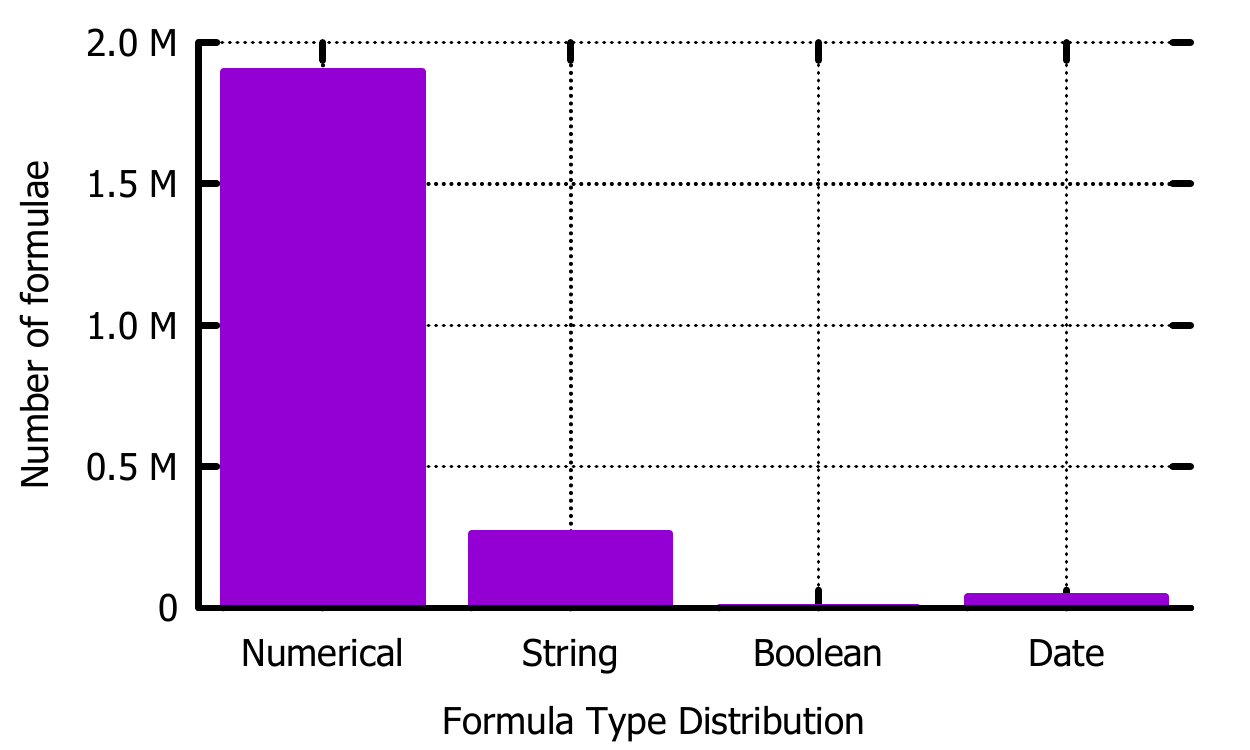}}}
	\hspace{5pt}
	\subfloat{}{\fbox{\includegraphics[width=0.22\textwidth,clip]{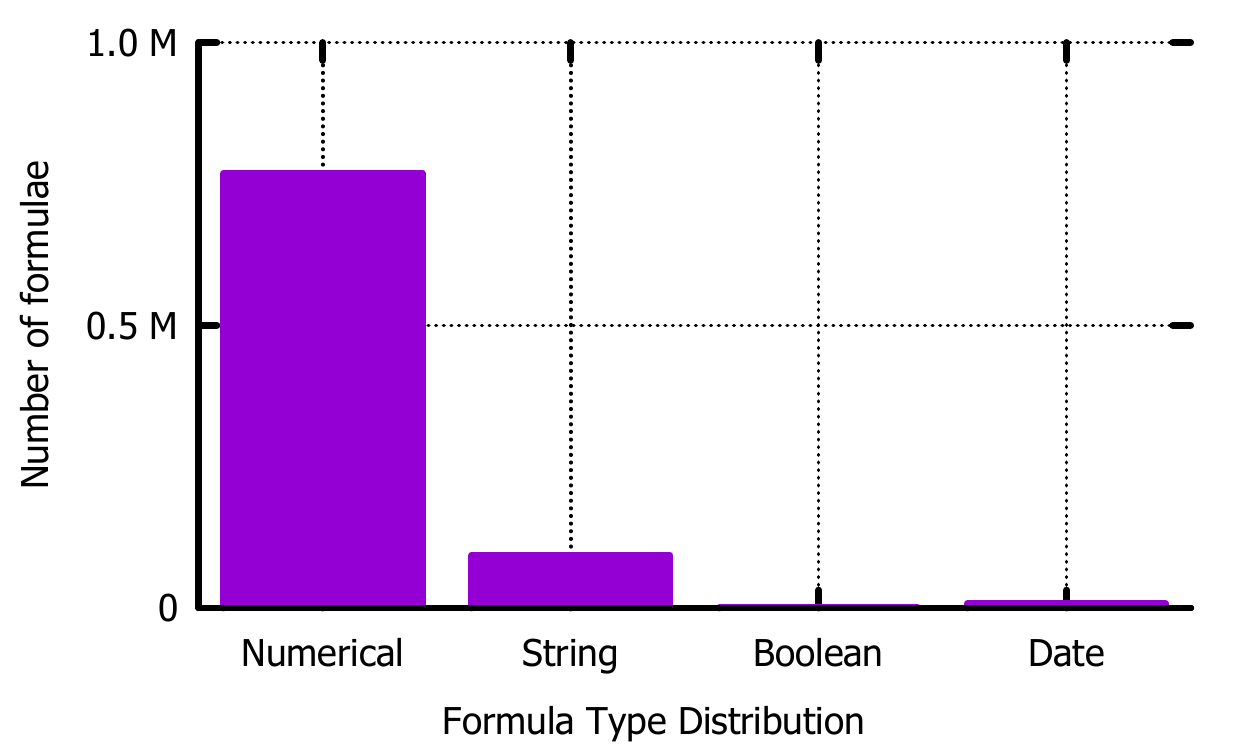}}}
	\hspace{5pt}
	\subfloat{}{\fbox{\includegraphics[width=0.22\textwidth,clip]{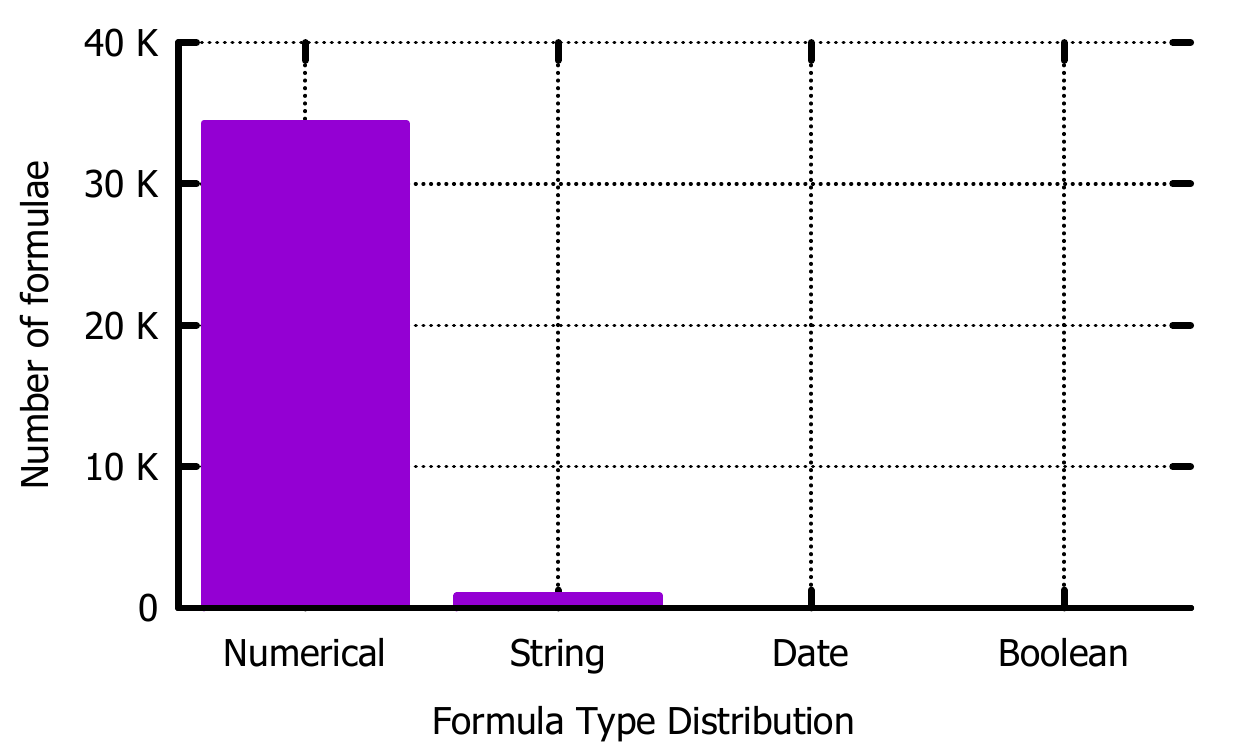}}}
	\vspace{-4pt}
	\caption{Formulae Type Distribution. - (a) Internet. (b) ClueWeb09. (c) Enron. (d) Academic.}
	\label{fig:formaula_type}
	\vspace{-12pt}
\end{figure*}}

\tr{
\begin{figure*}[t]
	\centering
	\subfloat{}{\includegraphics[width=0.23\textwidth,clip]{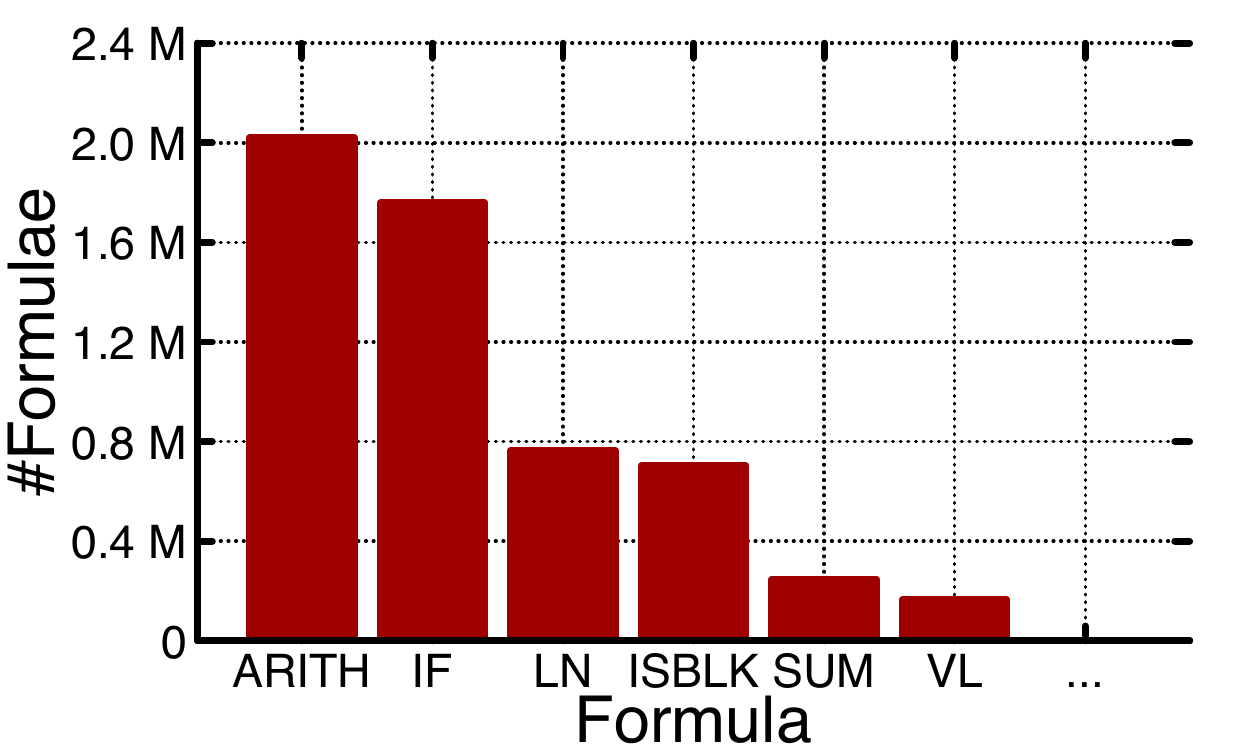}}
	\hspace{5pt}
	\subfloat{}{\includegraphics[width=0.23\textwidth,clip]{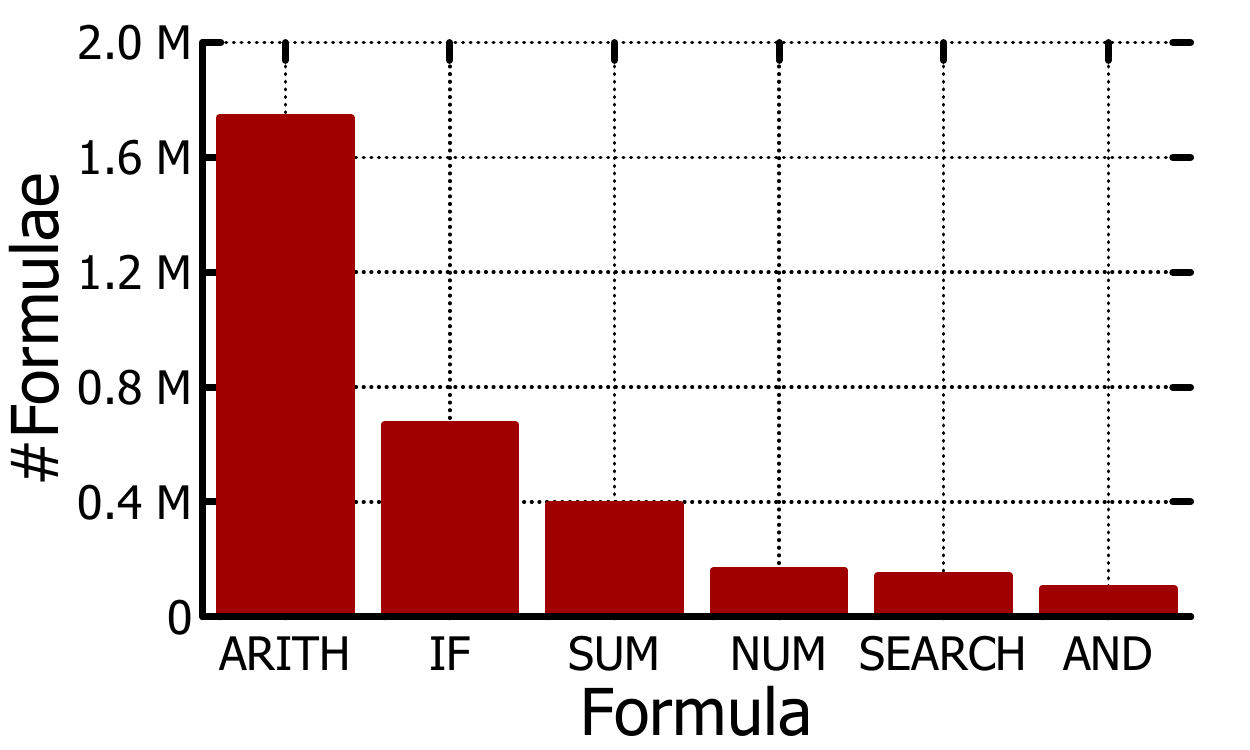}}
	\hspace{5pt}
	\subfloat{}{\includegraphics[width=0.23\textwidth,clip]{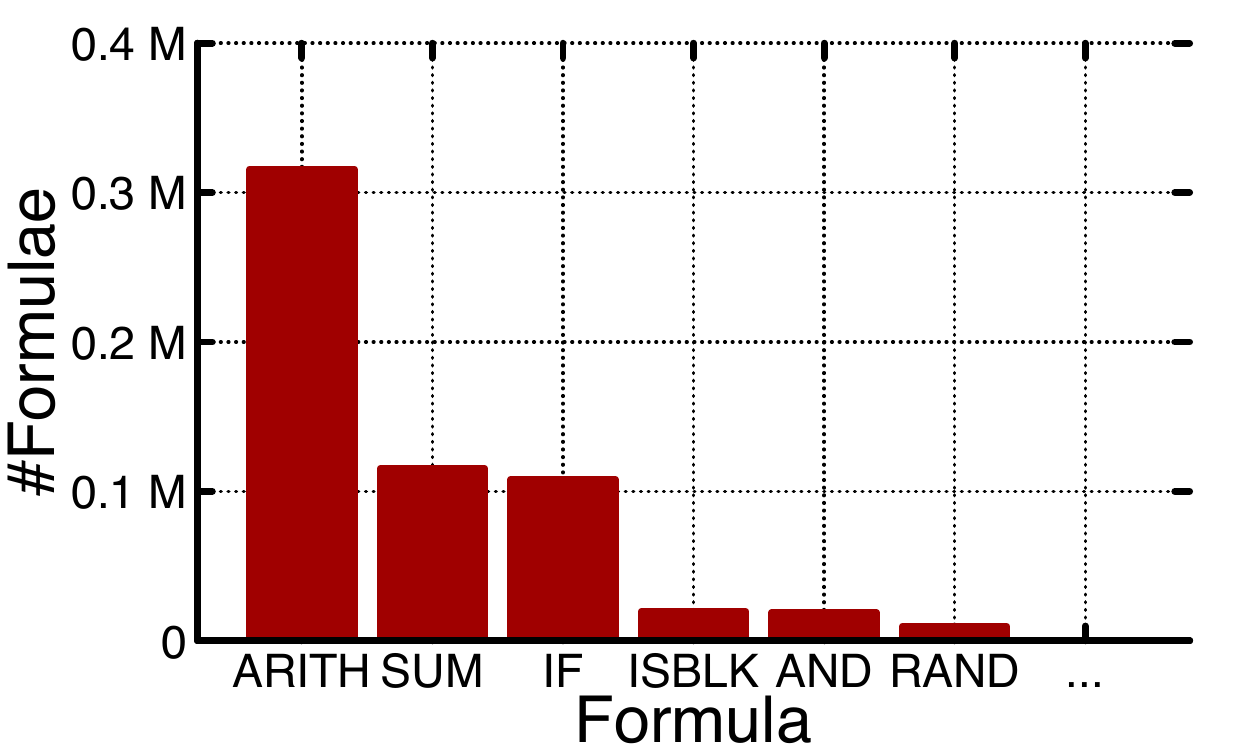}}
	\hspace{5pt}
	\subfloat{}{\includegraphics[width=0.23\textwidth,clip]{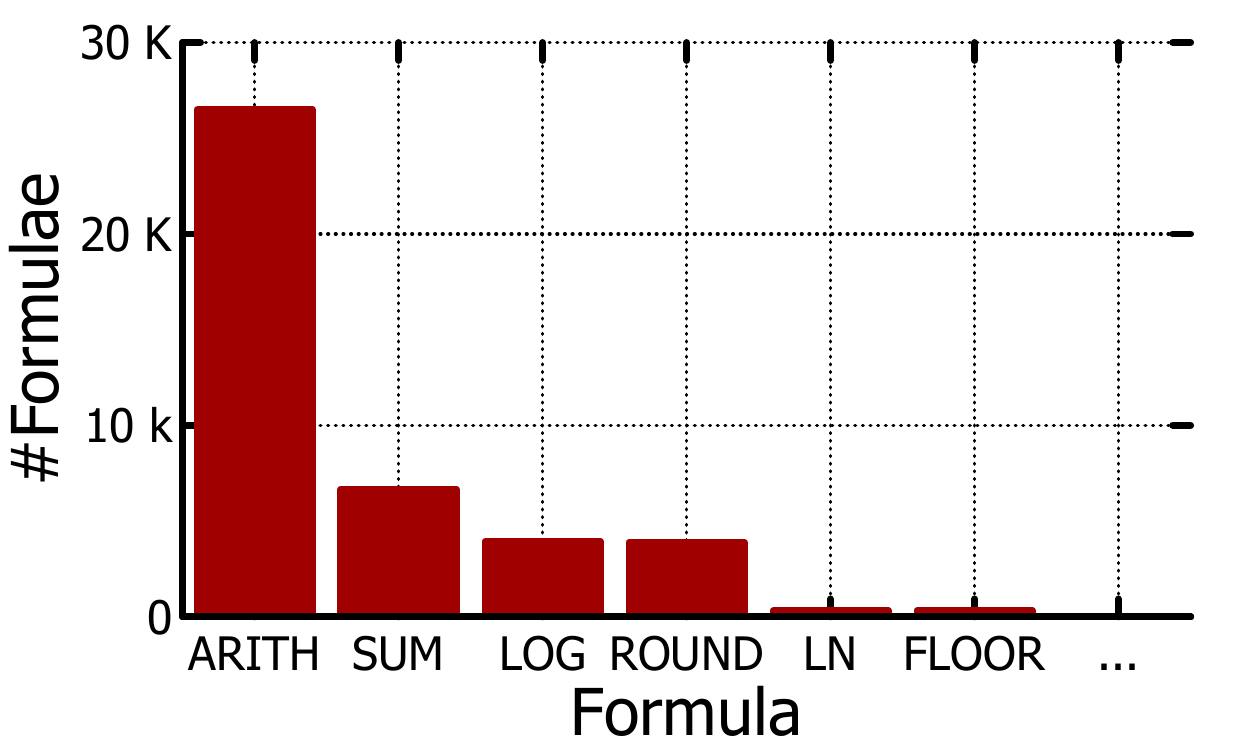}}
	\vspace{-4pt}
	\caption{Formulae Distribution --- (a) Internet. (b) ClueWeb09. (c) Enron. (d) Academic.}
	\label{fig:formula_dist}
	\vspace{-8pt}
\end{figure*}}

\section{Understanding Current Solutions for PDM}
\label{sec:empirical_study}

We now perform an empirical study to characterize
the functional requirements for a storage engine for PDM.
We hypothesize that understanding the present-day use of spreadsheets
for managing data is a suitable proxy for understanding 
the requirements of PDM;
of course, since current spreadsheets are limited
to those that can be manipulated in current software, they
are smaller than what we aim to support in \system.


We focus on two aspects: 
\begin{paraenum}
\item  identifying how users {\em structure}
data on the interface, and 
\item understanding common interface {\em operations}.
\end{paraenum} 
To do so, we first 
retrieve spreadsheets
from four sources and quantitatively analyze
them on different metrics. 
We supplement this analysis with a small-scale
user survey to understand the spectrum of operations frequently performed.
The latter is necessary since we do not have a readily available
trace of user operations
(\eg how often do users add rows).

We first describe our methodology for both these evaluations,
before diving into our findings.



\subsection{Methodology}
As described above, we have two forms of evaluation.

\subsubsection{Real Spreadsheet Datasets}
\label{sec:real_datasets}
\noindent For our evaluation of real spreadsheets,
we assemble the following four datasets\tr{
from a wide variety of sources}.

\stitle{Internet.} 
This dataset of 53k spreadsheets was generated by using Bing to search for \code{.xls} files, using a variety of keywords. 
\tr{As a result, these 53k spreadsheets vary widely
in content, ranging from tabular data to images.} 

\stitle{ClueWeb09.}
This dataset of 26k spreadsheets was generated by extracting \code{.xls} file URLs from the 
ClueWeb09~\cite{callan2009clueweb09} crawl.

\stitle{Enron.}
This dataset was generated by extracting  18k spreadsheets from the Enron email 
dataset~\cite{klimt2004introducing}.
These spreadsheets were used to exchange data within the Enron corporation.

\stitle{Academic.} 
This dataset was collected from an academic institution using spreadsheets to manage administrative data. 

\smallskip
\noindent
We list these four datasets in Table~\ref{tab:spreadsheet-datasets}.
The first two datasets are primarily meant
for {\em data publication}: thus, only about 29\% and 42\% of these sheets
(column 3) contain formulae, with the formulae occupying less than 3\% of
the total number of non-empty cells for both datasets (column 5). 
The third dataset is primarily meant for email-based
{\em data exchange}, with a similarly low fraction of 39\% of these sheets
containing formulae, and 3.35\% of the non-empty cells containing formulae.
The fourth dataset is primarily meant
for {\em data analysis}, with a high fraction of 91\% of the sheets
containing formulae, and 23.26\% of the non-empty cells containing
formulae.

\subsubsection{User Survey}
\noindent To evaluate the kinds of operations performed on spreadsheets, 
we solicited 30 participants from industry who exclusively used spreadsheets for data management,
for a qualitative user survey\paper{; these participants answered an online form}.
\tr{This survey was conducted via an online form, with the participants answering
a small number of multiple-choice and free-form questions,
followed by the authors aggregating the responses.}

\subsection{Structure Evaluation}
\noindent We begin by asking: {\em how do users structure data in PDM?}
Is the data typically organized and 
structured into tables, or is it largely unstructured? 
\tr{Does the type of structure depend on the intended use-case?}

\stitle{Across Spreadsheets: Data Density.} To evaluate whether
real spreadsheets are similar to structured relational data, we first
we estimate the
{\em density} of each sheet,
defined as the ratio of the number of filled-in cells
to the number of cells within the minimum bounding rectangular box enclosing the filled-in cells.
We depict the results in the last two columns of 
Table~\ref{tab:spreadsheet-datasets}: 
the spreadsheets in Internet, Clueweb09, and Enron are typically {\em dense}, 
\ie more than $50\%$ of the spreadsheets have density greater than $0.5$. 
On the other hand, for Academic, a high proportion (greater than $60\%$) of the spreadsheets have 
density values less than $0.2$.
This low density is because the latter dataset embeds many formulae and uses 
forms to report data in a user-accessible interface.

\ta{Presentational Awareness}{Structure in PDM can vary widely, 
from highly sparse to highly dense, necessitating data 
models that can adapt to such variations.}

\stitle{Within a Spreadsheet: Tabular regions.}
We further analyzed the sparse spreadsheets to 
evaluate whether there are regions within 
them with high density---essentially indicating
that these are structured tabular regions.
To do so, we first
constructed a graph of the filled-in cells within each spreadsheet,
where two cells (\ie nodes) have an edge between them if they are adjacent.
We then computed the connected components of this graph.
We declare a connected component to be a \emph{tabular region}
if it spans at least two columns and five rows,
and has a density of at least $0.7$,
defined as before.
In Table~\ref{tab:spreadsheet-datasets}, for each dataset,
we list the total number of identified tabular regions (column~8)
and the fraction of the total filled-in cells that 
are captured within these tabular regions (column~9).
\tr{In Figure~\ref{fig:table_dist} we plot the distribution of tables across our datasets.}
For Internet, ClueWeb09, and Enron,
we observe that greater than $60\%$ of the cells are part of tabular regions. 
For Academic, where the sheets are rather 
sparse, there still are a modest number of regions that are tabular (286 across 636 sheets). 

\tr{
We next characterize the connected components by understanding 
how they conform to a tabular structure. 
To study this, we estimate the
{\em density} of each connected component,
defined as the ratio of the number of filled-in cells
to the number of cells within the minimum bounding rectangular box enclosing the connected component.
Figure~\ref{fig:ccdensity} depicts the density distribution 
of connected components.
We note that
across all the four data sets the connected components are  {\em very dense}, specifically more than $80\%$ of the spreadsheets have density greater than $0.8$. 
}

\cut{
\begin{table}[!h]
\vspace{-5pt}
	\centering
	\scriptsize
	\begin{tabular}{|l|r|r|}
		\hline
		\bf{Dataset} & \multicolumn{1}{c|}{\bf{Tabular Regions}} &  \multicolumn{1}{c|}{\bf{\%Coverage}}  \\\hline
		Internet Crawl & 67,374  & 66.03 \\\hline
		ClueWeb09      & 37,164  & 67.68 \\\hline
		Enron          &  9,733  & 60.98 \\\hline
		Academic       &    286  & 12.10 \\\hline
	\end{tabular}
	\caption{Tabular Regions in Spreadsheets.}
	\label{tab:tabular_regions}    
	\vspace{-20pt}
\end{table}}

\ta{Presentational Awareness}{Even within a single spreadsheet, 
there is often high skew, with areas of high and low 
density, indicating the need for fine-grained data models that can treat these regions
differently.}


\subsection{Operation Evaluation}
We now ask: {\em What kinds of operations do
users naturally perform in PDM?} 
How often do users employ data manipulation operations? Or analysis operations, \eg formulae?
How do users refer to the portions of data in the operations?

\stitle{Popularity: Formulae Usage.}
Formulae use is common, but there is a high variance in the fraction of 
cells that are formulae (see column 5 in Table~\ref{tab:spreadsheet-datasets}), ranging from 1.3\% to 23.26\%.
We note that Academic embeds a high fraction of formulae since their spreadsheets are used primarily for
data management as opposed to sharing or publication.
Despite that, all of the datasets have a
substantial fraction of spreadsheets where the formulae 
occupy more than 20\% of the cells (column 4)---20.26\% 
and higher for all datasets.

\ta{Presentational Access}{Formulae are very common,
with over 20\% of the spreadsheets containing
a significant fraction of over $\frac{1}{5}$ of formulae.
Optimizing for the access patterns of 
formulae when developing data models is crucial.
}





\tr{
\stitle{Formulae Distribution and Patterns.} 
Next, we study the distribution of formulae used
within spreadsheets\tr{---see Figure~\ref{fig:formula_dist}}.
Not surprisingly, arithmetic operations (\code{ARITH}, \code{LN}, \code{SUM}) are very common,
along with conditional formulae (\code{IF}, \code{ISBLK}).
Overall, there is a wide variety of formulae that span both a small number of cell
accesses (\eg arithmetic), as well as a large number of them (\eg \code{SUM}, \code{VL} short for \code{VLOOKUP}).
Moreover, these formulae typically access a small number of rectangular region, \ie an area defined by a set of contiguous rows and columns, at a time (column 11). 
Many of the formulae used ended up reproducing relational
operations (\eg \code{VLOOKUP} for joins). 
}

\paper{
\stitle{Formulae Distribution and Patterns.} 
When we consider the distribution of formulae, and the 
access patterns thereof, we find  
that there is a wide variety of formulae that span both a small number of cell
accesses, \eg arithmetic, as well as a large number of them, \eg \code{SUM} (refer to column~10 of Table~\ref{tab:spreadsheet-datasets}).
Moreover, these formulae typically access a small number of rectangular regions, \ie an area defined by a set of contiguous rows and columns, at a time (column~11).
Many of the formulae used ended up reproducing relational
operations (\eg \code{VLOOKUP} for joins). 
We present more details in our extended technical report~\cite{techreport}.
}

\cut{
\begin{table}[!h]
	\centering
	\scriptsize
	\vspace{-5pt}
	\begin{tabular}{|l|r|r|}
		\hline
		\bf{Dataset}  &  \multicolumn{1}{c|}{\bf{Cells accessed}} & \multicolumn{1}{c|}{\bf{Connected Components}} \\
	    & \multicolumn{1}{c|}{\bf{per Formula}}  & \multicolumn{1}{c|}{\bf{per Formula}} \\\hline
		Internet  & 334.26 & 2.5  \\\hline
		ClueWeb09 & 147.99 & 1.92 \\\hline
		Enron     & 143.05 & 1.75 \\\hline
		Academic  &   3.03 & 1.54 \\\hline
	\end{tabular}
	\caption{Cells accessed by formulae.}
	\vspace{-20pt}
	\label{tab:accessed_cells}    
\end{table} }

\tr{
To gain a better understanding of how
much effort is necessary to execute these formulae,
we measure the number of cells accessed by each formula.
Then, we tabulate the average number of
cells accesses per formula in column 10 of Table~\ref{tab:spreadsheet-datasets}
for each dataset.
As we can see in the table, the average number of cells
accesses per formula is not small---with up to
300+ cells per formula for Internet,
and about 140+ cells per formula for Enron and
ClueWeb09.
Academic
has a smaller average number---many of these
formulae correspond to derived columns that
access a small number of cells at a time.
Next, we check if the accesses
made by these formulae were spread across the spreadsheet,
or could exploit spatial locality.
To measure this, we considered the set of cells
accessed by each formula, and then generated
the corresponding graph of these accessed cells 
as described in the previous
subsection for computing the number of tabular regions. 
We then counted the number of connected components, shown in column~$11$.
Even though the number of cells
accessed may be large, these cells stem from a small
number of connected components; as a result, we can 
exploit spatial locality to execute them more efficiently.}

\ta{Presentational Access and Awareness}{Formulae on spreadsheets 
access cells on the spreadsheet
by position; some common formulae such as \code{SUM} or \code{VLOOKUP} access a rectangular 
range of cells at a time.
The number of cells accessed by these formulae can be quite large,
and most of these cells stem from contiguous areas of the spreadsheet.}

\tr{
\begin{figure}[t]
\vspace{-10pt}
	\centering
	\includegraphics[scale=0.40]{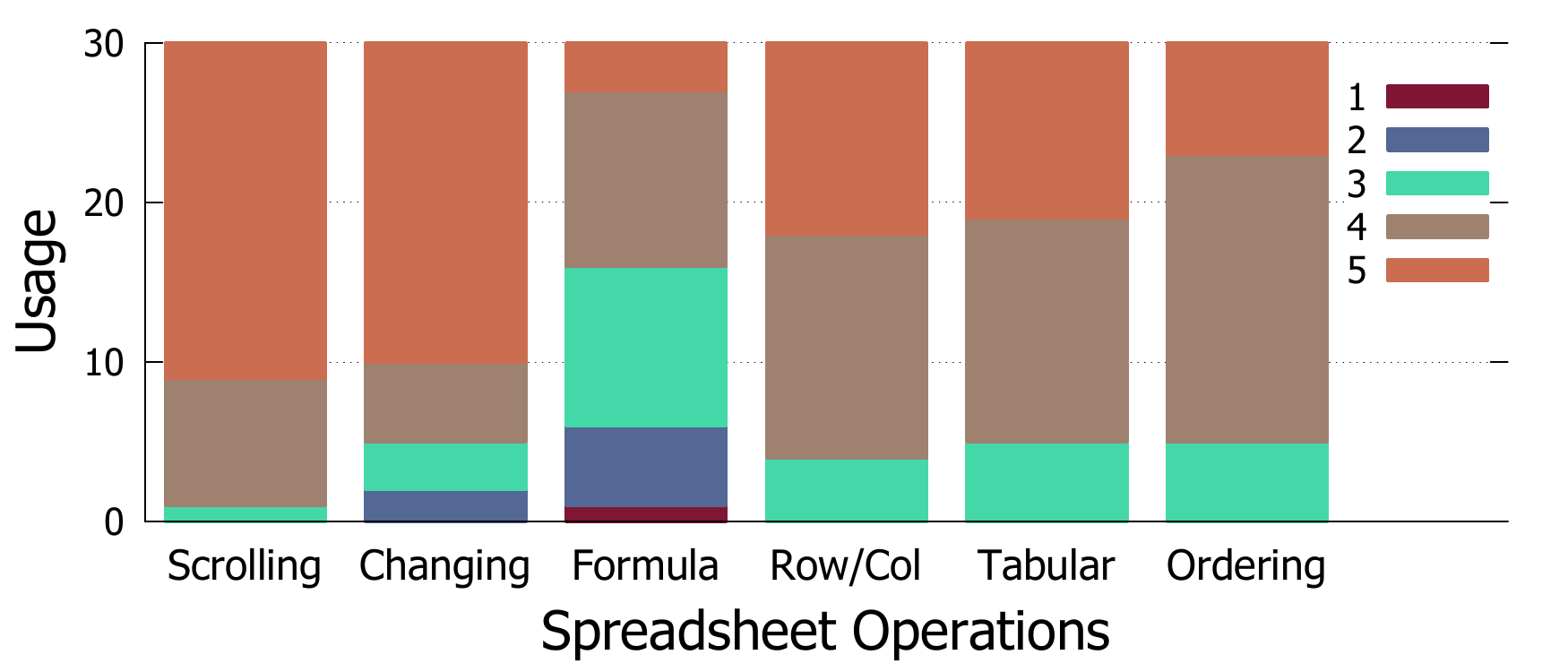}
	\vspace{-4pt}
	\caption{Operations performed on spreadsheets.}
	\label{fig:user_operations}
	\vspace{-15pt}
\end{figure}
}

\paper{
\stitle{User-Identified Operations.} Via a small-scale survey, 
we also identified common spreadsheet operations performed by users (beyond those
directly visible within spreadsheet datasets).
Detailed results can be found in our technical report~\cite{techreport}.
\ta{Presentational Access and Awareness}{There are several common interface operations 
performed by spreadsheet
users, including scrolling, row and column modification, and editing individual cells, all requiring access by position.}
}

\tr{
\stitle{User-Identified Operations.} We now analyze the common spreadsheet operations performed by users via a small-scale online survey of $30$ participants.
This
qualitative study is valuable since real spreadsheets
do not reveal traces of user operations. Our questions in this study were targeted 
at understanding
\begin{paraenum}
\item how users perform operations on spreadsheets and 
\item how users organize data on spreadsheets.
\end{paraenum}

We asked each participant to answer 
a series of questions where each question corresponded 
to whether they conducted the specific operation
under consideration on a scale of 1--5, where 1 corresponds
to ``never'' and 5 to ``frequently''.
For each operation, we plotted the results in a stacked bar chart in 
Figure~\ref{fig:user_operations}, with the higher numbers
stacked on the smaller ones.

We find that all the thirty participants perform {\em scrolling}, \ie
moving up and down the spreadsheet to examine the data, with 22 of them
marking 5 (column~1).
All participants reported to have performed editing of individual cells (column~2),
and many of them reported to have performed formula evaluation frequently (column~3).
Only four of the participants marked $< 4$ for some form of 
{\em row/column-level operations},
\ie deleting or adding one or more rows or columns at a time (column~4).

\ta{Presentational Access and Awareness}{There are several common operations performed by spreadsheet
users including scrolling, row and column modification, and editing individual cells.}
}

\tr{
Our second goal for performing the study was to understand 
how users organize their data. We asked each participant
if their data is organized in well-structured tables, or if the data 
scattered throughout the spreadsheet, on a scale of 1 (not organized)--5 (highly organized)---see Figure~\ref{fig:user_operations}. 
Only five participants marked $< 4$ which indicates that users do
organize their data on a spreadsheet (column 5). We also asked the importance of 
ordering of records in the spreadsheet on a scale of 1 (not important)--5 (highly important). Unsurprisingly, only five participants
marked $< 4$ for this question (column 6). We also provided a free-form textual
input where multiple participants mentioned that ordering comes naturally to them and is
often taken for granted while using spreadsheets.

\ta{Presentational Awareness}{Spreadsheet users typically try to organize their data as far as possible
on the spreadsheet, and rely heavily on the ordering and presentation of the data on
their spreadsheets.}
}


\section{Data Presentation Manager}
\label{sec:interface_model}

Given our findings on presentational awareness and access, 
we now abstract out
the functional requirements of the {\em data presentation manager},
the storage engine for PDM. 
We abstract out the presentational interface of a spreadsheet,
as a conceptual data model,
as well as the operations supported on it; concrete implementations
will be described in subsequent sections.
We will then describe our \system prototype to place these requirements
in context. 

\begin{figure}[hbt]
\centering
\includegraphics[scale=0.35]{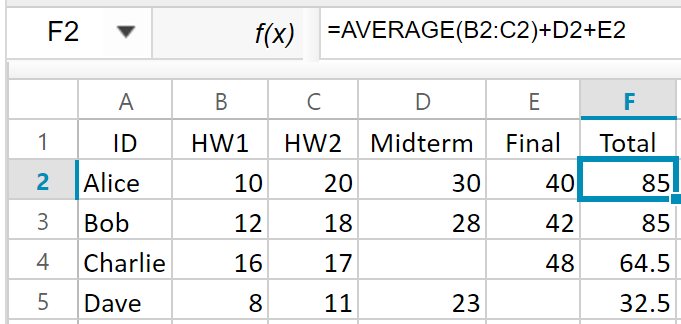}
\caption{Sample Spreadsheet (\system screenshot).}
\vspace{-12pt}
\label{fig:sample_spreadsheet}
\end{figure}

\stitle{Conceptual Data Model.}
A spreadsheet
consists of a collection of {\em cells}, referenced
by two dimensions: 
row and column.
Columns are referenced using letters \code{A}, $\ldots$, 
\code{Z}, \code{AA}, $\ldots$; while
rows are referenced using numbers \code{1},  $\ldots$
Each cell contains a {\em value},
or {\em formula}.
A value is a constant; \eg in Figure~\ref{fig:sample_spreadsheet}
(a \system screenshot), 
\code{B2} (column \code{B}, row \code{2}) 
contains the value \code{10}.
\tr{ 
In contrast, a formula is a mathematical expression 
that contains values 
and/or cell references as arguments, 
to be manipulated by operators or functions. 
For example, in Figure~\ref{fig:sample_spreadsheet}, cell \code{F2} contains the formula \code{=AVERAGE(B2:C2)+D2+E2}, which unrolls into the value \code{85}. 
In addition to a value or a formula, a cell could also
additionally have formatting associated with it; \eg
width, or font.
For simplicity, we ignore formatting aspects,
but these aspects can be easily captured 
without significant changes.}
\paper{
In contrast, a formula, \eg \code{F2}, is an expression 
operating on values 
and/or cell references. 
A cell could
additionally have formatting associated;
for simplicity, we ignore formatting,
but these aspects can be easily captured.
}

\stitle{Spreadsheet-Oriented Operations.}
We now describe the spreadsheet-like operations necessary for PDM, 
drawing from our survey (takeaway~3-5).

\emtitle{1. Retrieving a Range.} Our most basic read-only operation is 
\code{getCells(range)}, where
we retrieve a rectangular range of cells. This operation
is relevant in {\em scrolling}, where the user moves to a specific position and we need to retrieve the rectangular range of cells
visible at that position, \eg range \code{A1:F5}, 
is visible in Figure~\ref{fig:sample_spreadsheet}.
Similarly, {\em formula evaluation} also accesses one or more 
ranges of cells.

\emtitle{2. Updating an existing cell:}
The operation \code{updateCell(row, column, value)} corresponds to 
modifying the value of a cell. 


\emtitle{3. Inserting/Deleting row/column(s):} 
This operation corresponds to inserting/deleting row/column(s) 
at a specific position, followed by shifting subsequent row/column(s) appropriately:
\begin{paraenum}
\item \code{insertRowAfter(row)}
\item \code{insertColumnAfter(column)}
\item \code{deleteRow(row)}
\item \code{deleteColumn(column)}.
\end{paraenum}

\stitle{Database-Oriented Operations.}
We now describe the database-like operations for PDM,
enabling users to effectively use the interface to manage
and interact with database tables.
\cut{
Thus, a user can use traditional spreadsheet operations, 
\eg updating a cell's value, to perform table operations, \eg updating a tuple's attribute value.}

\emtitle{1. Link an existing table/Create a new table:}
This operation enables users to \emph{link} a region on a spreadsheet with an existing database relation,
establishing  
a two way correspondence 
between the 
spreadsheet interface and
the underlying table, such that any operations on the spreadsheet interface are translated by the data presentation manager 
into table operations on the linked table. 
Thus, a user can use traditional spreadsheet operations such as updating a cell's value to update a database table.
We introduce the function:
\code{linkTable(range, tableName)}.
If \code{tableName} does not exist, it will be created in the database, and then linked
to the spreadsheet interface. 

\emtitle{2. Relational Operators:}
Users can interact with the linked tables as well as other tabular regions
via relational operators, as well as SQL,
using the following spreadsheet functions: \code{
union,
difference,
intersection,
crossproduct,
join,
filter,
project,
rename,} and
\code{sql}.
These functions return a single composite table value;
to retrieve the individual rows and columns within that table value,
we have an \code{index(cell, i, j)} function that looks up the \code{(i, j)}th
row and column in the composite table value in location \code{cell}, and places it
in the current location.



\paper{
\stitle{Prototype Implementation.} 
We have implemented a fully functional \system prototype 
 as a web-based tool (using the open-source ZK Spreadsheet frontend~\cite{zkspread}) on top of a PostgreSQL database. 
The prototype along with its source code, documentation, and user guide can be found at \code{http://dataspread.github.io}.
Along with standard spreadsheet features, 
the prototype supports all the aforementioned spreadsheet-like and database-like operations. 
The prototype also addresses a 
number of additional challenges especially related to efficient formula evaluation, including
compact representation of formula dependencies, batched and lazy formula evaluation,
and optimizing for user attention---these are important issues, but
beyond the scope of this paper.
Screenshots of \system in action can be found in our qualitative evaluation
section (Section~\ref{sec:qual_eval}).
Additional architectural details can be found in our technical report~\cite{techreport}.}

\smallskip
\noindent Given the functional requirements for our data presentation manager,
in Section~\ref{sec:spreadsheet_representation},
we develop concrete mechanisms for representing our conceptual data model 
in a database back-end,
and in Section~\ref{sec:positional_mapping}, we develop data structures 
that enable efficient access in the presence of updates. 

\section{Presentational Awareness} \label{sec:spreadsheet_representation}
 
We now describe the high-level problem of representation
of spreadsheet data within a database; we will concretize
this problem later.
We focus
on one spreadsheet, but our techniques
seamlessly carry over to the multiple spreadsheet case.

\begin{figure*}
\centering
\scriptsize
\setlength\tabcolsep{3.5pt}
\vspace{-25pt}
	\begin{tabular}{|c|c|c|c|} 
		\multicolumn{4}{c}{} \\ \hline
		\textit{RowID} &  $Col_1$ & ... & $Col_6$\\ \hline \hline
		1 & ID, NULL & ... & Total, NULL \\ \hline
		2 & Alice, NULL & ... & 85, \textsf{AVERAGE(B2:C2)+D2+E2}\\ \hline
		... & ... & ... & ...\\ \hline
	\end{tabular}
	\hspace{10pt}
 	\begin{tabular}{|c|c|c|c|} 
 		\multicolumn{4}{c}{} \\ \hline
 		\textit{ColID} &  $Row_1$ & ... & $Row_5$\\ \hline \hline
 		1 & ID,NULL & ... & Dave,NULL \\ \hline
 		2 & HW1,NULL & ... & 8,NULL \\ \hline
 		... & ... & ... & ...\\ \hline
 	\end{tabular}
 	\hspace{10pt}
	\begin{tabular}{|c|c|c|} 
 		\multicolumn{3}{c}{} \\ \hline 
 		\textit{RowID} &  \textit{ColID} & Value\\ \hline \hline
 		1 & 1 & ID, NULL\\ \hline
 		... & ... & ..., ...\\ \hline
 		2 & 6 & 85, \textsf{AVERAGE(B2:C2)+D2+E2}\\ \hline
 		... & ... & ..., ...\\ \hline
 	\end{tabular}
 	\vspace{-5pt}
 	\caption{(a) Row-Oriented Model (b) Column-Oriented Model (c) Row-Column-Value Model for Figure~\ref{fig:sample_spreadsheet}.\label{fig:models}}
\end{figure*}

\subsection{High-level Problem Description}

The conceptual data model corresponds to a collection
of cells, represented as $C = \{C_1, C_2,\hdots,C_m\}$;
as described previously, each cell $C_i$
corresponds to a location (\ie a specific row and column),
and has some contents---either a value or a formula.
Our goal is to represent and store  $C$,
via one of the {\em physical data models}, $\datamodel$.
Each $T \in \datamodel$
corresponds to a collection of relational tables
$\set{T_1, \ldots, T_p}$.
Each table $T_i$ records the data in 
a certain portion of the spreadsheet. 
Given $C$,
a physical data model $T$ is said to be {\em recoverable} with respect to $C$
if for each $C_i \in C, \exists$ precisely one $T_j \in T$ such
that $T_j$ records the data in $C_i$.
Our goal is to identify physical 
data models that are recoverable. 

At the same time, we want to minimize
the amount of \emph{storage} required to 
record $T$, \ie
we would like to minimize $\size{T} = \sum_{i=1}^p \size{T_i}$.
Moreover, we would like to minimize the
time taken for accessing data using $T$, \ie the {\em access cost},
which is the cost of accessing a rectangular range of cells
for formulae (takeaway~4)
or scrolling (takeaway~5), both common operations.
And we would like to minimize the time taken to perform updates, \ie the {\em update cost},
which is the cost of updating cells,
and the insertion and deletion of rows and columns. 
\boxy{
\noindent \small 
Given a collection of cells $C$,
our goal is to identify a physical data model $T$
such that:
\begin{paraenum}
    \item $T$ is recoverable with respect to $C$, and
    \item $T$ minimizes a combination of storage, access, and update costs, among 
    all $T \in \datamodel$. 
\end{paraenum}
}
\noindent We begin by considering 
the setting where the physical data model $T$ 
has a single relational table, \ie $T = \set{T_1}$.
We develop three ways of representing this table---we call them {\em primitive data models}---drawn from prior work,
each of which works well for a specific
structure of spreadsheet (Section~\ref{sec:primitive_datamodels}).
Then, we extend this to the setting
where $|T| >1$ by defining
 {\em hybrid data models}
with multiple tables each of which uses
one of the three primitive data models
to represent a certain portion of the spreadsheet (Section~\ref{sec:hybrid_data_model}).
\tr{Given the high diversity of 
structure within spreadsheets and high skew
(takeaway 2), having multiple primitive
data models, and the ability to use
multiple tables, gives us substantial presentational awareness.}


\subsection{Primitive Data Models} \label{sec:primitive_datamodels}

Our primitive data models
represent trivial solutions for spreadsheet representation
with a single table.
Before we describe these data models,
we discuss a small wrinkle that affects all of these models.
To capture a cell's position
we need to
record a row and column number with each cell. 
Say we use an attribute to capture the row number for a cell.
Then, any insertion or deletion of rows
requires cascading updates to the row number attribute
for cells in all subsequent rows.
As it turns out,
all of the data models we describe here suffer from performance issues
arising from cascading updates, but the solution
to deal with this issue is similar for all of them, and will be described in 
Section~\ref{sec:positional_mapping}.
Thus, we focus here on \emph{storage} and \emph{access cost}.
Also, note that the access and update cost
of data models depends on whether
the underlying database is a row or a columnar store.
We use a row store, which 
our \system implementation employs, and
is
suitable for a hybrid read-write
setting. We now describe the primitive data models:

\stitle{Row-Oriented Model (ROM).}
The row-oriented data model 
is akin to the traditional
relational data model.
We represent each row from the sheet
as a separate tuple, with an 
attribute for each column \attname{$Col1$}, $\ldots$, \attname{$Col\cmax$},
where \attname{$Col\cmax$} is the largest non-empty column, 
and an additional attribute for explicitly capturing the row number, \ie \attname{$RowID$}.
The schema for ROM is:
\tabname{ROM}(\underline{\attname{$RowID$}}, \attname{$Col1$}, $\ldots$, \attname{$Col\cmax$})---we illustrate 
the ROM representation of Figure~\ref{fig:sample_spreadsheet} 
in Figure~\ref{fig:models}(a):
each entry is a pair corresponding to a value 
and a formula, if 
any.
For dense spreadsheets that are tabular (takeaways 1 and 2),
this data model can be quite efficient in storage 
and access, since 
each row number 
is recorded
only once, independent of the number of columns.
Overall, ROM shines
when entire rows are accessed at a time. 
\tr{It is also efficient for accessing a large range of cells
at a time.}


\stitle{Column-Oriented Model (COM).}
The second representation is the transpose of ROM.
Often, we find that certain spreadsheets have many columns
and relatively few rows, necessitating such a representation.
The schema for COM is: \tabname{COM}(\underline{\attname{$ColID$}}, \attname{$Row1$}, $\ldots$, \attname{$Row\rmax$}).
Figure~\ref{fig:models}(b) illustrates the COM representation of Figure~\ref{fig:sample_spreadsheet}.
Like ROM, COM shines for dense data;
while ROM shines for row-oriented operations, COM 
shines for column-oriented operations.
 

\stitle{Row-Column-Value Model (RCV).} 
The Row-Column-Value Model is inspired
by key-value stores, where
the Row-Column number  pair is treated as the key.
The schema for RCV is
\tabname{RCV}(\underline{\attname{$RowID$}}, \underline{\attname{$ColID$}}, \attname{$Value$}).
The RCV representation for Figure~\ref{fig:sample_spreadsheet} is provided in Figure~\ref{fig:models}(c).
For sparse spreadsheets often found in practice
(takeaway 1 and 2), this model
is quite efficient in storage and access since it
records only the filled in cells,
but for dense spreadsheets, it incurs the additional
cost of recording and retrieving the row and column numbers
for each cell as compared to ROM and COM\tr{, and has 
a much larger number of tuples}. 
RCV is also efficient when it comes
to retrieving specific cells at a time.

\stitle{Why these Data Models?}
Readers may be wondering  why 
we chose these data models (ROM, COM and RCV).
As it turns out, these data models represent
extremes in a space of data models that we
identify and refer to as {\em rectangular data models}.
We can further demonstrate that these three models
do not dominate each other, \ie there are settings
where each of them prevails and are {\em optimal}
within the space of rectangular data models.
\paper{See our technical report~\cite{techreport} for details.}
\tr{We refer the reader to Appendix~\ref{app:primitive_datamodels_optimality} for details.}

\stitle{Database-Linked Tables.}
Such tables are not represented using primitive data models,
instead stored as-is in the database.
We refer to this as a Table-Oriented Model (TOM). 
Our \mbox{\code{linkTable}} operation sets up a two-way synchronization between a database table and a spreadsheet table.


\subsection{Hybrid Data Model: Intractability}
\label{sec:hybrid_data_model}

\begin{figure}
	\centering
    \vspace{-10pt}
	\includegraphics[width=0.5\columnwidth]{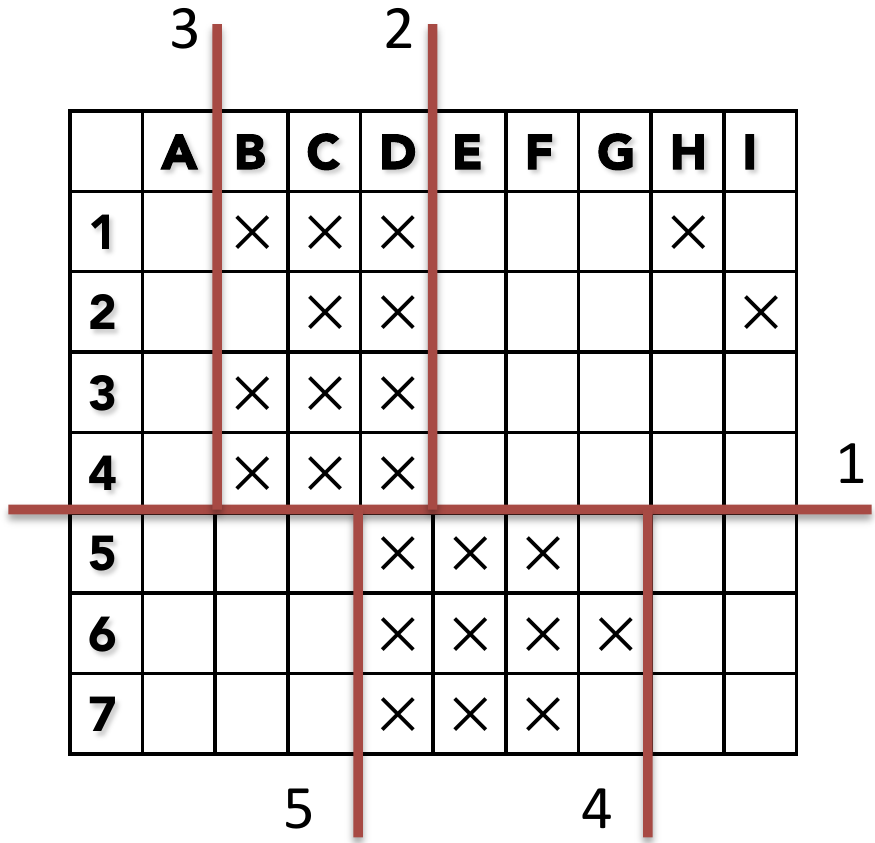}
	\caption{Hybrid Data Model: Recursive Decomposition.}
	\vspace{-15pt}
	\label{fig:hybrid}
\end{figure}

\tr{So far, we developed the
primitive data models to represent a spreadsheet using a single table in 
a database.}
We now develop better solutions by decomposing
a spreadsheet into multiple regions, each
represented by one of the primitive data models.
We call these {\em hybrid data models}.

\vspace{-5pt}
\begin{definition}[Hybrid Data Models]
Given a collection of cells $C$,
we define {\em hybrid data models} as the space of physical
data models that are formed using a collection of tables $T$
such that $T$ is recoverable with respect to $C$,
and further, each $T_i \in T$ is either a ROM, COM, RCV, or a TOM table.
\end{definition}
\vspace{-5pt}
As an example, for the spreadsheet in Figure~\ref{fig:hybrid}, 
we might want the
dense areas, \ie 
\code{B1:D4} and \code{D5:G7}, 
represented via a ROM table each
and the remaining area,
specifically, \code{H1} and \code{I2}
to be represented by an RCV table.

\stitle{Cost Model.}
As discussed earlier, the storage, access, and
update costs  impact our choice of data model.
For this section, we focus
exclusively on storage. 
\tr{We will generalize
to access cost in Appendix~\ref{app:extensions}.}
\paper{
Access cost is analogous and can be found in our
technical report~\cite{techreport}.
}
The update cost will be the focus of the next section.
Furthermore, we begin with ROM;
we will generalize to RCV and COM in Section~\ref{sec:extensions}.

Given a hybrid data model $T = \set{T_1, \ldots, T_p}$,
where each ROM table $T_i$ has $r_i$ rows and $c_i$ columns, the
cost of $T$ is
\vspace{-5pt}
\begin{align}
\cost{T} = &\sum_{i=1}^p s_1 + s_2\cdot (r_i\times c_i) + s_3\cdot c_i + s_4\cdot r_i.\label{eq:table-rom}
\vspace{-5pt}
\end{align}
Here, the constant $s_1$ is the cost of initializing a
new table, while
the constant $s_2$ is the cost of storing each individual cell (empty or not)
in the ROM table.
The non-empty cells that have content may require additional storage;
however, this is a constant cost that does
not depend on the data model.
The constant $s_3$ is the cost corresponding to each column,
while $s_4$ is the cost corresponding to each row.
The former is necessary to record schema information per column,
while the latter is necessary to record the row information in the \attname{$RowID$} attribute.
Overall, while the specific costs $s_i$ may differ 
across databases, what is clear is that all of
these costs matter. 

\stitle{Formal Problem.} 
We now state our formal problem below. 
\vspace{-5pt}
\begin{problem}[Hybrid-ROM]\label{prob:hybrid-rom}
Given a spreadsheet with a collection of cells $C$,
identify the hybrid data model $T$ with only ROM tables
that minimizes $\cost{T}$.
\end{problem}
\vspace{-5pt}

\tr{
Unfortunately, Problem~\ref{prob:hybrid-rom} is {\sc NP-Hard},
via a reduction from the minimum edge length 
partitioning of rectilinear polygons problem~\cite{Lingas198253} 
of finding a partitioning of a polygon whose edges are aligned to the X and Y axes,
into rectangles, 
while minimizing the total perimeter; see Appendix~\ref{sec:np-hard}.}

\paper{
Unfortunately, Problem~\ref{prob:hybrid-rom} is {\sc NP-Hard},
via a reduction from {\em minimum edge length 
partitioning of rectilinear polygons}~\cite{Lingas198253}: finding a partitioning of a polygon whose edges are aligned to the X and Y axes,
into rectangles, 
while minimizing the total perimeter; see our technical report~\cite{techreport}.
}

\vspace{-5pt}
\begin{theorem}[Intractability]\label{thm:hybrid-rom}
Problem~\ref{prob:hybrid-rom} is {\sc NP-Hard}.
\end{theorem}
\vspace{-5pt}

\subsection{Optimal Recursive Decomposition}
\label{sec:optimal-recursive-decomposition}
Instead of directly solving Problem 1, which is intractable,
we instead aim to make it tractable, by reducing the search space of solutions.
In particular, we focus on hybrid data models that can be obtained by {\em recursive decomposition}.
Recursive decomposition is a process where we repeatedly subdivide
the spreadsheet area from $[1 \ldots \rmax, 1 \ldots \cmax]$
by using a vertical cut between two columns or a horizontal cut between two rows,
and then recurse on the resulting areas. 
As an example, in Figure~\ref{fig:hybrid},
we can make a cut along line 1 horizontally, 
giving us two regions from rows 1 to 4
and rows 5 to 6.
We can then cut the top portion along line 2 vertically, followed
by line 3, separating out one table \code{B1:D4}.
By cutting the bottom portion along line 4 and line 5, we can separate
out the table \code{D5:G7}.
Further cuts can help us carve out tables out of \code{H1} or \code{I2},
not depicted here.

As the example illustrates, recursive decomposition is very powerful,
since it captures a broad space of hybrid models;
basically, anything that can be obtained via recursive cuts along the $x$ and $y$ axis.
Unfortunately, there is an \emph{exponential number of such models}. 
Now, a natural question is: what sorts of hybrid data models cannot be composed
via recursive decomposition?

\begin{figure}
	\centering
	\vspace{-10pt}
	\includegraphics[width=0.4\columnwidth]{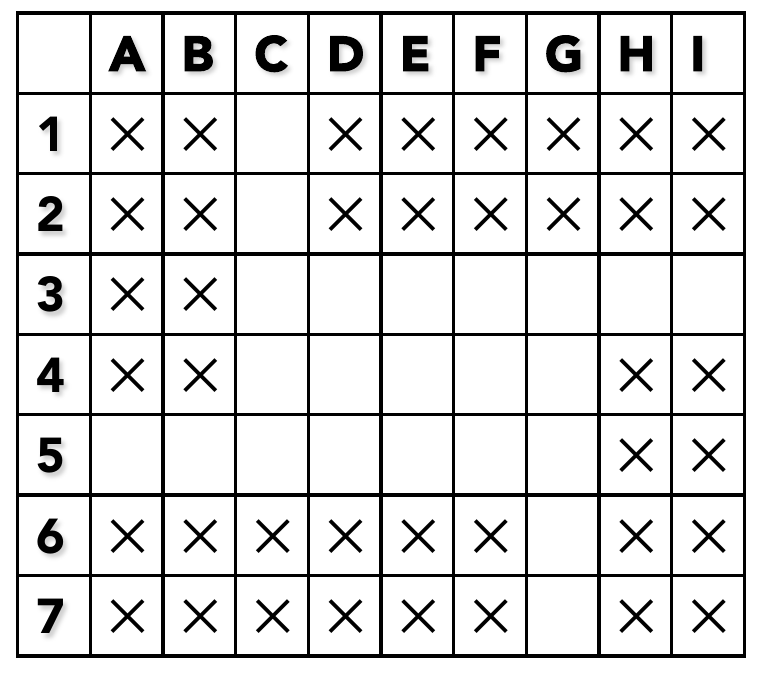}
	\hspace{10pt}
	\includegraphics[width=0.3\columnwidth]{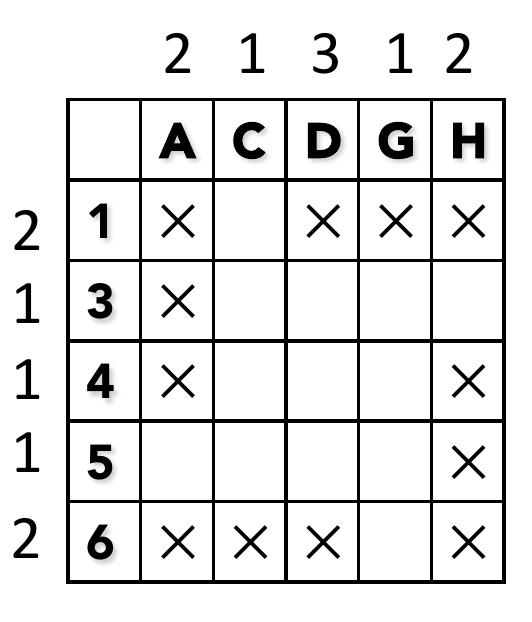}
	\vspace{-5pt}
     \caption{(a) Counterexample. (b) Weighted Representation.}
     \vspace{-15pt}
  \label{fig:dp-counter-example}
\end{figure}

\vspace{-5pt}	
\begin{observation}[Counterexample]
In Figure~\ref{fig:dp-counter-example}(a), the tables: \code{A1:B4}, \code{D1:I2},
\code{A6:F7}, and \code{H4:I7} cannot be obtained via recursive decomposition.
\end{observation}
\vspace{-5pt}	
To see this, note that any vertical or horizontal cut that one would make at the start
would cut through one of the four tables, making the decomposition impossible.
Nevertheless,
we expect this form of construction to not be frequent, whereby
the hybrid data models obtained via recursive decomposition
form a natural class of data models. 

Despite the space of recursively decomposed hybrid data models being exponential,  
as it turns out, identifying the optimal data model in this space 
to Problem~\ref{prob:hybrid-rom}
is {\sc PTIME}.
We use dynamic programming; our algorithm makes the most optimal ``cut'' 
horizontally or vertically at every step, and proceeds recursively; details below.

Consider a rectangular area formed from $x_1$ to $x_2$ 
as the top and bottom row numbers respectively, both inclusive,
and from $y_1$ to $y_2$ as
the left and right column numbers respectively, both inclusive, for some $x_1, x_2, y_1, y_2$.
Now, the optimal cost of representing this rectangular area,
\ie $\Opt{(x_1, y_1), (x_2, y_2)}$, is the minimum of the following
possibilities:
\squishlist           
	\item If there is no filled cell in the area $(x_1, y_1), (x_2, y_2)$, then we do not use any data model, and the cost is $0$.
    \item Do not split, \ie store as a ROM model ($\romCost{}$):
    \vspace{-10pt}
    \begin{multline}
    \romCost{(x_1, y_1), (x_2, y_2)} = s_1 + s_2\cdot (r_{12}\times c_{12}) \\ + s_3\cdot c_{12} + s_4\cdot r_{12},\label{eq:romcost}
    \end{multline}
    where number of rows $r_{12} = (x_2-x_1+1)$, and the number of columns $c_{12} = (y_2-y_1+1)$.
    \item Perform a horizontal cut ($C_H$):
    \vspace{-10pt}
    \begin{multline}    
    C_H = \min_{i\in \{x_1,\hdots,x_2\}} \!\!\!\!\! \Opt{(x_1, y_1), (i, y_2)} \\ +  \Opt{(i+1, y_1), (x_2, y_2)}\label{eq:horiz-cut}
    \end{multline}
    \item Perform a vertical cut ($C_V$):
    \vspace{-10pt}
    \begin{multline}        
    C_V = \min_{j\in \{y_1,\hdots,y_2\}} \!\!\!\!\! \Opt{(x_1, y_1), (x_2, j)} \\ + \Opt{(x_1, j+1), (x_2, y_2)}\label{eq:vert-cut}
    \end{multline}        
\squishend
Therefore, when there are filled cells in the rectangle,
\vspace{-10pt}
\begin{multline*}
\Opt{(x_1, y_1), (x_2, y_2)}  = \\ \min\big(\romCost{ \left(x_1, y_1\right), \left(x_2, y_2\right)}, C_H, C_V \big).
\end{multline*}

\tr{
The base case is when the rectangular area is of dimension $1\times 1$. 
Here, we store the area as a ROM table if it is filled. 
Hence, we have, 
$\Opt{(x_1, y_1), (x_1, y_1)} = c_1 + c_2 + c_3 + c_4$, if filled, and $0$ if not.}

\noindent We have the following theorem:
\vspace{-5pt}
\begin{theorem}[Dynamic Programming Optimality]
\label{th:dp_optimality}
For the exponential space of ROM-based hybrid data models based on recursive decomposition, we can obtain the optimal solution via dynamic programming in {\sc PTIME}.
\end{theorem}
\vspace{-5pt}

\stitle{Time Complexity.}
\paper{Our dynamic programming algorithm runs in time \O{n^5}, 
where $n$ is the length of the larger side of the 
minimum enclosing rectangle of the spreadsheet.}
\tr{
Our dynamic programming algorithm runs in polynomial time with respect to the size of the spreadsheet. Let the length of the larger side of the minimum enclosing rectangle of the spreadsheet be of size $n$. Then, the number of candidate rectangles is \O{n^4}. For each rectangle, we have \O{n} ways to perform the cut. Therefore, the running time of our algorithm is \O{n^5}. However, this number could be very large if the spreadsheet is massive--which is typical of the use-cases we aim to tackle.}

\stitle{Approximation Bound.}
Even though our dynamic programming algorithm only identifies the best recursive decomposition based hybrid data model, we can derive a bound for its cost relative to the best hybrid data model overall. 
\tr{We refer the reader to Appendix~\ref{sec:dp-bound} for the proof.}

\vspace{-5pt}
\begin{theorem}[Approximation Bound~\paper{\cite{techreport}}]
\label{th:dp_bound}
Say there are $k$ rectangles in the optimal decomposition with cost $c$. Then, the recursive decomposition algorithm identifies a decomposition with cost at most $c+s_1\times\frac{k(k-1)}{2}$.
\end{theorem}
\vspace{-5pt}
\noindent Typically $k$ is small, so this is a small additive approximation.
\paper{
In our experiments, we compare the solution obtained from recursive decomposition 
with a lower bound of the optimal solution (Section~\ref{sec:exp_impact_hybrid}),
and show that it is near-optimal.} 
\tr{
As observed from Figures~\ref{fig:table_dist} and~\ref{fig:ccdensity}, typically spreadsheets have a small number of highly dense connected components.
By deriving an upper bound for the number of tables in the optimal solution for each connected component, we can get an upper bound for $k$. 
The following theorem implies that the high density of the connected components makes it sub-optimal to split them in the optimal decomposition, thereby suggesting that the number of tables in the optimal decomposition, \ie $k$, is small.

\begin{theorem}[Connected Component Solution]
\label{th:cc_tables}
The optimal solution to Problem~\ref{prob:hybrid-rom} for a minimum bounding rectangle of a connected component will have at most 
$\left\lfloor\frac{e \times s_2}{s_1} + 1 \right\rfloor$ 
rectangles, where $e$ is the number of empty cells in the bounding rectangle.
\end{theorem}

We empirically show in Section~\ref{sec:exp_impact_hybrid} that the upper bound of $k$ is small by obtaining the distribution of $\sum\left\lfloor\frac{e \times s_2}{s_1} + 1 \right\rfloor$ across our datasets. 
Additionally, we compare the solution obtained from our recursive decomposition 
with a lower bound of the optimal solution (Section~\ref{sec:exp_impact_hybrid}),
and demonstrate that it is in fact, near-optimal.
} 

\paper{
\stitle{Weighted Representation.}
Our algorithm, with its \O{n^5} complexity, can be rather expensive.
To reduce the effective
size $n$, we can collapse rows/columns with identical structure down
to a single weighted row/column respectively as in Figure~\ref{fig:dp-counter-example}(b).
Our technical report~\cite{techreport} has more details.
}
\tr{
\stitle{Weighted Representation.}
Notice that in many real spreadsheets, there are many rows
and columns that are very similar to each other in structure,
\ie they have the same set of filled cells. 
We exploit this property to reduce the effective
size $n$ of the spreadsheet.
Essentially, we collapse rows that have identical structure down
to a single weighted row, and similarly collapse columns.
Consider Figure~\ref{fig:dp-counter-example}(b) which shows
the weighted version of Figure~\ref{fig:dp-counter-example}(a).
Here, we can collapse column B down into column A, which is now
associated with weight~2; similarly, we can collapse row~2 into
row~1, which is now associated with weight~2.
\tr{The effective area of the spreadsheet now
 becomes 5$\times$5 as opposed to 7$\times$9.}
Now, we apply the same dynamic programming algorithm to
the weighted representation: in essence, we are avoiding
making cuts ``in-between'' the weighted edges,
thereby reducing the search space.
This does not sacrifice optimality.
\vspace{-5pt}
\begin{theorem}[Weighted Optimality]
The optimal hybrid data model obtained by recursive decomposition
on the weighted spreadsheet is no worse than the optimal hybrid data model
obtained by recursive decomposition on the original spreadsheet.
\end{theorem}
\vspace{-5pt}
}

\subsection{Greedy Decomposition Algorithms}
\label{sec:greedy}

\stitle{Greedy Decomposition.} 
To improve the running time even further, we propose 
a greedy heuristic that avoids the high complexity
but sacrifices somewhat on optimality.
The greedy algorithm essentially repeatedly 
splits the spreadsheet area 
in a top-down manner
identifying the operation that results in the lowest local cost.
We have three alternatives for an area $(x_1, y_1), (x_2, y_2)$:
Either we do not split, with cost 
from Equation~\ref{eq:romcost}, \ie  $\romCost{(x_1, y_1), (x_2, y_2)}$.
Or we split horizontally (vertically), 
with cost $C_H$ ($C_V$)
from Equation~\ref{eq:horiz-cut} (Equation~\ref{eq:vert-cut}), 
but with $\Opt{}$ replaced with $\romCost{}$,
since we are making a locally optimal decision.
The smallest cost decision is followed,
and then we continue recursively decomposing using the same rule
on the new areas, if any.

\stitle{Complexity.}
\paper{This algorithm has a complexity of \O{n^2}.}
\tr{This algorithm has a complexity of \O{n^2},
since each step takes \O{n} and there are \O{n} steps.}
While the greedy algorithm is sub-optimal,
its local decision is {\em optimal assuming the worst case about the decomposed areas}\tr{,
\ie with no further information about the decomposed areas this
is the best decision to make at each step}.

\stitle{Aggressive Greedy Decomposition.}
Since it is based on the worst case, the greedy algorithm may halt prematurely,
even though further decompositions may have helped to
reduce cost.
An alternative is one where we don't stop subdividing,
\ie we always choose to use the best horizontal or vertical
cut, until we end up with rectangular
areas where all of the cells are non-empty.
After this point, we backtrack up the tree 
of decompositions, assembling the best solution 
that was discovered, 
considering whether to not split, or perform a horizontal or vertical split.

\stitle{Complexity.} The aggressive greedy approach also
has complexity \O{n^2}, but takes longer since it considers
a larger space of data models than the greedy approach.

\subsection{Extensions: Models, Costs, Maintenance}\label{sec:extensions}
\tr{In Appendix~\ref{app:extensions},}
\paper{In our technical report~\cite{techreport},} we describe
a number of extensions\tr{to the cost model and the
dynamic programming, greedy, and aggressive greedy algorithms,}
including:
\begin{paraenum}
\item cost model extensions to COM, RCV, and TOM tables;
\item maintaining the decompositions incrementally over time;
\item incorporating access cost along with storage; and
\item incorporating the costs of indexes.
\end{paraenum}

\section{Presentational Access for Updates}\label{sec:positional_mapping}

For all of our data models,
storing the row and/or column numbers 
may result in substantial overheads 
due to cascading updates---this makes
working with large spreadsheets infeasible.
To eliminate
the overhead of cascading updates, we introduce 
{\em positional mapping}.
For our discussion we focus on row numbers; 
the techniques can be analogously applied to columns---we
use the term {\em position} to refer to this number. 
\tr{In addition, row and column numbers can be dealt with independently.}


\stitle{Problem.} We require a data structure to capture a specific ordering 
among the items (here, tuples) and efficiently support: 
\begin{paraenum}
	\item {\em fetch} items based on a position,
	\item {\em insert} items at a position, and
	\item {\em delete} items from a position.
\end{paraenum} 
\paper{The insert and delete operations require updating 
the positions of the subsequent items.}
\tr{The insert and delete operations require updating 
the positions of the subsequent items, \eg inserting an item at the $n$\nth position requires us to first increment by one the positions of all the items that have a position greater than or equal to $n$, and then add the new item at the $n$\nth position.} 
\tr{Due to the interactive nature of \system, our goal is to perform these operations
within a few hundred milliseconds.}

\tr{\begin{table}[htb]
\centering \small
\begin{tabular}{  l | r | r  } 
Operation & \multicolumn{1}{c|}{RCV} &  \multicolumn{1}{c}{ROM}\\ 
\hline
Insert & 87,821 ms & 1,531 ms\\ 
\hline
Fetch & 312 ms &  244 ms\\
\end{tabular}
\caption{The performance  of storing  Position-as-is.} \label{tab:row_num_asis}
\vspace{-8pt}
\end{table}}

\stitle{Na\"ive Solution: Position as-is.}
The simplest approach is to store the position
along with each tuple: this makes fetch efficient 
at the expense of  insert/delete operations. 
With a traditional index, \eg a  \mbox{B+ tree} index, 
the complexity to access
an arbitrary row identified by a position
is \O{\log N}.
On the other hand, insert and delete operations 
require updating the positions
of subsequent tuples. 
These updates also need to be propagated 
in the index, and therefore it results in a 
worst case complexity of \O{N \log N}.
\tr{
To illustrate the impact of these complexities in practice,
in Table~\ref{tab:row_num_asis}, we display
the performance of storing the positions as-is
for two operations---fetch and insert---on 
a spreadsheet containing $10^6$ cells. 
We note that irrespective of the 
data model used, the performance of inserts 
is beyond our acceptable threshold
whereas that of the fetch operation is acceptable. 
}

\cut{
\begin{table}[htb]
\centering \small
\vspace{-5pt}
\begin{tabular}{  l | r | r  } 
\multicolumn{3}{c}{Position as is} \\
\hline
Operation & RCV & ROM\\ 
\hline
Insert & 87,821 & 1,531\\ 
\hline
Fetch & 312 &  244\\
\end{tabular}
\hspace{10pt}
\begin{tabular}{  l | r | r } 
\multicolumn{3}{c}{Positional Mapping} \\
\hline
Operation & RCV & ROM\\ 
\hline
Insert & 9.6 & 1.2\\ 
\hline
Fetch & 30,621 & 273\\
\end{tabular}
\caption{The performance of (in ms) (a) storing  Row Number as is 
(b) \msb{to remove} Monotonic Positional Mapping. } \label{tab:row_num_asis}
\vspace{-5pt}
\end{table}}

\stitle{Hierarchical Positional Mapping.} 
To improve the performance of inserts and deletes 
for ordered items, 
we introduce the idea of \emph{positional
mapping}. 
At its core, the idea is simple:
we do not store positions explicitly but instead obtain them on the fly.
Formally, positional mapping~$\posmapping$ is a bijective function that maintains
the relationship between the position $r$ and tuple pointers $p$,
\ie $\posmapping(r) \rightarrow p$.


\begin{figure}[t]
    \centering
 	\includegraphics[width=0.75\columnwidth]{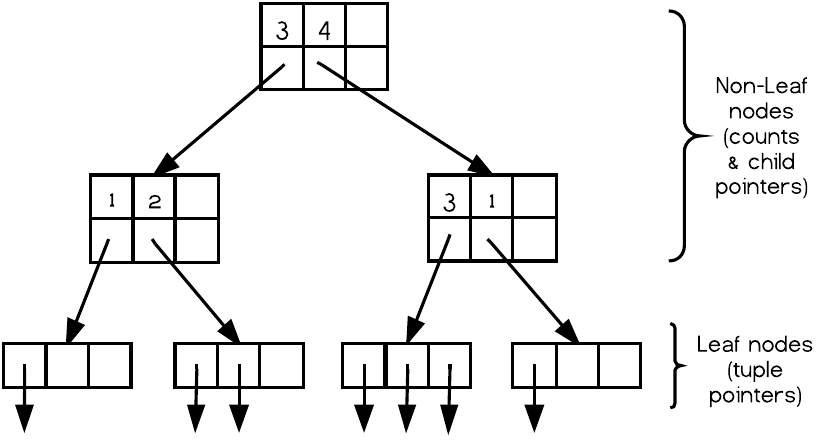}
 	\vspace{-10pt}
 	\caption{ Hierarchical Positional Mapping.}\label{fig:pos_mapping}
 	\vspace{-12pt}
\end{figure}

We now describe {\em hierarchical positional mapping},
which is an indexing structure that
adapts classical work on
{\em order-statistic trees}~\cite{into-algo}.
Just like a typical \mbox{B+ tree} 
is used to capture the mapping from keys to 
records, 
we can use the same structure to map positions 
to tuple pointers. 
Here, instead of storing a key 
we store the count of elements 
stored within the entire sub-tree. 
The leaf nodes store tuple pointers, while 
the remaining store children pointers along with counts.

We show an example hierarchical positional mapping
structure in Figure~\ref{fig:pos_mapping}.
Similar to a \mbox{B+ tree} of order $m$, 
our structure satisfies the following invariants. 
\begin{paraenum}
	\item Every node has at most $m$ children.
	\item Every non-leaf node (except root) as at-least  $\left\lceil\frac{m}{2}\right\rceil$ children.
	\item All leaf nodes appear at the same level.
\end{paraenum}
Again similar to \mbox{B+ tree}, 
we ensure the invariants by either splitting or merging nodes,
ensuring that the height of the tree is at most $\log_{\left\lceil m/2\right\rceil} N$.

Our hierarchical mapping structure makes
accessing an item at the $n$th position efficient, by
starting from the root node with $n' = n$, and traversing downwards;
at each node, given our current count $n'$, we subtract
the counts of as many of the children nodes from left-to-right (representing counts of sub-trees)
as long as $n'$ stays positive, and then follow the pointer to that child node,
and repeat the process until we reach a leaf node and access a pointer to a tuple. 
Overall, the complexity of this operation is \O{\log N}.
Insert/delete is similar, where we start at the appropriate leaf node (as before),
insert or delete appropriate tuple pointers, and then
update the counts of all nodes on the path to the modified leaf node. 
Once again, the complexity of this operation is  \O{\log N}.

\smallskip
Overall, we find that the complexity of the hierarchical positional mapping is \O{\log N} for
all operations, while the Position-as-is approach has \O{\log N} for access,
but \O{N \log N} for insert/delete. We empirically evaluate the impact of
the difference in complexities in 
Section~\ref{sec:experiments}.
 

\tr{

\section{{\large \system} Architecture}\label{sec:architecture}

\begin{figure}
\vspace{-10pt}
	\centering
	\includegraphics[width=0.8\columnwidth]{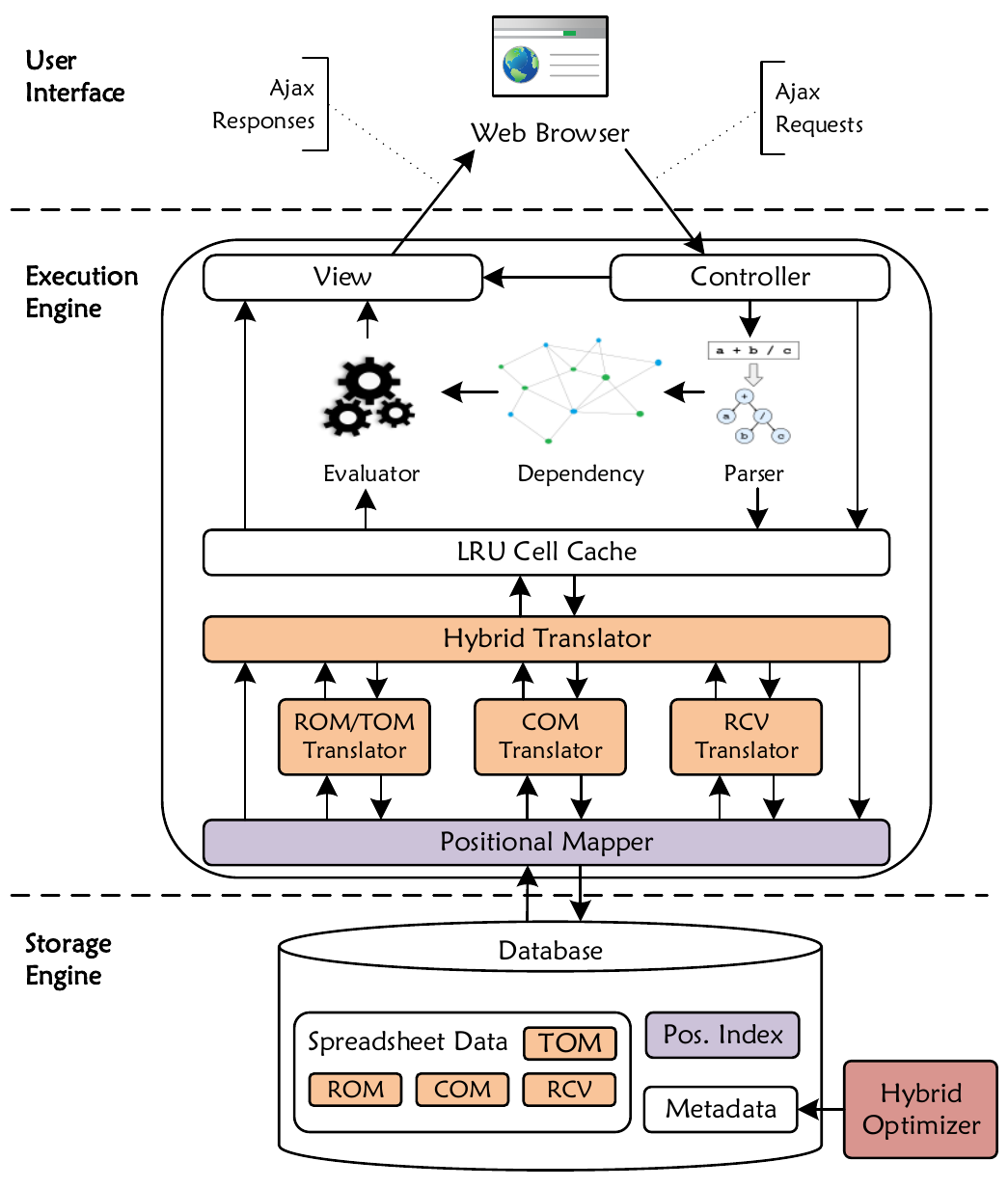}
	\vspace{-6pt}
	\caption{\system Architecture.}
	\vspace{-20pt}
	\label{fig:arch}
\end{figure}

We have implemented a fully functional \system prototype 
as a web-based tool (using the open-source ZK Spreadsheet frontend~\cite{zkspread}) on top of a PostgreSQL database. 
The prototype along with its source code, documentation, and user guide can be found at \code{http://dataspread.github.io}.
Along with standard spreadsheet features, 
the prototype supports all the spreadsheet-like and database-like operations discussed in Section~\ref{sec:interface_model}. 
Screenshots of \system in action can be found in our qualitative evaluation
section (Section~\ref{sec:qual_eval}).

Fig.~\ref{fig:arch} illustrates \system's architecture, 
which at a high level 
can be divided into three main layers,
\begin{paraenum}
\item user interface,
\item execution engine, and
\item storage engine. 
\end{paraenum}
The \textit{user interface} consists 
of a \textit{spreadsheet widget}, which presents a spreadsheet 
on a web-based interface 
and handles the interactions on it. 
The \textit{execution engine} is a Java web application 
that resides on an application server.
\tr{The \textit{controller} accepts user interactions 
in form of events and identifies the corresponding actions, 
\eg a formula update is sent to the formula parser, 
an update to a cell is sent to the cell cache.}  
The \textit{dependency graph} captures the formula dependencies 
between the cells and aids in triggering the computation of dependent cells. 
The \textit{positional mapper} translates the row and column numbers 
into the corresponding stored identifiers and vice versa. 
The \textit{ROM/TOM, COM, RCV, and hybrid translators} 
use their corresponding spreadsheet representations 
and provide a ``collection of cells'' abstraction to the upper layers.
The TOM data model is handled as a special case of ROM.
The hybrid translator is responsible for mapping the different regions on a spreadsheet to corresponding data models.
ROM/TOM, COM, and RCV translators service \mbox{\code{getCells}} 
by using the tuple pointers, obtained from the \emph{positional mapper}, to fetch required tuples. 
For a hybrid model, the mapping from a range to model is stored as metadata. The hybrid translator services \code{getCells}
by identifying the responsible data model and delegating the call to it. 
Other operations such as cell updates are performed by the hybrid model in a similar fashion.
This collection of cells are then cached in memory via an \textit{LRU cell cache}.
The storage engine consists of a database responsible for persisting data.
This data is persisted using a combination of \textit{ROM, COM, RCV, and TOM} 
(Section~\ref{sec:spreadsheet_representation}) 
along with \textit{positional mapping indexes}, which map row/column numbers to tuple pointers (Section~\ref{sec:positional_mapping}), 
and \textit{metadata}, which records information about the hybrid data model.  
The \textit{hybrid optimizer} determines the optimal 
hybrid data model and is responsible for migrating data 
across data models.  

\stitle{Formula Evaluation.}
\cut{Formula evaluation in \system is triggered by updates 
to cells from the \textit{user interface}. 
In brief, mappings between cells and dependent cells are
maintained in a {\em dependency graph};
when users add a formula, the dependent cells
are fetched from the \textit{LRU cell cache} in a read-through manner,
with the result being written in a write-through manner.
When users update cells, computation of 
the cells dependent on it via formulae
are triggered, with prioritization given to the cells visible to the user.}
When a formula is entered a cell, the \textit{parser} interprets it and identifies the cells required for its computation.
The cell containing the formula is then said to be \emph{dependent} on the cells required for its computation. The \textit{parser} captures this  dependency, \ie the mapping between a cell and its dependent cells, 
in a so-called \emph{dependency graph}.
The \textit{evaluator} fetches the cells required for computing the formula from the \textit{LRU cell cache} in a read-through manner, \ie the cache fetches the cells that are not present on demand from the underlying layer, and then computes the result of the formula. Finally, it persists the computed result by passing it back to the \textit{LRU cell cache} in a write-through manner, \ie the cache pushes its updates to the database via the  \textit{hybrid translator}.  
Whenever a user updates a cell, the \textit{evaluator} uses the 
dependency graph to identify the cells that are dependent on the updated cell and triggers their computations.  
The triggered computations are performed by the \textit{evaluator} and resultant values are persisted  in the database by passing them to the \textit{LRU cell cache}.   
If the resultant values are different from the old ones, then the \textit{evaluator} triggers computation of the cells dependent on them, if any.
The prototype also addresses a 
number of additional challenges especially related to efficient formula evaluation, including
compact representation of formula dependencies, batched and lazy formula evaluation,
and optimizing for user attention---these are important issues, but
beyond the scope of this paper, which focuses on the storage engine.


\stitle{Relational Operations.} Since \system
is built on top of a traditional relational database,
it leverages the SQL engine of the database
and seamlessly supports SQL queries, via \code{sql} function, on the front-end
spreadsheet interface. 
In addition, we support relational operators via the following spreadsheet functions:
\code{
union,
difference,
intersection,
crossproduct,
join,
filter,
project,} and
\code{rename}.
These functions return a single composite table value;
to retrieve the individual rows and columns within that table value,
we have an \code{index(cell, i, j)} function that looks up the \code{(i, j)}th
row and column in the composite table value in location \code{cell}, and places it
in the current location.
See Appendix~\ref{app:rel-ops-support} for details.
}

\section{Experimental Evaluation}\label{sec:experiments}

In this section, we present an evaluation of the storage engine of \system.

\begin{figure*}[t]
	\centering
	\vspace{-75pt}
		\includegraphics[height=4.5cm]{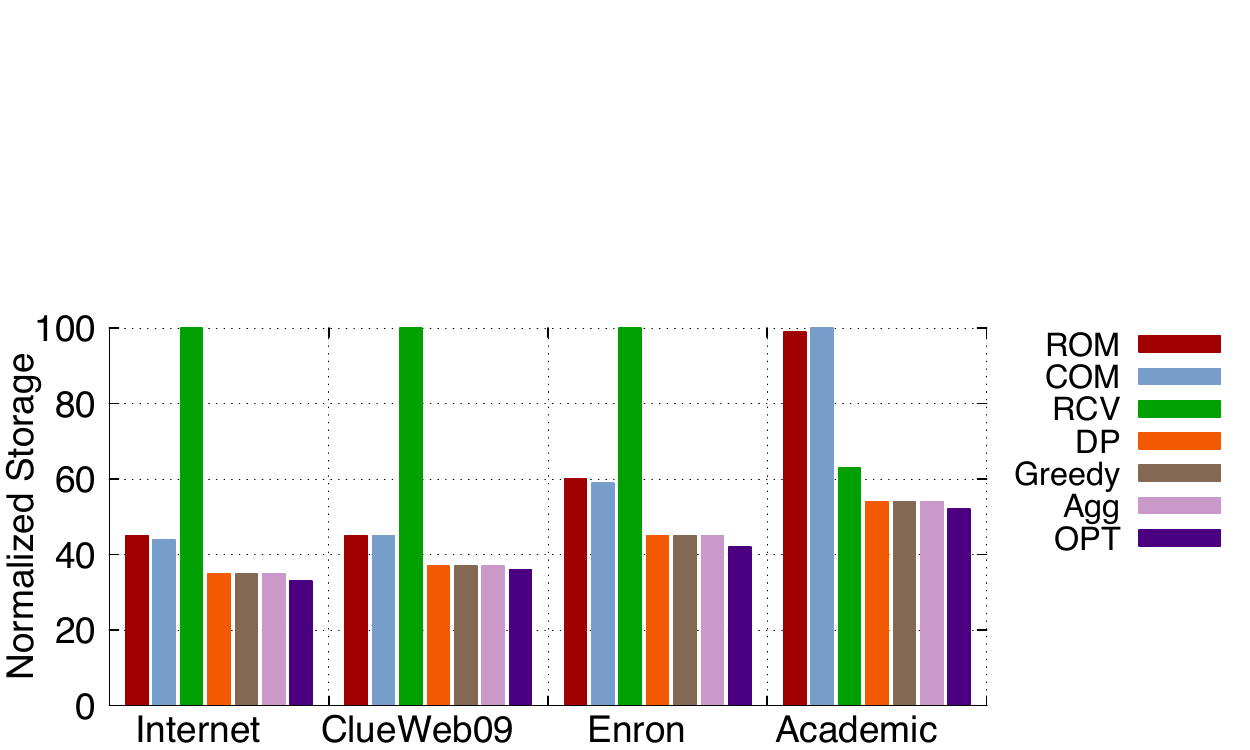}
		\includegraphics[height=4.5cm]{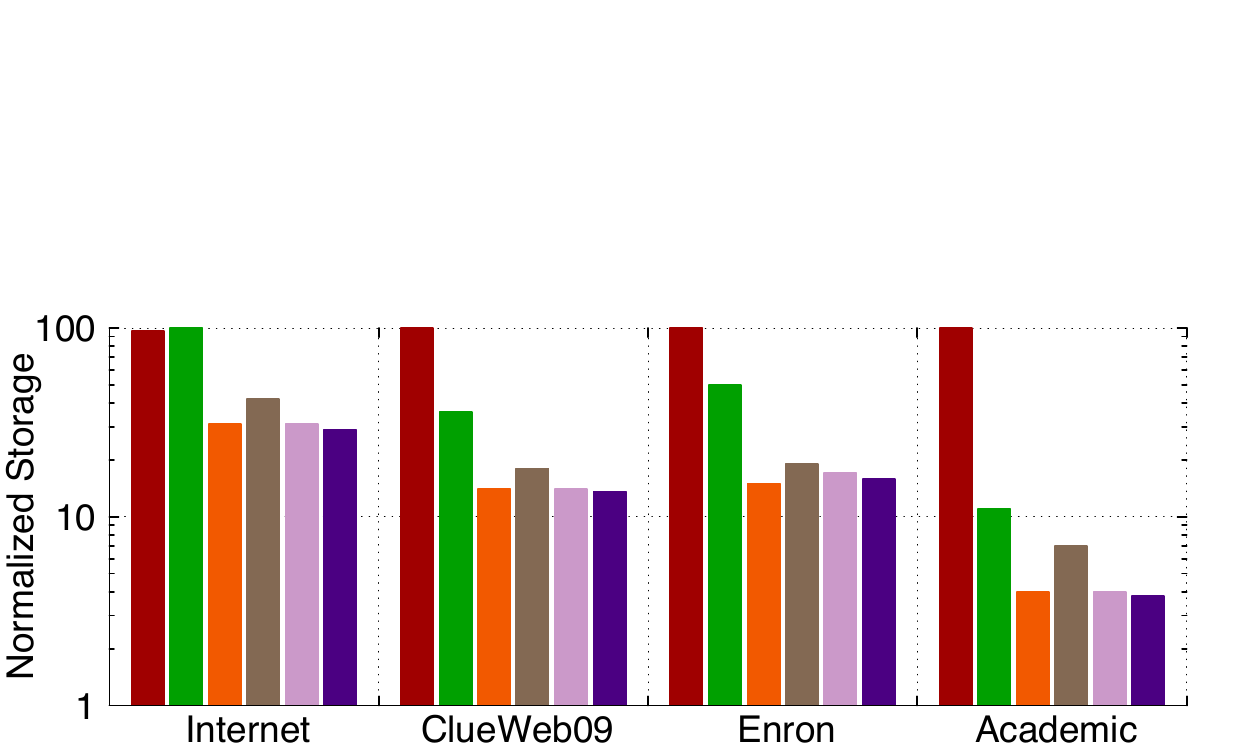}
		\vspace{-5pt}
	\caption{(a) Storage Comparison for PostgreSQL. (b) Storage Comparison on an Ideal Database.}
	\label{fig:hybrid_actual_storage_cost}
	\vspace{-10pt}
\end{figure*}

\tr{
\begin{figure*}[t]
	\centering
	\subfloat{}{\includegraphics[width=0.23\textwidth,clip]{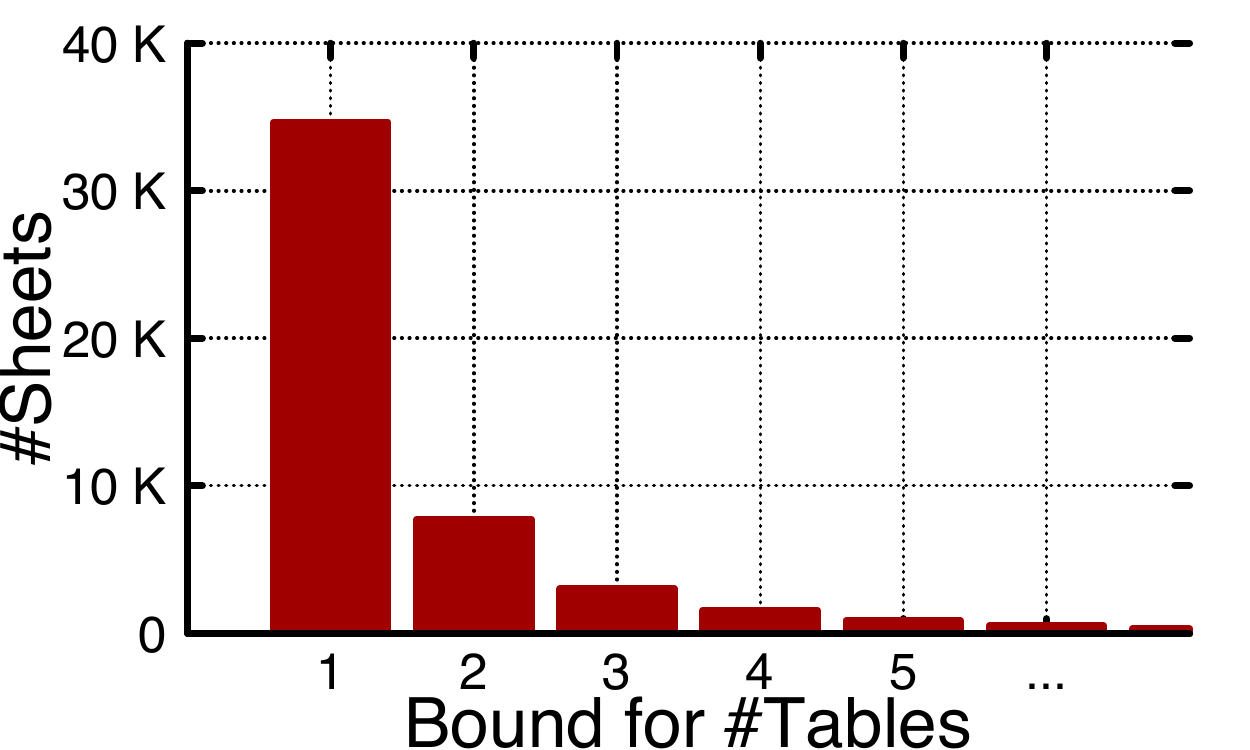}}
	\hspace{5pt}
	\subfloat{}{\includegraphics[width=0.23\textwidth,clip]{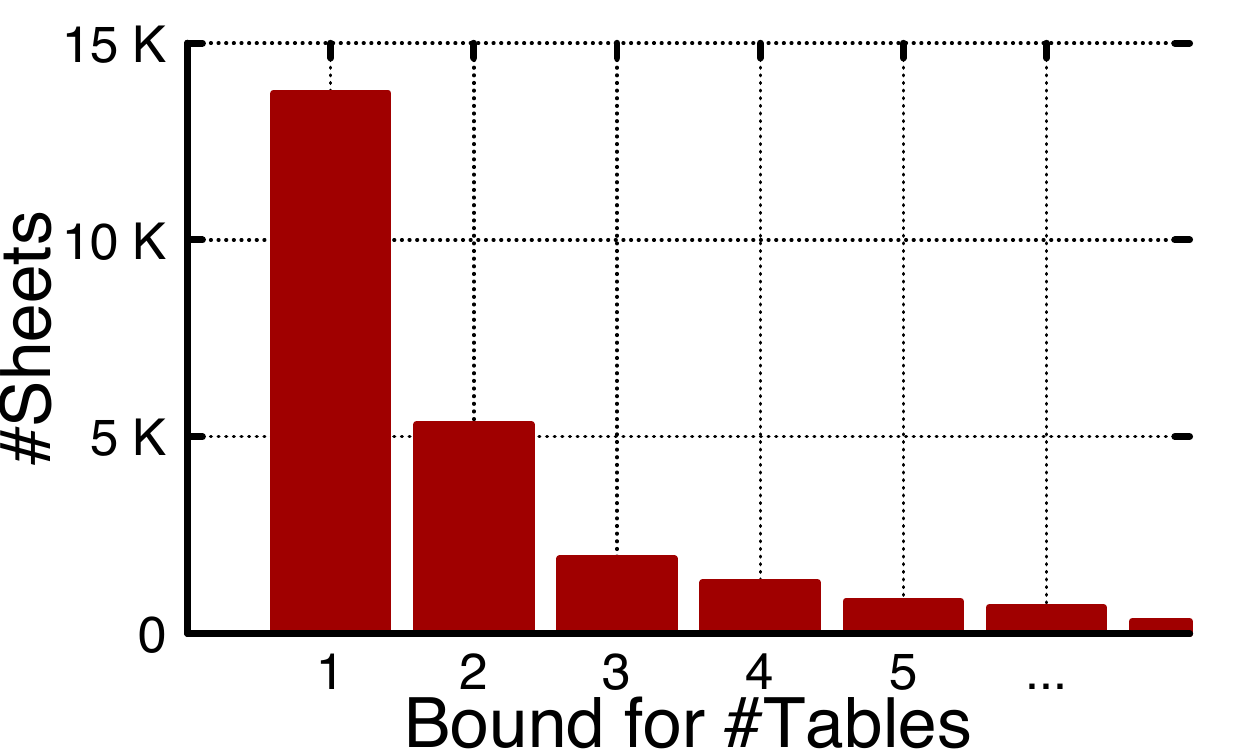}}
	\hspace{5pt}
	\subfloat{}{\includegraphics[width=0.23\textwidth,clip]{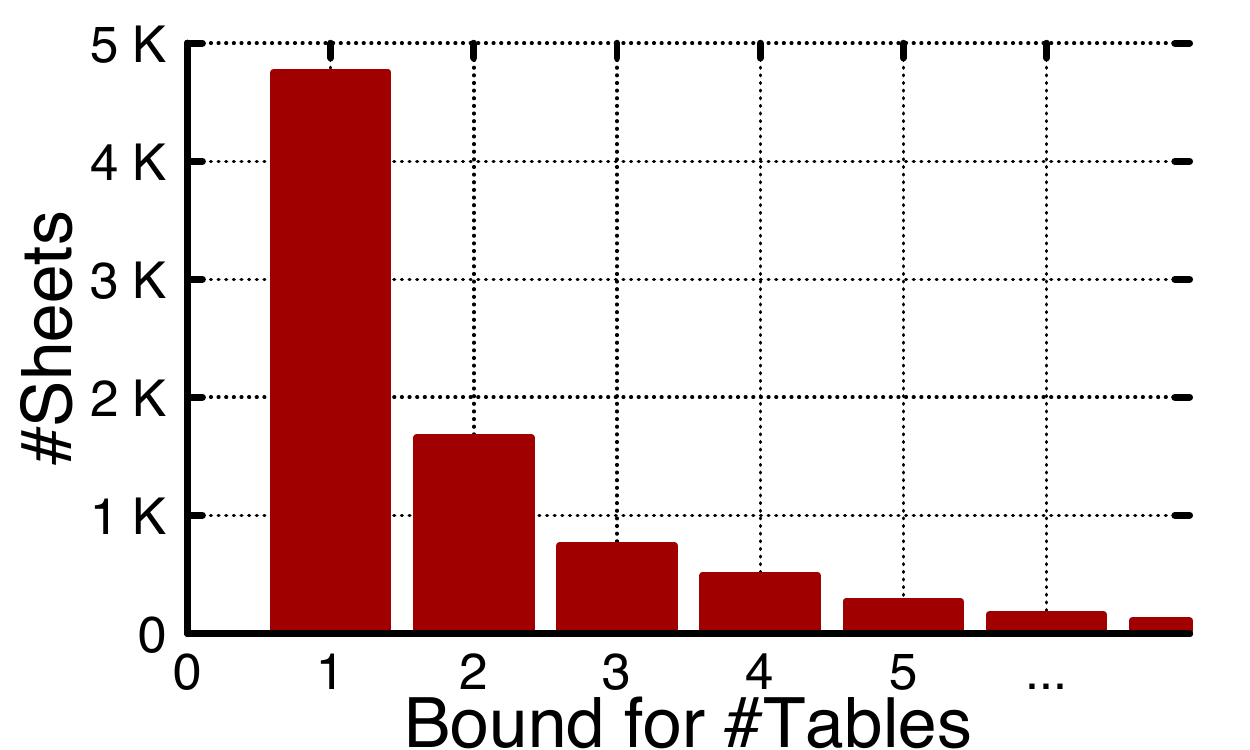}}
	\hspace{5pt}
	\subfloat{}{\includegraphics[width=0.23\textwidth,clip]{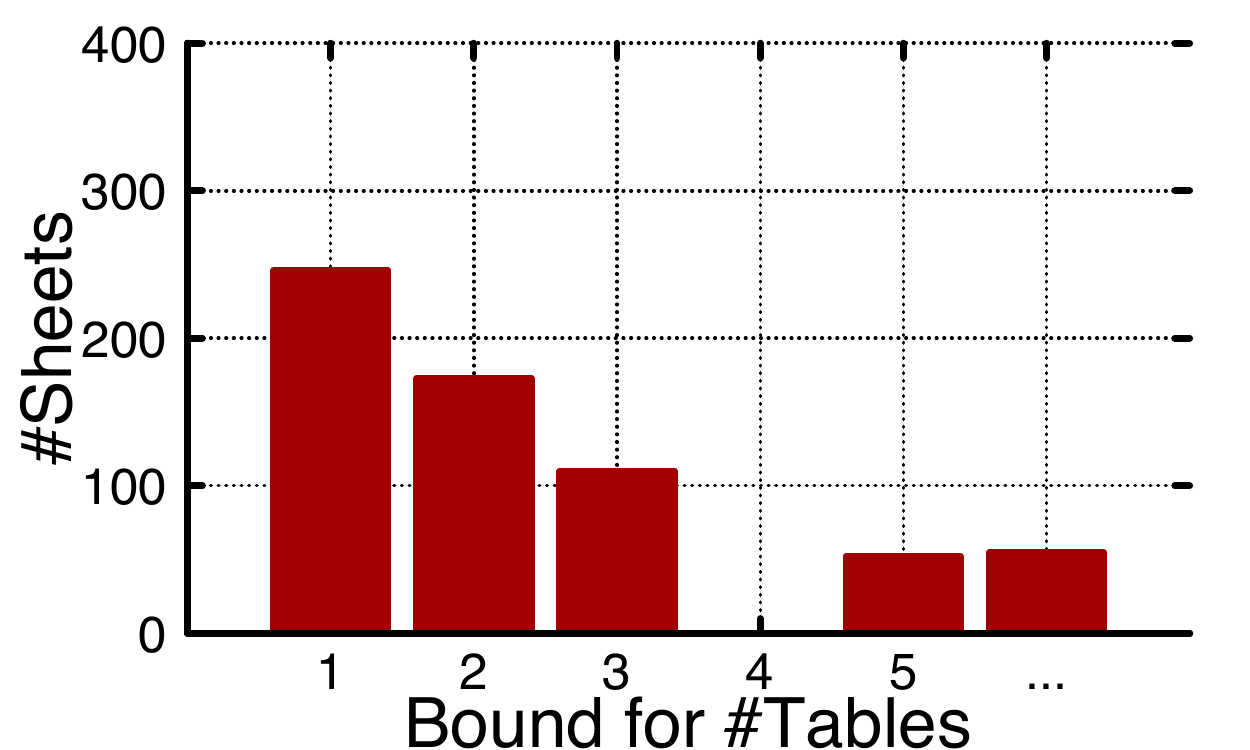}}
	\caption{Upper bound for \#Tables in the optimal decomposition --- (a) Internet. (b) ClueWeb09. (c) Enron. (d) Academic.}
	\label{fig:kbound}
	\vspace{-8pt}
\end{figure*}
}

\subsection{Experimental Setup}

\stitle{Environment.} 
We have implemented the data models and positional mapping techniques using PostgreSQL 9.6, 
configured with default parameters.
We run all of our experiments on a workstation 
running Windows 10 on an Intel Core i7-4790K 4.0 GHz
with 16 GB RAM.
Our test scripts are single-threaded applications developed in Java. 
\tr{While we have a fully functional prototype, 
our test scripts are independent of 
it, so that we can isolate the back-end performance 
implications.}
We ensured fairness by clearing the 
appropriate cache(s)
before every run.

\stitle{Datasets.}
We evaluate our algorithms on a variety of real 
and synthetic datasets.
Our real datasets are the ones listed
in Table~\ref{tab:spreadsheet-datasets}.
To test scalability, since our real-world datasets limited in scale by what
current spreadsheet tools can support, we constructed large synthetic
spreadsheet datasets.
We identify several goals for our experimental evaluation:

\stitle{Goal 1: Presentational Awareness and Access on Real and Synthetic Datasets.} 
We evaluate the hybrid data models selected
by our algorithms against the primitive data models,
when the cost model is optimized for storage.
We compare our algorithms: {\em DP} (Section~\ref{sec:optimal-recursive-decomposition}), and 
{\em Greedy} and {\em Agg} (greedy and aggressive-greedy from Section~\ref{sec:greedy})
against {\em ROM, COM, and RCV}, which represent
our best current database approach. 
We evaluate these data models on both {\em storage},
as well as {\em formulae access cost}, based on the 
spreadsheet formulae.
In addition, we evaluate the {\em running time}
of the hybrid optimization algorithms for {\em DP, Greedy, and Agg}.

\stitle{Goal 2: Presentational Access (With Updates) on Synthetic Datasets.}
We evaluate the impact of our positional mapping
schemes in aiding access on the spreadsheet.
We focus on {\em Position-as-is,
Monotonic, and Hierarchical} positional mapping schemes (introduced later)
applied on the ROM primitive model,
and evaluate the performance of {\em fetch, insert, and delete}
operations on varying the {\em number of rows}.

\stitle{Goal 3: Qualitative Evaluation.}
We evaluate the user experience of \system relative to Excel, and study
whether \system's storage engine enables users to effectively work with large datasets in two different
scenarios. 

\paper{
\stitle{Other Experiments.} In our technical report~\cite{techreport},
we present other experiments,
including
\begin{paraenum}
\item a drill-down into the performance of hybrid data models;
\item an investigation of incremental maintenance of hybrid data models; and
\item a study on varying spreadsheet parameters on positional mapping.
\end{paraenum}}

\tr{
\stitle{Other Experiments.} In the Appendix,
we present a number of other complementary experiments,
including
\begin{paraenum}
\item Goal 1: a drill-down into the performance of hybrid data models (Appendix~\ref{sec:app-drill-down});
\item Goal 2: an investigation of incremental maintenance of hybrid data models (Appendix~\ref{sec:app-inc-hybrid-exp});
\item Goal 3: a study on varying parameters of synthetic spreadsheets on positional mapping (Appendix~\ref{sec:app-vary-parameters}).
\end{paraenum}}

\subsection{Presentational Awareness and Access}
\label{sec:exp_impact_hybrid}

\boxyta{Hybrid data models provide substantial benefits
over primitive data models, with up to {\bf \em 20\% reductions
in storage, and up to 50\% reduction in formula evaluation
time} on PostgreSQL on real and synthetic spreadsheet datasets, compared to 
the best primitive data model. 
While DP has better performance on storage than Greedy and Agg,
it suffers from high running time; {\bf \em Agg 
bridges the gap between Greedy and DP}, while taking 
only marginally more running time than Greedy; both Agg and Greedy are
within 10\% of the optimal storage. 
Lastly, if we were to design a database storage engine from
scratch, the hybrid data models would provide {\bf \em up to 50\% reductions
in storage} compared to the best primitive data model.
Overall, {\bf \em our hybrid data models bring scalability to spreadsheets:
efficiently support storage across a range of spreadsheet structures, and access
data via position in an efficient manner.} 
}

\noindent The goal of this section is to evaluate
presentational access and awareness (without updates)
by evaluating our data models\tr{---on real and synthetic datasets}. 

\stitle{a. Real Dataset: Storage Evaluation on PostgreSQL.}
We begin with an evaluation of storage for different
data models on PostgreSQL.
The costs for storage on PostgreSQL as measured by us is as follows: 
$s_1$ is $8$~KB, $s_2$ is $1$~bit, $s_3$ is $40$~bytes,
$s_4$ is $50$~bytes, and $s_5$~(RCV's tuple cost) is $52$ bytes. 
We plot the results in Figure~\ref{fig:hybrid_actual_storage_cost}(a):
here, we depict the average normalized storage across sheets;
in addition to the aforementioned data models, we also plot a lower bound
for the optimal hybrid data model (denoted OPT)---the cost of storing only non-empty cells in a single ROM, 
\ie the cost ignoring the overhead of extra tables and empty cells.
For Internet, ClueWeb09, and Enron,  
we found RCV to have the worst performance, and hence normalized it to a cost of 100,
and scaled the others accordingly;
for the Academic datasets,
we found COM to have the worst performance, and hence normalized it to a cost of 100,
and scaled the others.
The first three datasets
are primarily used for data sharing, and as a result are quite {\em dense}.
As a result, ROM and COM do well, using about 40\%
of the storage of RCV.
At the same time, DP, Greedy and Agg perform roughly similarly, 
and better than the primitive data models, providing
an additional reduction of 15-20\%.
On the other hand, the last dataset\tr{, which is primarily used
for computation, and} is very sparse, RCV does better 
than ROM and COM, while DP, Greedy, and Agg once again
provide additional benefits.
We finally observe that DP, Greedy, and Agg are all very close 
(within $10\%$) of OPT. 
\tr{From this we conclude that the solution give by Agg is close to the optimal in terms of cost.}

\tr{
We next show that the error bound of using a recursive decomposition based algorithms (DP, Greedy, and Agg) is small as compared to the optimal solution. 
For this we plot the upper bound for the number of tables in the optimal solution, \ie $\sum\left\lfloor\frac{e \times s_2}{s_1} + 1 \right\rfloor$, for the four data sets in Figure~\ref{fig:kbound}.
Here, we observe the the number of tables in the optimal solution is typically small -- $90\%$ of spreadsheets have fewer than $10$ tables in the optimal decomposition. 
From the above observation and Theorem~\ref{th:dp_bound}, we conclude that 
the error bound of using the search space of recursive decomposition for practical purposes is small.
}

\stitle{b. Real Dataset: Storage Evaluation on an Ideal Database.}
Note that the reason why RCV does so poorly for the first three datasets
is because PostgreSQL imposes a high overhead per tuple, of 50 bytes,
considerably larger than the amount of storage per cell. 
So, to explore this further, we investigated the scenario 
if we could redesign our database storage
engine from scratch. 
We consider a theoretical ``ideal''
cost model, where the cost of a ROM or COM table is equal to the 
number of cells, plus the length and breadth of the table
(to store the data,  the schema, as well as position),
while the cost of an RCV row is simply 3 units 
(to store the data, as well as the row
and column number).
We plot the results in Figure~\ref{fig:hybrid_actual_storage_cost}(b)
in log scale for each of the datasets---we exclude COM for this chart
since it is identical to ROM.
Here, we find that ROM has the worst cost since
it no longer leverages benefits from minimizing the number of
tuples. \tr{(For Internet, ROM and RCV are similar, but RCV is slightly worse.)}
As before, we normalize the cost of the worst model to 100 for each sheet, 
and scaled the others accordingly.
As an example, we find that for the ClueWeb09 corpus, 
RCV, DP, Greedy and Agg have normalized
costs of about 36, 14, 18, and 14 respectively---with the hybrid data models 
more than halving the cost of RCV, and getting $\frac{1}{7}^{th}$ the cost
of ROM. 
Furthermore, DP provides additional
benefits relative to Greedy, and Agg ends up bringing us close to DP performance; finally,
we find that Agg and DP are both very close to OPT (within $10\%$).


\begin{figure}[t]
	\vspace{-10pt}
		\includegraphics[width=0.24\textwidth,clip]{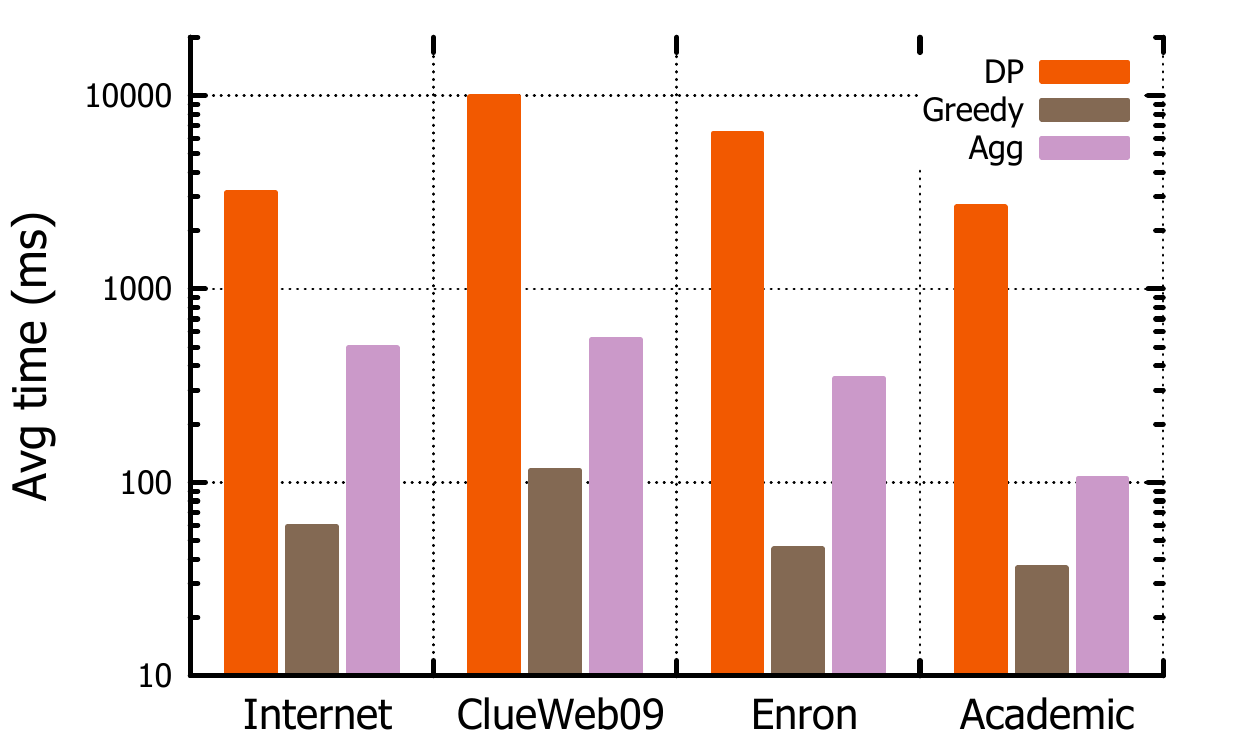}
		\hspace{-10pt}
		\includegraphics[width=0.24\textwidth]{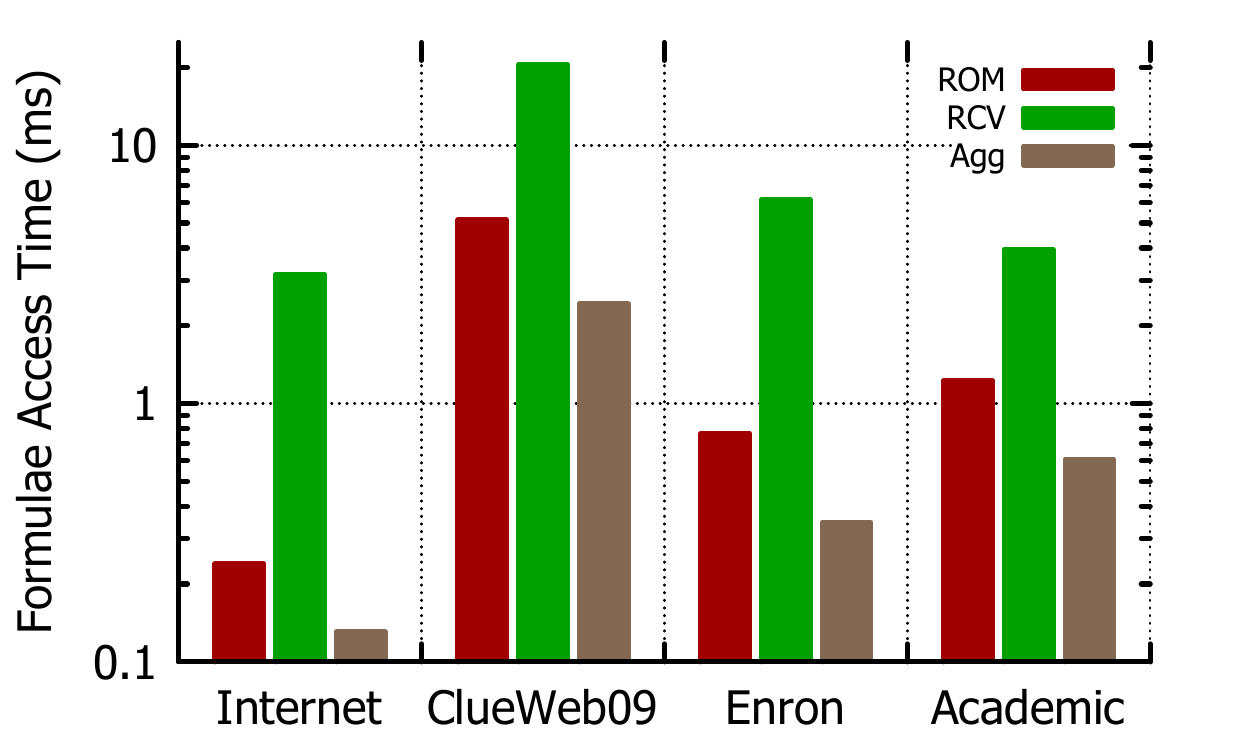}
		\vspace{-5pt}
		\caption{(a) Hybrid optimization algorithms: Running time. (b)~Average access time for formulae.}
	\label{fig:running-formulae}
	\vspace{-12pt}
\end{figure}

\begin{figure}[t]	
\centering
\includegraphics[width=0.35\textwidth,clip]{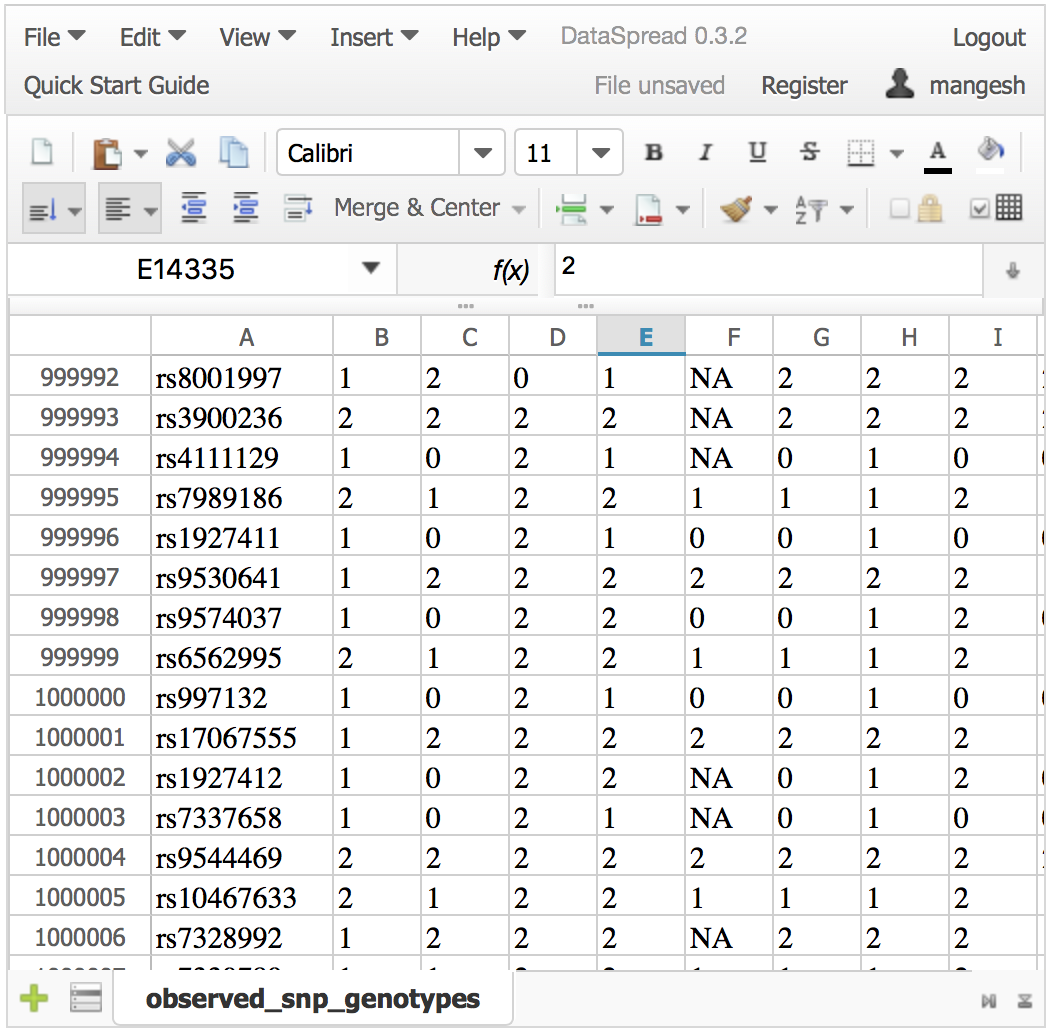}
\caption{Genomics Use Case: VCFs in \system.}
\label{fig:vcf_loaded}
\vspace{-12pt}
\end{figure}

\stitle{c. Real Dataset: Running Time of Hybrid Optimization Algorithm.}
Our next question is how long our hybrid data model 
optimization algorithms for DP, Greedy, and Agg, 
take on real datasets.
In Figure~\ref{fig:running-formulae}(a),
we depict the average running time
on the four real datasets.
The results for all datasets are similar, \eg for Enron, DP took 
6.3s on average, Greedy took 45ms (a 140$\times$ reduction), 
while Agg took 345ms (a 20$\times$ reduction).
Thus DP has the highest running time for all datasets, 
since it explores the entire space of models that can be obtained by 
recursive partitioning. 
Between Greedy and Agg, Greedy
turns out to take less time. 
Note that these observations are consistent with
our complexity analyses from Section~\ref{sec:greedy}.
That said, Agg allows us to trade off running time
for improved performance on storage (as we saw earlier).
Greedy takes less time than Agg; but Agg allows us to trade off running time
for improved performance on storage.
\tr{
Between Greedy and Agg, Greedy
turns out to take less time. 
Note that these observations are consistent with
our complexity analyses from Section~\ref{sec:greedy}.
That said, Agg allows us to trade off running time
for improved performance on storage (as we saw earlier).
We note that for the cases where the spreadsheets were large,
we terminated DP after about 10 minutes, 
since we want our optimization
to be relatively fast. (Note that using
a similar criterion for termination, Agg and Greedy did
not have to be terminated for any of the real datasets.)
To be fair across all the algorithms, 
we excluded all of these spreadsheets
from this chart---if we had included them, 
the difference between DP and the other algorithms
would be even more stark.}

\stitle{d. Real Dataset: Formulae Access Evaluation on PostgreSQL.}
We next evaluate if our hybrid data models,
optimized only on storage, have any impact on 
the access cost for spreadsheet formulae.
Our hope is that spreadsheet formulae
focus on ``tightly coupled''
tabular areas, which our hybrid
data models are able to capture and store in separate tables.
For this evaluation, we focus on Agg, since it provided
the best trade-off between running time and storage costs.
Given a sheet in a dataset, for each data model,
we measured the time taken to evaluate the formulae
in that sheet, and averaged this time across all sheets and all formulae.
We plot the results in Figure~\ref{fig:running-formulae}(b)
in log scale in ms.
As a concrete example, on Internet,
ROM has a formula access time of 0.23, RCV has 3.17, 
and Agg has 0.13.
Thus, Agg provides a substantial reduction of 96\%
over RCV and 45\% over ROM---{\em even though Agg was optimized
for storage and not for formula access.}
This validates our design of hybrid data models\tr{
to store spreadsheet data}.
Note that while the performance numbers 
for the real spreadsheet datasets are small for all 
data models (due to the size limitations in present spreadsheets),
when scaling up to large datasets, and formulae on them, 
these numbers will increase in proportionally,
at which point it is even more important to opt for hybrid data models, as we will see next.

\begin{figure}[htb]
\vspace{-10pt}
	\centering
	\subfloat{}{\includegraphics[width=0.24\textwidth,clip]{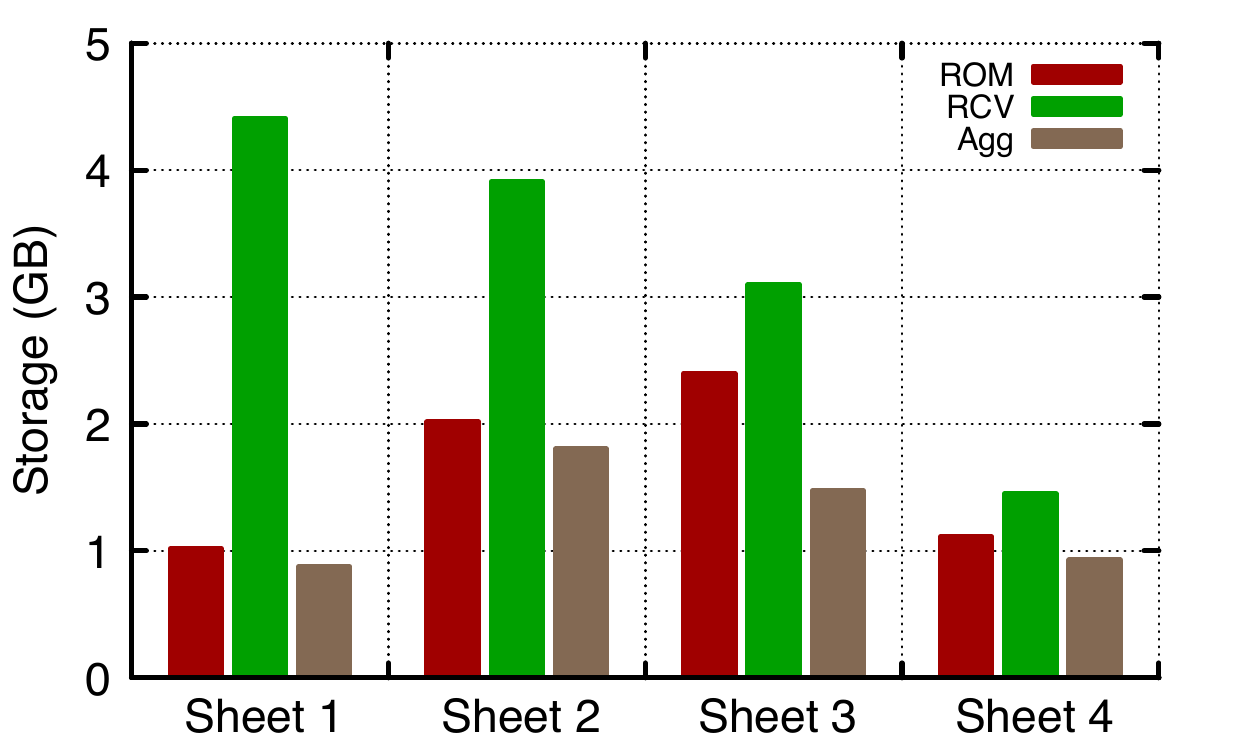}}
	\hspace{-10pt}
	\subfloat{}{\includegraphics[width=0.24\textwidth,clip]{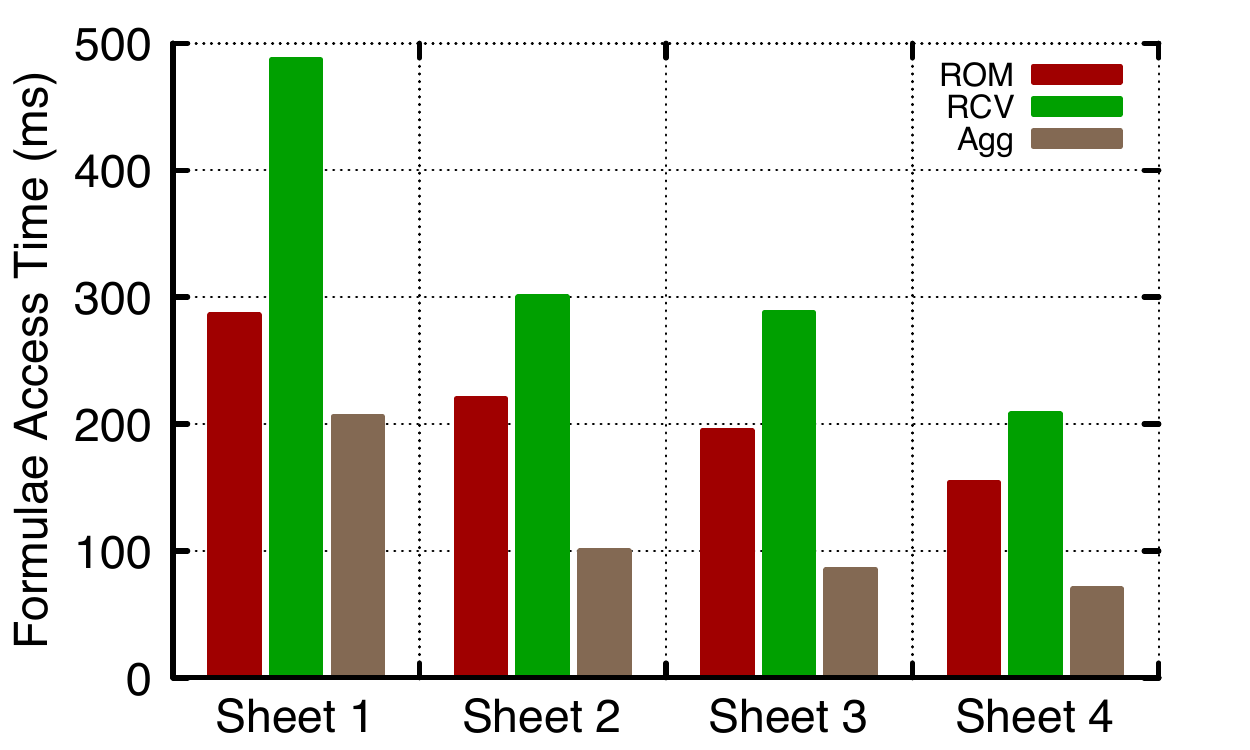}}	
	\vspace{-6pt}
	\caption{
	Synthetic sheets
	(a) Storage.
	(b) Access time.}
	\label{fig:syn_sheets}
	\vspace{-12pt}
\end{figure}

\stitle{e. Synthetic Dataset: Storage and Formula Access Evaluation}
We now run our tests on large synthetic spreadsheets with 100+ million cells 
to evaluate our techniques in large dataset scenarios.
We create synthetic spreadsheets by populating an empty sheet with twenty dense rectangular regions to simulate randomly placed tables.    
We add $100$ randomly generated formulae that access rectangular ranges of these tables. 
Figures~\ref{fig:syn_sheets}(a) and~\ref{fig:syn_sheets}(b)
depict the storage requirements and the formulae access time respectively 
for four synthetic spreadsheets, which are in the decreasing order of density (the fraction of cells that are filled-in in the minimum bounding rectangle). 
For both storage and access, we find that Agg is better than ROM, which is better than RCV; as
density is decreased, RCV's performance becomes closer to ROM. 
Agg performs the best, providing substantial reductions of up
to 50-75\% of the time taken for access with ROM or RCV.


\begin{figure*}[t]
  \centering
  \subfloat{}{\includegraphics[width=0.3\textwidth,clip]{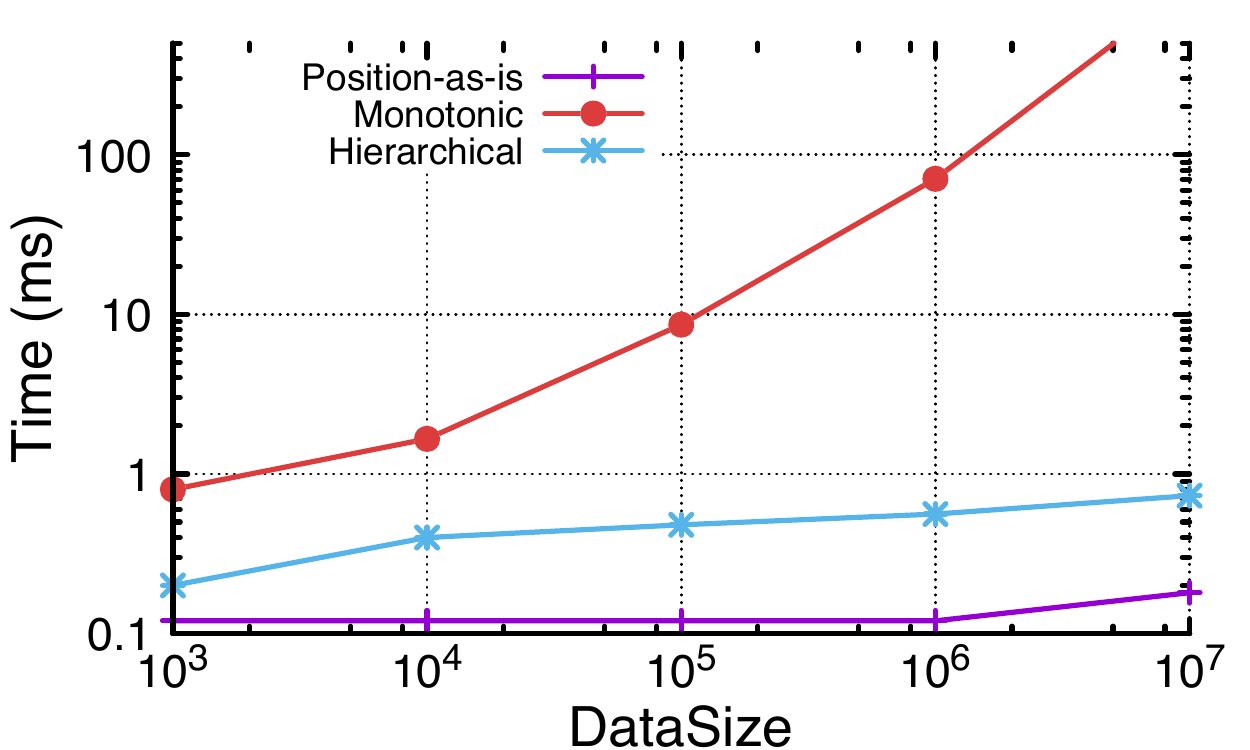}}
  \hspace{5pt}
  \subfloat{}{\includegraphics[width=0.3\textwidth,clip]{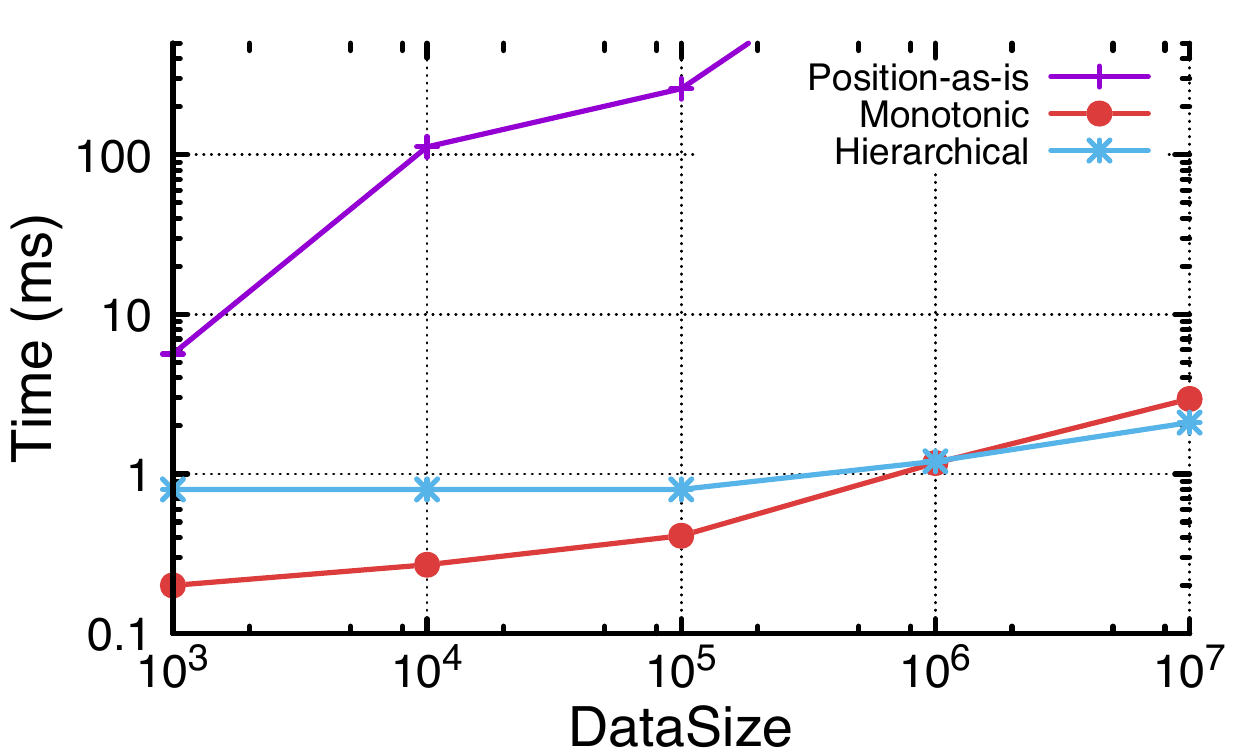}}
  \hspace{5pt}
  \subfloat{}{\includegraphics[width=0.3\textwidth,clip]{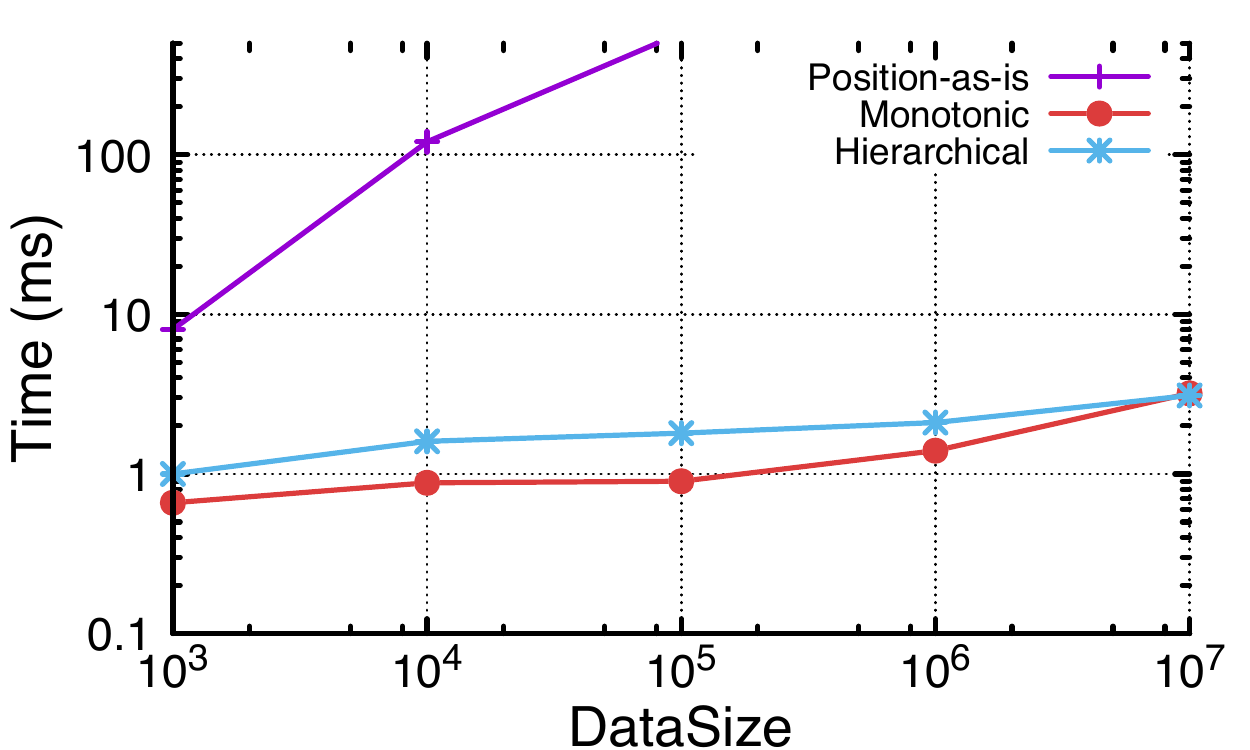}}
  \vspace{-5pt}
  \caption{Positional mapping performance for (a) Select. (b) Insert. (c) Delete.}
  \label{fig:pos_index_performance}
  \vspace{-12pt}
\end{figure*}

\subsection{Presentational Access with Updates}
\label{sec:exp_pos_mapping}

\boxyta{Hierarchical positional mapping retains
the rapid fetch benefits of position-as-is, while
also providing rapid inserts and updates. 
Thus, {\bf \em hierarchical positional mapping is able to 
perform positional operations within a few milliseconds}\tr{, while the other schemes often take
seconds on large datasets}.
\tr{Overall, {\bf \em our hierarchical positional
mapping schemes support presentational access with updates,
validating the fact that our storage engine can support interactivity}.}}

We now  evaluate
presentational access (with updates)
by studying our positional mapping methods (Section~\ref{sec:positional_mapping}) 
on  synthetic datasets. 
We compare our hierarchical positional mapping scheme (denoted hierarchical),
with position as-is (denoted position-as-is): this is the approach
a traditional database with a \mbox{B+ tree} would use.
In addition, motivated by the online
dynamic reordering technique in Raman et al.~\cite{raman1999online},
we consider another baseline (denoted monotonic), where
we store a monotonically increasing sequence of identifiers (with gaps) to capture the position. 
Using this sequence we dynamically order the tuples
at run-time (by sorting); whereas the gaps in the sequence 
enable efficient insert/delete operations.
The dynamic reordering sacrifices the performance of the 
fetch operation as it needs to discard $n-1$ tuples to fetch $n^{th}$ tuple. 

We operate on a dense synthetic dataset ranging from
$10^3$ to $10^7$ rows, with $100$ columns with all cells filled;
and repeat this $1000$ times. 
We evaluate the performance of a single ROM table with
all of the data; evaluations for other data models are similar.
Figure~\ref{fig:pos_index_performance} displays
the average time taken to perform a fetch, insert, and delete
of a single (random) row.

\begin{figure}[t]	
\centering
\includegraphics[width=0.35\textwidth,clip]{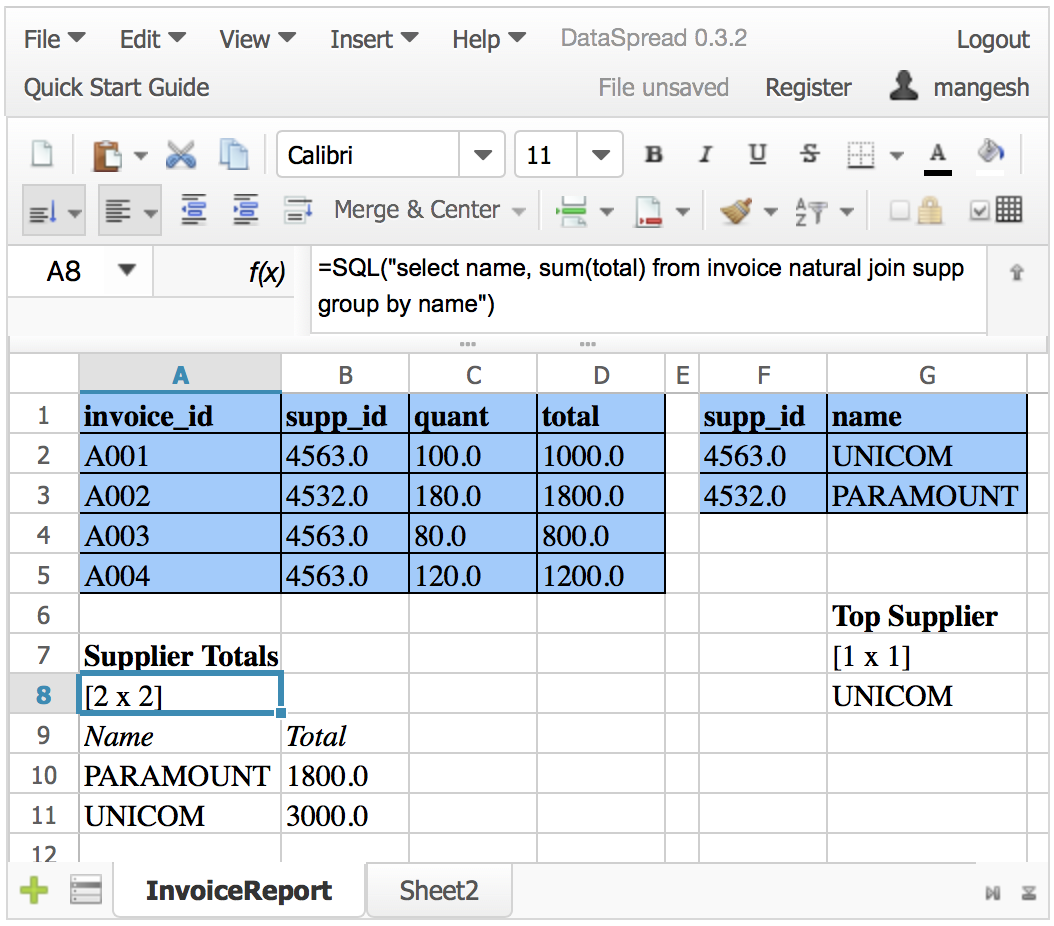}
\caption{Customer Management in \system.}
\label{fig:tables_and_sql}
\vspace{-15pt}
\end{figure}

We see that position-as-is 
performs well for fetch. 
However, the insert and delete time 
increases rapidly with 
the data size,
due to cascading updates;
thus, beyond a data size of $10^5$, position-as-is
is no longer interactive ($>500ms$) for
insert and delete.
Conversely, the response time of monotonic 
for fetch increases rapidly with data size\tr{. 
This is again expected}, as we need to linearly search through 
the monotonic keys to retrieve the required records---making it infeasible
for large datasets. 
Lastly, we find that hierarchical 
performs well \emph{for all operations} 
and performance does not get degrade 
even with data sizes of $10^9$ tuples.
In comparison with the other schemes, 
hierarchical performs all of the three aforementioned operations 
in few milliseconds\tr{, which makes it the practical choice 
for presentational access with updates}. 

\subsection{Qualitative Evaluation}
\label{sec:qual_eval}

We now evaluate \system to see how it can handle 
the use cases described in Section~\ref{sec:introduction}. 
With our genomics use case, we evaluate
the {\em scalability} of \system, and with our customer management
use case, we evaluate the {\em functionality}. 

\stitle{a. Evaluating Scale for Genomics:}
For this evaluation, we used a VCF file provided by our biology
collaborators, as described in Example~$1$, and used it to perform basic exploration.
We contrast the performance of \system with Excel.
Our VCF file has 1.3M rows and 284 columns. 
Unfortunately, we were unable to load this file in Excel since it exceeds
Excel's limits. 
Importing the file in \system takes about a minute.
On the other hand, even after reducing the VCF file to 1M rows, Excel is unable to import the file within an hour. 
After substantially reducing the file size to 130K rows, we were able to import it into Excel in about $10$ minutes.  
After loading the 1.3M VCF file, we were able to take advantage
of \system's efficient positional access to scroll up and down to
explore the data with interactive (sub-second) response times.
Figure~\ref{fig:vcf_loaded} shows a screenshot of the 
file in \system, having scrolled to the millionth row.

\boxyta{
\system enables users to {\bf \em interactively work on large spreadsheets} that 
main-memory based spreadsheet applications are unable to handle.}

\stitle{b. Evaluating Functionality for Customer Management:}
For evaluating functionality, as described in Example~$2$, 
we leverage the database-oriented operations 
discussed in Section~\ref{sec:interface_model}.
Using \code{linkTable}, we first establish a 
two-way synchronization between the spreadsheet regions and 
the \code{invoice} and \code{supp} tables in the database (Figure~\ref{fig:tables_and_sql}). 
These linked regions enable us to directly manipulate 
the underlying tables via spreadsheet operations such as cell updates; 
this is not possible in spreadsheet tools that only allow one-way import 
of data from a backend database to a spreadsheet.
We used the \code{sql} function in cell \code{A8} to join the two tables 
and perform grouping and aggregation; this is less cumbersome 
and more efficient compared to Excel's \code{vlookup} and pivot tables,
and indexed into the composite value in \code{A8} to display the results in \code{A9:B11}. 
Finally, we use the \code{project} and \code{select} functions to get the top supplier in cell \code{G8}; 
any updates to the underlying tables are automatically reflected in the function's output. 
\boxyta{
The direct manipulation and database-oriented features of \system enable {\bf \em interactive management of data via database tables on a spreadsheet interface}.}


\section{Related Work} \label{sec:related}

Our work draws on related work from multiple areas: 
\begin{paraenum}
\item that enhance database usability, 
\item those that attempt to merge spreadsheet and database 
functionalities, but without a holistic integration, and 
\item order or array-based database management systems.
\end{paraenum}
\anon{The only work sharing a similar vision of integrating
spreadsheets and databases holistically is a recent demo 
paper~\cite{dbspread}---we draw from and extend that
vision to develop optimal representation and indexing techniques, 
and evaluate these techniques extensively.}{We described our 
vision for \system in an earlier demo paper~\cite{dbspread}.}


\stitle{1. Making databases more usable.} There has been a lot of 
recent work on making database interfaces more user friendly~\cite{lowell, jagadish_making_2007}. 
This includes recent work on gestural query and scrolling interfaces~\cite{mandel_gestural_2013, nandi_guided_2011, nandi_querying_2013, singh_skimmer:_2012,idreos2013dbtouch}, 
visual query builders~\cite{catarci_visual_1997,abouzied2012dataplay},
query sharing and recommendation tools~\cite{khoussainova_case_2009, cetintemel_query_2013, khoussainova_snipsuggest:_2010},
schema-free databases~\cite{qian_crius:_2010},
schema summarization~\cite{yu_schema_2006},
and visual analytics tools~\cite{callahan_vistrails:_2006,mackinlay2007show, stolte2002polaris, gonzalez_google_2010}.
However, none of these tools can replace spreadsheet software
which has\tr{ the ability to analyze, view, and modify data
via} a direct manipulation interface~\cite{shneiderman_direct_1983} and \tr{has} 
a large user base---this paper aims to make this interface available to manipulate databases.

\stitle{2a. One-way import of data from databases to spread\-sheets.}
There are various mechanisms for 
importing data from databases to spreadsheets, and then
analyzing this data within the spreadsheet.
This approach is followed by Excel's Power BI tools\tr{, including Power Pivot}~\cite{powerpivot},
with Power Query~\cite{powerquery} for exporting data
from databases and the web\tr{ or deriving additional columns} and Power View~\cite{powerquery}
to create presentations; 
and Zoho~\cite{zoho} and ExcelDB~\cite{exceldb}\tr{(on Excel)}, and
Blockspring~\cite{blockspring}
enabling the
import from a variety of sources including the databases and the web.
Typically, the import is one-shot, with the data residing
in the spreadsheet from that point on, negating the scalability
benefits from the database.
Indeed, Excel 2016 specifies a limit of 1M records that can be
imported, illustrating that the scalability benefits are lost.
\tr{Zoho specifies a limit of 0.5M records.} 
Furthermore, the connection to the base data is lost: modifications made at either end are not propagated.

\stitle{2b. One way export of operations from spreadsheets to data\-bases.}
There has been some work on exporting spreadsheet operations
into database systems, such as Oracle~\cite{witkowski_advanced_2005, witkowski_query_2005}, 1010Data~\cite{1010data} and AirTable~\cite{airtable},
to improve the performance of spreadsheets.
However, the database itself has no awareness of the existence of the
spreadsheet, making the integration superficial.
In particular, positional and ordering aspects
are not captured, and user operations on the front-end, 
\eg inserts, deletes, and
adding formulae, are not supported.
\tr{
Indeed, the lack of awareness makes the integration one-shot, with
the current spreadsheet being exported to the database, with
no future interactions supported at either end:
thus, in a sense, the {\em interactivity} is lost.
Other efforts in this space include
that by Cunha et al.~\cite{cunha_spreadsheets_2009} to
recognize functional dependencies in spreadsheets.}
Other work has examined the extraction of structured relational data from spreadsheets~\cite{autowebextract, chen2013senbazuru}.

\stitle{2c. Using a spreadsheet to mimic a database.}
There has been some work on using a
spreadsheet to pose as traditional database.
\Eg Tyszkiewicz~\cite{tyszkiewicz_spreadsheet_2010}
describes how to 
simulate database operations in a spreadsheet. 
However, this approach loses the scalability benefits of relational databases.
Bakke et al.~\cite{bakke2016expressive, bakke_spreadsheet-based_2011, bakke_schema-independent_2011} 
support joins by depicting relations
using a nested relational model.
Liu et al.~\cite{liu_spreadsheet_2009}
use spreadsheet operations
to specify single-block SQL queries; 
this effort is essentially a replacement for visual query builders. 
Recently, Google Sheets~\cite{google-sheets} has provided
the ability to use single-table SQL on its frontend, without 
availing of the scalability benefits of database integration.
Excel, with its Power Pivot and Power Query~\cite{powerquery} functionality
has made moves towards supporting SQL in the front-end, with
the same limitations.
Like this line of work, we support SQL queries on the spreadsheet frontend\tr{,
but our focus for this paper is on representing and operating on spreadsheet
data within a database}.

\stitle{3. Order-aware database systems.}
Some limited aspect of presentational awareness, in particular, order, has been studied.
The early work of online dynamic reordering \cite{raman1999online} supports data reordering based on user preference, citing a spreadsheet-like interface \cite{raman1999scalable} as an application.
More recently, there has been work on array-based databases, but most of these systems do not support edits, \eg SciDB~\cite{Brown:2010:OSL:1807167.1807271}\tr{ supports an append-only, no-overwrite data model}. 


\section{Conclusions}\label{sec:conclusion}

We introduced our vision of {\em presentational data management}: 
building a system that holistically integrates spreadsheets and databases. 
We focused on developing a storage engine for our PDM prototype \system, 
characterizing key requirements in the form of presentational awareness and access.
We addressed presentational awareness by proposing three primitive data models
for representing spreadsheet data, along with algorithms for identifying optimal hybrid data models
from recursive decomposition. Our hybrid data models provide substantial reductions
in terms of storage (up to 20--50\%) and formula evaluation (up to 50\%)
over the primitive data models.
For presentational access, we couple our hybrid data models 
with a hierarchical positional mapping scheme, making
working with large spreadsheets interactive.
Overall, \system emerges as a promising solution for interactively analyzing,  manipulating, and managing
large datasets.

\balance
{
\scriptsize
\bibliographystyle{abbrv}
\bibliography{dbspread}
}
\tr{
\appendices
\later{

\section{Primitive Data Models: Generalization and Extreme Points}

The three primitive data models we described in Section~\ref{sec:primitive_datamodels}
are natural extensions of models
common in the database literature.
A natural question is whether these models
are reasonable choices,
and whether any of them dominate the others.
We now introduce the notion of {\em rectangular
data models}, a generalized class of data models
that all of these primitive data models belong to,
and show that these primitive data models
represent optimal extreme points in that space.

\subsection{Rectangular Data Models}
We define the space of {\em rectangular data models}
as those that store a spreadsheet in a single table, where 
each tuple corresponds to a rectangular region of equal size,
comprising of $a \times b$ cells, with $a$ rows of cells vertically,
and $b$ rows of cells horizontally, laid out as $a \times b$ attributes.
In addition, there are two attributes representing the row
and column number of the cell that is at the top left corner.
For example, if we had a $8 \times 8$ spreadsheet,
one such data model within the space of rectangular
data models would have just four tuples,
formed by dividing the region into two horizontally
as well as vertically:
one corresponding to rows 1:4 and columns 1:4,
one corresponding to rows 5:8 and columns 1:4,
and so on.
Another data model within this space would have sixteen tuples,
formed by dividing the region using four cuts horizontally
and four vertically, with a separation of two cells between
the cuts.

Now, it is easy to see that ROM, COM and RCV are extremes
within the space of rectangular data models:
ROM is formed from using only horizontal cuts,
one per row of the spreadsheet;
COM from using only vertical cuts;
and RCV from using horizontal and vertical cuts
corresponding to each row and column of the spreadsheet.
Note that in ROM and COM, we can additionally avoid
storing the column and row
identifiers respectively, since they can be inferred 
from the attribute names, saving 
some additional space.

\subsection{Valuable Extremes}
\label{sec:opt_tuple_size}

In this section, we demonstrate that the three primitive
data models represent valuable extreme points
and dominate other rectangular data models 
under certain settings that occur naturally in practice.
Our focus in this section is the use of a single table
to represent the entire area of the spreadsheet.
In Section~\ref{sec:hybrid_data_model}, we will consider
the setting when there are multiple tables to represent
a spreadsheet.

We consider the cost of storage in order to perform this analysis.
Our cost model will differ slightly from 
Section~\ref{sec:hybrid_data_model} since we're focusing
on one table.

\agp{start}
In terms of storage, we use the following cost model:
$$\cost{T} = (2 + a \times b) \times  \textrm{(number of tuples)}$$
Essentially, the cost of a table $T$ represented via tuples of size $a \times b$
is the number of tuples multiplied by 2 (to record the row and column information)
plus $a \times b$.
The cost of recording the schema is a constant cost which is ignored.
For the special case of ROM and COM,
the constant $2$ becomes $1$ since we can record that information in the attribute names.
As before, we aim to identify the best $T$ to represent a certain collection
of cells $C$: let $\rmax$ and $\cmax$ represent the largest row and column number
attained by any of the cells in $C$.
We have the following cases:

\stitle{Sparse Spreadsheet.}
Consider the case when $C$ is very sparse: \ie for every cell, there is
a small probability $\alpha\ll 1$ with which a neighboring cell is filled.
In such a case, when compared to the RCV data model,
with high probability every rectangular data model will have
the same or similar number of tuples, but at least twice as much for $a \times b$,
since at least one of $a$ or $b$ would be $\geq 2$;
thus, the cost of the RCV data model will be the lowest in such a case.

\stitle{Dense, but Wide Spreadsheet.}
Consider the case when $C$ is very dense: \ie the probability of a cell
being empty is a small probability $\alpha\ll 1$,
and let $\cmax > \rmax$.
In such a setting, with high probability
all data models will have roughly equal values
for $a \times b \times \textrm{number of tuples}$, 
since the tuples will all be densely packed across the spreadsheet 
area in $C$.
So the only thing that will differ is the initial constant
multiplied by the number of tuples. 
When $\cmax > \rmax$, 

\stitle{Dense, but Narrow Spreadsheet.}
\agp{end}

In terms of access cost, we As noted in Section~\ref{sec:desiderata}, spreadsheet operations, \eg scrolling, formula evaluation, are rectangular, \ie access a rectangular range of cells. 
This motivates us to consider a workload comprising rectangular accesses, specifically $n_1$, $n_2$, and $n_3$ of cell, row, and column lookups respectively.
We formalize our claim as below.

We quantify the optimality of a data model based on its storage requirements and its performance in terms of response time for typical spreadsheet operations.  
Inspired by traditional relational databases' cost models, to quantify the response time, we consider the following two factors
\begin{paraenum}
	\item the number random accesses $D_S$ and
	\item the amount of data transfer $D_T$.
\end{paraenum} 
To quantify the storage cost, we consider the following factors   
\begin{paraenum}
	\item table creation cost $c_1$,
	\item cell storage cost $c_2$,
	\item schema storage cost $c_3$, and 
	\item key storage cost $c_4$.
\end{paraenum}

We claim that depending on the spreadsheet density and workload, one proposed primitive data models, \ie ROM, COM and RCV, is optimal for representing a spreadsheet~$\mathds{S}$, when the operations are restricted to rectangular regions.
As noted in Section~\ref{sec:desiderata}, spreadsheet operations, \eg scrolling, formula evaluation, are rectangular, \ie access a rectangular range of cells. 
This motivates us to consider a workload comprising rectangular accesses, specifically $n_1$, $n_2$, and $n_3$ of cell, row, and column lookups respectively.
We formalize our claim as below.

\begin{proposition}{\textbf{Primitive Data Model Optimality:}}
	\label{prep:optimal_tuple}
	For a spreadsheet~$\mathds{S}$, with respect to minimizing
	\begin{paraenum}
	\item \emph{response time}, \ie	
		the number of random accesses $D_S$ and
		the amount of data transfer $D_T$
	for a workload comprising of $n_1$, $n_2$, and $n_3$ of cell, row, and column lookups respectively and
	\item \emph{storage cost}, \ie	
		table creation cost $c_1$,
		cell storage cost $c_2$,
		schema storage cost $c_3$, and 
		key storage cost $c_4$;
	\end{paraenum}
	 we have:
	\begin{paraenum}
		\item RCV is optimal when $\mathds{S}$ is \emph{sparse} and $n_1 \gg n_2$ and $n_1 \gg n_3$.
		\item ROM is optimal when $\mathds{S}$ is \emph{dense} and $n_2 \gg n_1$ and $n_2 \gg n_3$.
		\item COM is optimal when $\mathds{S}$ is \emph{dense} and $n_3 \gg n_1$ and $n_3 \gg n_2$.
	\end{paraenum}
\end{proposition}

\begin{IEEEproof}[for proposition~\ref{prep:optimal_tuple}]

Consider a spreadsheet of dimensions $m\times n$, which can 
be visualized as a rectangle.
We store this spreadsheet as a table in a relational database 
by {\em dividing} the sheet into non-overlapping rectangles, 
where each rectangle corresponds to a tuple in the table.
Let the tuples be of dimension $p\times q$.
The question we address in this section is the following: how to 
determine $p$ and $q$, given an instance of the spreadsheet, along 
with the associated workload of formulae we wish to perform on the 
said spreadsheet.

We factor in the workload of formulae while determining the data
model (finding the optimal tuple dimensions $p\times q$)
because formulae are quite common in spreadsheets, and optimizing for
access patterns while designing data models is crucial---recall 
takeaway 3 from Section \ref{sec:empirical_study}. Consider a general 
workload where:
\begin{itemize}
	\item Number of cell-based formulae: $n_1$ (\eg \code{B2=A2+5})
	\item Number of row-based formulae: $n_2$ (\eg  \code{E1=AVERAGE(B1:B30)})
	\item Number of column-based formulae: $n_3$ (\eg  \code{F5=SUM(A1:A100)})
\end{itemize}

We consider the aforementioned workload because typical formulae in 
spreadsheets access a rectangular range of cells at a time (takeaway 4). To compute a formulae, we need to
consider the following two classical metrics: 1) number of disk seeks required
to execute the formulae and 2) the amount of data transferred. These 
two metrics trade-off because as we increase the size of the tuple ($p\times q$),
the number of disk seeks reduces while the amount of data transferred
increases, and vice-versa. We shall formalize this interplay soon.

We assume that for a given formulae, the disk seeks are independent
of each other. Throughout this proof, we assume that each lookup operation
to execute a formulae corresponds to a disk seek. Note that this assumption is 
representative of the worse-case scenario when data accesses are truly random.




\stitle{Problem Formulation.}
We shall denote the number of disk seeks as $D_S$, and the amount of data 
transferred as $D_T$. We assume a general workload where the number of cell-based
formulae, row-based formulae and column-based formulae are $n_1$, $n_2$ and $n_3$
respectively.
\begin{align}
D_S &= n_1 + n_2 \cdot \frac{n}{q} + n_3 \cdot \frac{m}{p} \text{\qquad and}\\
D_T &= n_1 \cdot p \cdot q + n_2 \cdot p\cdot n + n_3 \cdot q \cdot m. 
\end{align}
We represent the relative importance of $D_S$ over $D_T$ as $k$, where $0\leq k\leq 1$. 
Therefore, we want to minimize $k\cdot D_S + (1-k)\cdot D_T$.
Depending on whether we want to give more importance to minimizing $D_S$ or $D_T$, we have the value of $k$ less $0.5$ or more than $0.5$ respectively. 
Therefore, we want to minimize\footnote{We show optimizing for storage costs as intractable in Section \ref{sec:np-hard}}:
\begin{align}
f(p,q) &= k \cdot D_S + (1-k)\cdot D_T\\
&= k \cdot \left(n_1 + n_2\cdot \frac{n}{q} + n_3\cdot\frac{m}{p}\right) \nonumber \\
& + (1-k)\cdot(n_1\cdot p\cdot q + n_2\cdot p\cdot n + n_3\cdot q\cdot m) \label{eg:pq_opt}
\end{align}

Since we have $1\leq p\leq m$, $1\leq q\leq n$, and $p,q\in \mathbb{N}$, we can enumerate all admissible values of $p, q$ and find the optimal value in \O{m\cdot n} time.




Since a spreadsheet's density impacts the values of $p$ and $q$, we consider two cases based on the density of the spreadsheet. We formally show that our primitive data models are {\em optimal} in these considerations.

\stitle{Case 1: Sparse spreadsheets.}
\label{sec:sparseblocksize}
Spreadsheets that are relatively sparse are typically meant for presentation of data and hence we presume that individual cell lookups are important and dominate the access in contrast with row-based and column-based accesses.
Therefore, we have $n_1 \gg n_2$ and $n_1 \gg n_3$. From Equation~\ref{eg:pq_opt}, we have: 

\begin{align}
f(p,q) &= k\cdot\left(n_1 + n_2\cdot\frac{n}{q} + n_3\cdot\frac{m}{p}\right) \nonumber \\
&+ (1-k)\cdot(n_1\cdot p\cdot q + n_2\cdot p\cdot n + n_3\cdot q\cdot m)\\
&= n_1 \cdot k \cdot \left(1 + \frac{n_2}{n_1}\cdot\frac{n}{q} + \frac{n_3}{n_1}\cdot\frac{m}{p}\right) \nonumber \\ 
&+ n_1 \cdot(1-k)\cdot\left(p\cdot q + \frac{n_2}{n_1}\cdot p\cdot n + \frac{n_3}{n_1}\cdot q\cdot m\right).\\
\intertext{Since we have $\frac{n_2}{n_1}\approx 0, \frac{n_3}{n_1}\approx 0$,}
f(p,q) &\approx n_1  \cdot (k + (1-k) \cdot p\cdot q).
\end{align}

Therefore, when the spreadsheet is sparse, $f(p,q)$ is minimized when $p = 1, q = 1$. 
Hence, restricting each tuple to a {\em single cell} of the spreadsheet minimizes disk seeks and data transfer. 
Note that this is independent of the value of $k$, for $k < 1$.
This justifies a data model where a cell corresponds to a tuple is an optimum choice for sparse spreadsheets.  

\stitle{Case 2: Dense spreadsheets.}
Dense spreadsheets typically correspond to logical tables that contain lots of data. 
We consider the two fundamental ways of arranging two dimensional data in a spreadsheet: row-oriented  and column-oriented. 
We posit that for data laid out in a row-oriented fashion, row lookups would dominate column lookups and cell lookups. Hence in this case, $n_2 \gg n_1$ and $n_2 \gg n_3$. From Equation~\ref{eg:pq_opt}, an approximation similar to the sparse case  leads us to:

\begin{align}
f(p,q) &= k\cdot\left(n_1 + n_2\cdot\frac{n}{q} + n_3\cdot\frac{m}{p}\right) \nonumber \\
&+ (1-k)\cdot(n_1\cdot p\cdot q + n_2\cdot p\cdot n + n_3\cdot q\cdot m) \\
&= n_2\cdot k\left(\frac{n_1}{n_2} + \frac{n}{q} + \frac{n_3}{n_2}\cdot\frac{m}{p}\right) \nonumber \\
&+ n_2\cdot (1-k)\cdot \left(\frac{n_1}{n_2}\cdot p\cdot q + p\cdot n + \frac{n_3}{n_2}\cdot q\cdot m\right). \\
\intertext{Since we have $\frac{n_1}{n_2}\approx 0, \frac{n_3}{n_2}\approx 0$,}
f(p,q) &\approx n_2 \cdot \left(\frac{k\cdot n}{q} + (1-k) \cdot p\cdot n \right)
\end{align}

Therefore, when the spreadsheet is dense, $f(p,q)$ is minimized when $q$ is maximized and $p$ is minimized. Since $q\leq n, p\geq 1$, the optimal value of $(p,q) = (1,n)$. Hence, restricting each tuple to a {\em single row} of the spreadsheet minimizes disk seeks and loading extra cells for $0<k<1$.

Similarly, if the data is laid out in a column-oriented fashion, column lookups would dominate row lookups and cell lookups. Hence in this case, $n_3\gg n_1$ and $n_3\gg n_2$. An argument similar to the row-oriented case  would result in the optimal value of $(p,q)$ to be $(m,1)$, which corresponds to having each tuple as a {\em single column} of the spreadsheet.\end{IEEEproof}

}

\section{Hybrid Decomposition Details}
\subsection{Hybrid Decomposition: Hardness}
\label{sec:np-hard}

\noindent 
Here, we demonstrate that Problem~\ref{thm:hybrid-rom} is {\sc NP-Hard};
for the decision version of the problem, a value $k$ is provided,
and the goal is to test if there is a hybrid data model
with $\cost{T} \leq k$.

We use a reduction from the minimum edge length 
partitioning problem
of rectilinear polygons~\cite{Lingas198253}.
A rectilinear polygon is one
in which all edges are either aligned with the $x$-axis or the $y$-axis.
The minimality criterion is 
the total length of the edges (lines) used 
to form the internal partition. 
Notice that this doesn't necessarily correspond to 
the minimality criterion of reducing the 
number of components. 
We  illustrate this in Figure~\ref{fig:min-edge-example}, 
which is borrowed from the original paper~\cite{Lingas198253}.
The following decision problem was shown to be 
NP-Hard in \cite{Lingas198253}: 
Given any rectilinear polygon $P$ and a number $k$, 
is there a rectangular partitioning whose total edge length does not exceed $k$?
	
\begin{figure}
		\centering
		\includegraphics[width=0.75\columnwidth]{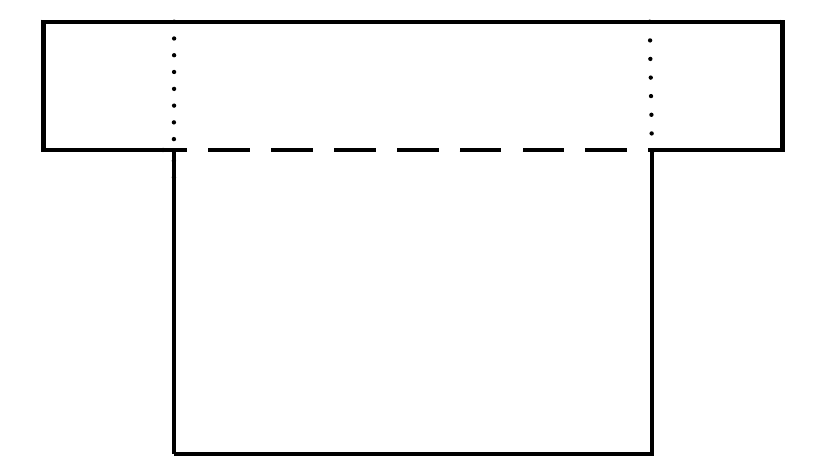}
		\vspace{-5pt}
		\caption{Minimum number of rectangles ({\bf --} {\bf --} {\bf --}) does not coincide with minimum edge length ($\cdots$).}
		\label{fig:min-edge-example}
		\vspace{-12pt}
	\end{figure}	

\begin{IEEEproof}	
Consider an instance $P$ of the polygon partitioning problem 
with minimum edge length required to be at most $k$. 
We now represent the polygon $P$ in a spreadsheet 
by filling the cells interior of the polygon with arbitrary values, 
and not filling any other cell in the spreadsheet. 
Let $C = \{C_1, C_2,\hdots,C_m\}$ represent the set 
of all filled cells in the spreadsheet. 
We claim that a minimum edge length partition of 
the given rectilinear polygon $P$ of length at 
most $k$ exists iff there is a solution for the following setting 
of the optimal hybrid data model problem: $s_1 = 0, s_2 = 2|C|+1, s_3 = s_4 = 1$, 
where the storage cost should not exceed 
$k' = k + \frac{\operatorname{Perimeter}(P)}{2} + (2|C|+1)|C|$ 
for some decomposition of the spreadsheet.

\smallskip	
\noindent $\Rightarrow$ Say the spreadsheet 
we generate using $P$ has a decomposition of 
rectangles whose storage cost is less than $k' = 
k + \frac{\operatorname{Perimeter}(P)}{2} + s_2|C|$. We have to show that there exists a partition with minimum edge length of at most $k$.

First, notice that there exists a valid decomposition that doesn't store any blank cell. 
Say there is a decomposition that stores a blank cell. Since we are now storing $|C| + 1$ cells at minimum, 
$k' > s_2(|C| + 1) = |C|s_2 + s_2 = |C|s_2 + 2|C| + 1$ and thus $k' > |C|(s_2 + 1 + 1)$, which is the cost of storing each cell in a separate table.
Therefore, if we have a decomposition that stores a blank cell, we also have a decomposition 
that does not store any blank cell and has lower cost.
Second, there exists a decomposition of the spreadsheet where all the tables are disjoint. 
The argument is similar to the previous case since storing the same cell twice 
in different tables is equivalent to storing an extra blank cell.
	
From our above two observations, we conclude that there exists a 
decomposition where all tables are disjoint, and no table stores a blank cell. 
Therefore, this decomposition corresponds to {\em partitioning} the given spreadsheet into rectangles. 
We represent this partition of the spreadsheet by $T = \{T_1, T_2,\hdots,T_p\}$. 
We now show that this partition of the spreadsheet corresponds to a partitioning of the rectilinear polygon $P$ with edge-length less than $k$.
On setting $s_1 = 0, s_2 = 2|C | + 1, s_3 = s_4 = 1$, we get:
	\begin{align}
	\operatorname{cost}(T) &=\sum_{i=1}^p 0 + s_2|C| + 1\cdot \left(\sum_{i=1}^p c_i + \sum_{i=1}^p  r_i\right)\\
	\intertext{since $\operatorname{cost}(T) \leq k' = k + \frac{\operatorname{Perimeter}(P)}{2} + s_2|C|$,}
	& \sum_{i=1}^p (r_i + c_i) \leq k + \frac{\operatorname{Perimeter}(P)}{2}\\
	\tr{& \sum_{i=1}^p \frac{\operatorname{Perimeter}(T_i)}{2} \leq k + \frac{\operatorname{Perimeter}(P)}{2}\\}
	\implies &\sum_{i=1}^p \operatorname{Perimeter}(T_i) \leq 2\times k + \operatorname{Perimeter}(P)
	\end{align}
Since the sum of perimeters of all the tables $T_i$ counts the boundary of $P$ exactly once, 
and the edge length partition of $P$ exactly twice, the partition of the spreadsheet $T = \{T_1, T_2,\hdots,T_p\}$ corresponds to an edge-length partitioning of the given rectilinear polygon $P$ with edge-length less than $k$.

\smallskip	
\noindent $\Leftarrow$ Let us assume that the given rectilinear polygon $P$ has a minimum edge length partition of length at most $k$. We have to show 
that there exists a decomposition of the spreadsheet whose storage cost is at most $k' = k + \frac{\operatorname{Perimeter}(P)}{2} + s_2|C|$. 
\paper{This part of the proof is straightforward, and can be found in our technical report~\cite{techreport}.}
\tr{Let us represent the set of rectangles that corresponds to an edge length partition of $P$ of at most $k$ as $T = \{T_1, T_2,\hdots,T_p\}$. We shall use the partition $T$ of $P$ as the decomposition of the spreadsheet itself:

\begin{multline}
\operatorname{cost}(T) =\sum_{i=1}^p s_1 + s_2\cdot (r_i\times c_i) + s_3\cdot c_i + s_4\cdot r_i
	\end{multline}	
\begin{multline}
\operatorname{cost}(T) =\sum_{i=1}^p s_1 + s_2\sum_{i=1}^p \cdot (r_i\times c_i) \\
	+ s_3\sum_{i=1}^p c_i 
+ s_4\sum_{i=1}^p r_i
\end{multline}	
substituting $s_1 = 0, s_2 = 2|C | + 1, s_3 = s_4 = 1$, we get:
\begin{flalign}
\operatorname{cost}(T) = \sum_{i=1}^p 0 + s_2|C| + 1\cdot \left(\sum_{i=1}^p c_i + \sum_{i=1}^p  r_i\right) &&
\end{flalign}
\begin{flalign}
\operatorname{cost}(T) = s_2|C| + \sum_{i=1}^p (r_i + c_i) &&
\end{flalign}
\begin{flalign}
\operatorname{cost}(T) = s_2|C| + \sum_{i=1}^p \frac{\operatorname{Perimeter}(T_i)}{2} &&
\end{flalign}
since $\sum_{i=1}^p \operatorname{Perimeter}(T_i) = 2\times k + \operatorname{Perimeter}(P)$, we have:
\begin{flalign}
	\operatorname{cost}(T) = s_2|C| + k + \frac{\operatorname{Perimeter}(P)}{2} = k' &&
\end{flalign}
$\implies \operatorname{cost}(T) = k'$

	Therefore, the decomposition of the spreadsheet using $T$ corresponds to a decomposition whose storage cost equals $k'$. Note that our reduction can be done in polynomial time. Therefore we can solve the minimum length partitioning problem in polynomial time, if we have a polynomial time solution to the optimal storage problem. However, since the minimum length partitioning problem is \mbox{NP-Hard}~\cite{Lingas198253}, the optimal hybrid data model problem is \mbox{NP-Hard}.
This completes our proof.
	}
\end{IEEEproof}

\stitle{Hardness Results for Extensions.}
So far, we have only considered the ROM data model for our tables, 
and assume that data is duplicated when two tables overlap. 
It can be shown that even when we extend to the case when RCV is permitted, and when
we allow data to be represented in multiple tables, the problem continues to be {\sc NP-Hard}.
\paper{These proofs can be found in our technical report~\cite{techreport}.}
\tr{
For each case we consider, we restrict our argument to an outline of the proof 
since it largely mimics what is presented for the basic case.

\emtitle{ROM and RCV; without data overlap.}
Now, we consider the following problem. Given a spreadsheet with data, our goal is to decompose the spreadsheet into tables (using ROM and RCV) such that the overall storage cost is minimized, with no two tables overlapping with each other.

\noindent (Proof sketch) Our basic idea is to achieve two goals:
\begin{paraenum}
\item ensure that the blank cells in the spreadsheets aren't stored in any table and
\item ensure that we always prefer ROM over RCV, since we can then employ our proof for Problem~\ref{prob:hybrid-rom}.
\end{paraenum}

We achieve goal~1 by setting $s_2=\infty$ in our cost model, and goal~2 by setting $s_1 = 0$, $s_3=1$, and $s_4=1$ as before. We now use a reduction from
\emph{minimum edge length 
partitioning problem
of rectilinear polygons}~\cite{Lingas198253}
as before to show hardness. Note that if we allow tables to overlap, and when they do, if we duplicate the data in the overlapping cells in both tables, setting $s_1=0, s_2=\infty, s_3=1$, and $s_4=1$ can lead to a reduction from the \emph{minimum edge length 
partitioning problem
of rectilinear polygons} to show hardness. The same proof as for Problem~\ref{prob:hybrid-rom} continues to hold.

\emtitle{ROM with overlap; without data duplication.}
Now, we consider a variant of the previous setting: when the tables overlap, we do not duplicate the same cell's value in two different tables. Note that this is different from earlier formulations, where we indeed duplicated data across overlapping tables.

\noindent (Proof sketch) Our basic idea is to achieve to achieve two goals:
\begin{paraenum}
\item ensure that blank cells aren't stored in any table and
\item only the number of rectangles, \ie tables, we have matters, not their dimensions.
\end{paraenum}

We achieve goal~1 by setting $s_2 = \infty$ in our cost model, and goal~2 by setting $s_3 =0$ and $s_4=0$. We have $s_1=1$.  We now use a reduction from the rectangle covering problem~\cite{Culberson19942} to show hardness. The reduction is straightforward since we are essentially minimizing the number of tables to cover all of the filled cells in the sheet, which is exactly what the rectangle covering problem aims to minimize.
}

\tr{
\subsection{Hybrid Decomposition: DP Bound}
\label{sec:dp-bound}

\stitle{Approximation Bound for DP.}
Here, we obtain an approximation bound for our dynamic programming formulation discussed in Section~\ref{sec:optimal-recursive-decomposition}. Say we have $k$ rectangles in the optimal decomposition; with storage cost $c$. Then, we show that the dynamic programming algorithm, which explores the entire space of 
models obtained from
recursive decomposition, identifies a decomposition with a cost that is at most 
$c+s_1\times\frac{k(k-1)}{2}$, where $s_1$ is the cost of storing a new table as in Equation~\ref{eq:table-rom}. 

\begin{figure}
		\centering
		\includegraphics[width=0.70\columnwidth]{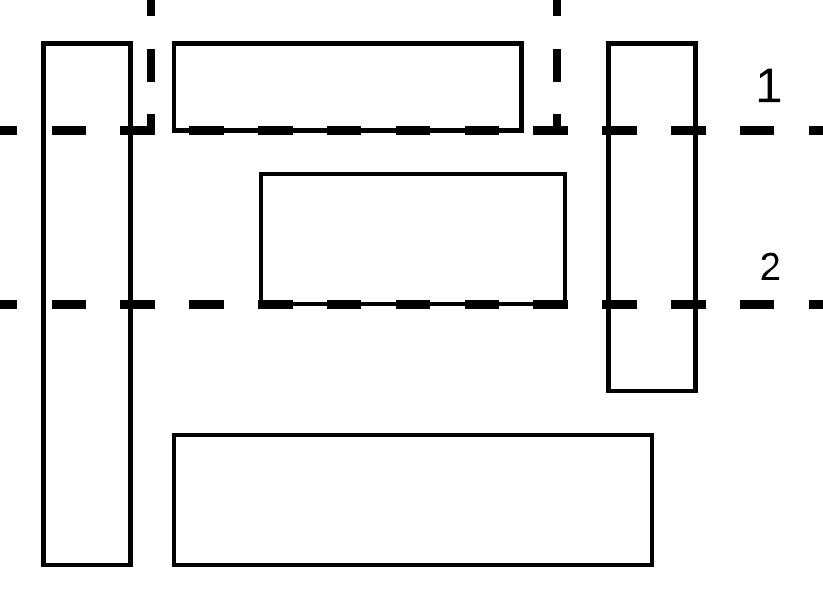}
		\caption{Obtaining a recursive decomposition from the optimal solution.}
		\label{fig:fig-dp-bound}
         \vspace{-12pt}
\end{figure}

\begin{IEEEproof}
Let the optimal decomposition consist of a set of five rectangles $R$ as in Figure~\ref{fig:fig-dp-bound}. 
Starting from $R$, we will construct a recursive decomposition solution with cost $c+s_1\times\frac{k(k-1)}{2}$, denoted as $R'$, using the following steps.
Sort the rectangles from the optimal solution in the increasing order of their bottom edge. 
Pick the first rectangle, and use 
a line through its 
 bottom edge to cut or partition the remaining rectangles.
This is the first ``partitioning'' step, denoted as~1 in Figure~\ref{fig:fig-dp-bound}. 
This partitioning step leads to two portions. 
We handle the top portion with vertical partitions, while for the bottom portion we recurse. 
This partition introduces at most $k-1$ new rectangles in the top half and eliminates one rectangle. 

Thus, at every step, we have $k-1$ new rectangles and reduce the total number of rectangles by~$1$. 
That is, the next partition will introduce at most $k-2$ rectangles; and so on. 
So, we in total we $(k-1)+(k-2)+\ldots+1=\frac{k(k-1)}{2}$ new rectangles. 
Since the dynamic programming algorithm explores the entire space of recursive decomposition based data models, it also considers $R'$ as one of the candidates. Thus; its solution must be at least as good.
Hence proved.
\end{IEEEproof}

\stitle{Bound for number of Tables.}
Since the bound given by Theorem~\ref{th:dp_bound} and proved above is an additive bound in terms of number of tables, we use Theorem~\ref{th:cc_tables} to show that practically the number of tables in the optimal decomposition is small; thereby enabling us to claim that our dynamic programming solution is not too far from optimal. Here, we show that the optimal solution to Problem~\ref{prob:hybrid-rom} for a minimum bounding rectangle of a connected component will have at most 
$\left\lfloor\frac{e \times s_2}{s_1} + 1 \right\rfloor$ 
rectangles, where $e$ is the number of empty cells in the bounding rectangle.

\begin{IEEEproof}
Let the optimal decomposition for a minimum bounding rectangle of a connected component $C$ have $k'$ tables. 
Therefore, we have the cost representing the minimum bounding rectangle using a single table is higher than the optimal decomposition into $k'$ tables, \ie 
\begin{multline}
\label{eq:fullcost_compare}
\sum_{i=1}^{k'} s_1 + s_2\cdot (r_i\times c_i) + s_3\cdot c_i + s_4\cdot r_i
\leq
\\
s_1 + s_2\cdot (r_0\times c_0) + s_3\cdot c_0 + s_4\cdot r_0,
\end{multline}
where $r_1,..,r_{k'}$ and $c_1,..,c_{k'}$ are the number of rows and columns respectively for each of the tables in the optimal decomposition
and $r_0$ and $c_0$ are number of rows and columns for the     minimum bounding rectangle. 

Since our region of focus is a minimum bounding rectangle encapsulating a connected component, we do not have any empty rows or columns. Thus, each row and column should be captured by at least one rectangle in the optimal decomposition. Hence, we have, 
\begin{align}
\label{eq:rowcol_cost_compare}
\sum_{i=1}^k s_3\cdot c_i + s_4\cdot r_i
\geq s_3\cdot c_0 + s_4\cdot r_0
\end{align}
Subtracting Equation~\ref{eq:rowcol_cost_compare} from \ref{eq:fullcost_compare} we have, 
\begin{align}
\label{eq:s1s2cost_compare}
\sum_{i=1}^k s_1 + s_2\cdot (r_i\times c_i)
\leq
s_1 + s_2\cdot (r_0\times c_0).
\end{align}
Since the optimal solution should represent all the filled-in cells at least once, we have  $\sum_{i=1}^k (r_i\times c_i) \geq r_0\times c_0 - e$, where $e$ is the number of empty cells in the bounding box. Subtracting this from Equation~\ref{eq:s1s2cost_compare} and simplifying, we get
\begin{align}
\label{eq:s1s2cost_compare2}
k\cdot s_1 &\leq s_1 + e\times s_2.\\
k &\leq 1 + \frac{e\times s_2}{s_1}.
\end{align} 
Since $k$ is an integer, we have 
\begin{align}
k &\leq \left\lfloor 1 + \frac{e\times s_2}{s_1} \right\rfloor.
\end{align} 
Hence proved.
\end{IEEEproof}
} 

\subsection{Hybrid Data Model: Extensions}\label{app:extensions}

In this section, we discuss a number of extensions 
to the cost model of the hybrid data model.
We will describe these extensions to the cost model,
and then describe the changes to the 
basic dynamic programming algorithm;
modifications to the greedy and aggressive greedy
decomposition algorithms are straightforward.

\subsubsection{RCV, COM and TOM}
The cost model can be extended
in a straightforward manner to allow each rectangular
area to be a ROM, COM, or an RCV table. (We deal with the TOM case later.)
First, note that it doesn't benefit us to have multiple RCV
tables---we can simply combine all of these tables into one,
and assume that we're paying a fixed up-front cost to have one RCV table.
Then, the cost for a table $T_i$, if it is stored as a COM table is:
\begin{align*}
\comCost{T_i} = s_1 + s_2\cdot (r_i\times c_i) + s_4\cdot c_i + s_3\cdot r_i.
\end{align*}
This equation is the same as Equation~\ref{eq:table-rom}, 
but with the last two constants transposed.
And the cost for a table $T_i$, if it is stored as an RCV table is simply:
\begin{align*}
\rcvCost{T_i} = s_5 \times \#\text{cells}.
\end{align*}
where $s_5$ is the cost incurred per tuple.
Once we have this cost model set up, it is straightforward
to apply dynamic programming once again to identify the optimal
hybrid data model encompassing ROM, COM, and RCV.
The only step that changes in the dynamic programming equations is Equation~\ref{eq:romcost},
where we have to consider the COM and RCV alternatives in addition to ROM.
To handle TOM tables, we assume that the corresponding cells are empty; while also setting the $\romCost$ and $\comCost$ of any tables overlapping with these cells as $\infty$. 
We have the following theorem.
\begin{theorem}[ROM, COM, TOM, and RCV]
The optimal ROM, COM, TOM, and RCV-based hybrid data model 
based on recursive decomposition 
can be determined in {\sc PTIME}.
\end{theorem}

\subsubsection{Incremental Decomposition}\label{sec:incr-decomp}
So far, we have focused on finding an optimal decomposition given a static spreadsheet. 
We now consider how we can support incremental maintenance of the decomposition across updates. 
Here, along with the storage cost, we also consider the cost of migrating cells from an existing 
decomposition $T_o$ to a new decomposition $T$. We define the migration cost as
$\migCost{(x_1, y_1), (x_2, y_2)} = \#\text{cells}$, where $\#\text{cells}$ denotes the number of populated cells in the rectangular region defined by $(x_1, y_1), (x_2, y_2)$.
To migrate a region of a spreadsheet into a new decomposition, we 
assume that we only use an existing table if it exactly covers the region;  
for all other cases we migrate all of the populated cells within the region to the new decomposition. 
\tr{In other words, we do not consider the cases when an existing table needs to be modified either to accommodate  or eliminate rows or columns.}
We introduce a factor $\eta$ to enable users to balance the trade-off between the migration cost and storage cost;
our objective is thus find a data model $T$ such that $\cost{T} + \eta\cdot\migCost{}$ is minimized. 



For incremental decomposition, we update the  dynamic programming formulation 
by adding an additional case that retains the decomposition as is and updates 
the $\romCost{}$ to include the migration cost in terms of the number of cells 
that need to be migrated from the existing model into a new model. 
As the migration cost for a region is defined as the number of populated cells, the migration cost of a region can be computed independently of the remaining regions. 
This enables us to employ dynamic programming
once again.

\squishlist 
    \item Keep the decomposition as-is. 
    This is permissible only if the there exists a ROM model at $(x_1, y_1), (x_2, y_2)$ in $T_o$.
    \begin{multline}
    \romCost{(x_1, y_1), (x_2, y_2)} = \\ s_1 + s_2\cdot (r_{12}\times c_{12})  + s_3\cdot c_{12} + s_4\cdot r_{12},
    \end{multline}   
    \item Store the area as ROM by migrating the non-empty cells into the new model.  
       \begin{multline}
    \romCost{(x_1, y_1), (x_2, y_2)} = s_1 + s_2\cdot (r_{12}\times c_{12})  + \\                          s_3\cdot c_{12}  + s_4\cdot r_{12} + \eta \cdot \migCost{(x_1, y_1), (x_2, y_2)}
    \end{multline}        
\squishend
Note that we may be able to migrate a region more efficiently
by leveraging existing tables that partially cover the region; however,
this will lead to complications in leveraging tables that span more than one region,
and any reorganization costs involved. 
For simplicity, we do not consider such migrations. 
As we will demonstrate in Appendix~\ref{sec:app-exp}, this still leads
to adequate performance. 


\subsubsection{Access Cost.}
So far, we have only been focusing on storage.
Our cost model can be extended in a straightforward
manner to handle access cost---both scrolling-based operations,
and formulae, and our dynamic programming algorithms
can similarly be extended to handle access cost without any substantial changes.
We focus on formulae since they are often the more substantial cost of
the two; scrolling-based operations can be similarly handled.
For formulae, there are multiple aspects that contribute
to the time for access: the number of tables accessed,
and within each table, since data is retrieved at a tuple level,
the number of tuples that need to be accessed, and the size of these tuples.
Once again, each of these aspects can be captured within the
cost model via constants similar to $s_1, \ldots, s_5$, and 
can be seamlessly incorporated into the dynamic programming algorithm.
Thus, we have:

\begin{theorem}[Optimality with Access Cost]
The optimal ROM, COM, and RCV-based hybrid data model 
based on recursive decomposition, 
across both storage and access cost,
can be determined via dynamic programming.
\end{theorem}

\paper{\subsubsection{Size Limitations of Present Databases and Indexes}
In our technical report~\cite{techreport}, 
we include extensions to deal with the fact that current databases
pose limitations on the number of columns,
and also to include the costs of indexes for each of the
tables that comprise the hybrid data model.
}

\tr{
\subsubsection{Size Limitations of Present Databases.}
Current databases impose limitations on the number of columns
within a relation\footnote{\scriptsize Oracle column number limitations: \url{https://docs.oracle.com/cd/B19306_01/server.102/b14237/limits003.htm\#i288032}; MySQL column limitations: \url{https://dev.mysql.com/doc/mysql-reslimits-excerpt/5.5/en/column-count-limit.html}; PostgreSQL column limitations: \url{https://www.postgresql.org/about/} }; since spreadsheets often have an arbitrarily large
number of rows and columns (sometimes 10s of thousands each),
we need to be careful when trying to capture
a spreadsheet area within a collection of tables
that are represented in a database.

This is relatively straightforward to capture in our context:
in the case where we don't split (Equation~\ref{eq:romcost}),
if the number of columns is too large to 
be acceptable, we simply return $\infty$
as the cost. 

\begin{theorem}[Optimality with Size Constraints]
The storage optimal ROM, COM, and RCV-based hybrid data model, with
the constraint that no tables violate size constraints,
based on recursive decomposition,
can be determined via dynamic programming.
\end{theorem}

\subsubsection{Incorporating the Costs of Indexes.}
Within our cost model, it is straightforward to incorporate
the costs associated with storage of indexes,
since the size of the indexes are typically proportional to
the number of tuples for a given table,
and the cost of instantiating an index is another fixed constant
cost.
Since our cost model is general, by suitably re-weighting
one or more of $s_1, s_2, s_3, s_4$, we can capture
this aspect within our cost model, and apply 
the same dynamic programming algorithm.

\begin{theorem}[Optimality with Indexes]
The storage optimal ROM-based hybrid data model, with
the costs of indexes included,
based on recursive decomposition,
can be determined via dynamic programming.
\end{theorem}
}


\tr{
\section{Relational Operations Support}\label{app:rel-ops-support}


In addition to supporting standard spreadsheet functions, \system leverages the SQL engine of the 
underlying  
database to
seamlessly supports SQL queries and relational operators on the front-end spreadsheet interface.

\system supports executing of SQL queries via a spreadsheet function \code{sql(query, [param1], \ldots)}, which takes a SQL statement along with parameters values as arguments. 
The \code{query} parameter is a single SQL \code{SELECT} statement, possibly containing `?'s.
When one or more `?'s exists in the query, \system treats the query like a SQL prepared statement, where each `?' is substituted by the values \code{param1, \ldots} in order. 
The number of parameters must match the number of `?'s in the query. 
Each parameter must evaluate to a single value, \ie it cannot refer to a range.

The \code{sql} function and the other functions that we discuss in this section return a single composite table value;
to retrieve the individual rows and columns within that composite table value,
we have an \code{index(table, row, [column])} function that looks up the \code{(row, column)}th
cell in the composite table value in location \code{table}, and places it
in the current location.

In addition, \system supports relational operators via the following spreadsheet functions:
\code{
union(table1, table2),
difference(table1, table2),
intersection(table1, table2),
crossproduct(table1, table2),
select(table, filter),
join(table1, table2, [filter]),
project(table, attribute1, [attribute2], ...),} and
\code{rename(table, oldAttribute, newAttribute)}.

The arguments \code{table1} and \code{table2} can either refer to a composite table value or a (contiguous) range of non-table values, which is treated as a table. 
The \code{filter} argument must be a Boolean expression which may utilize standard spreadsheet functions and can refer to attributes in tables.
}
\begin{figure*}[t]
  \centering
  \subfloat{}{\includegraphics[width=0.3\textwidth,clip]{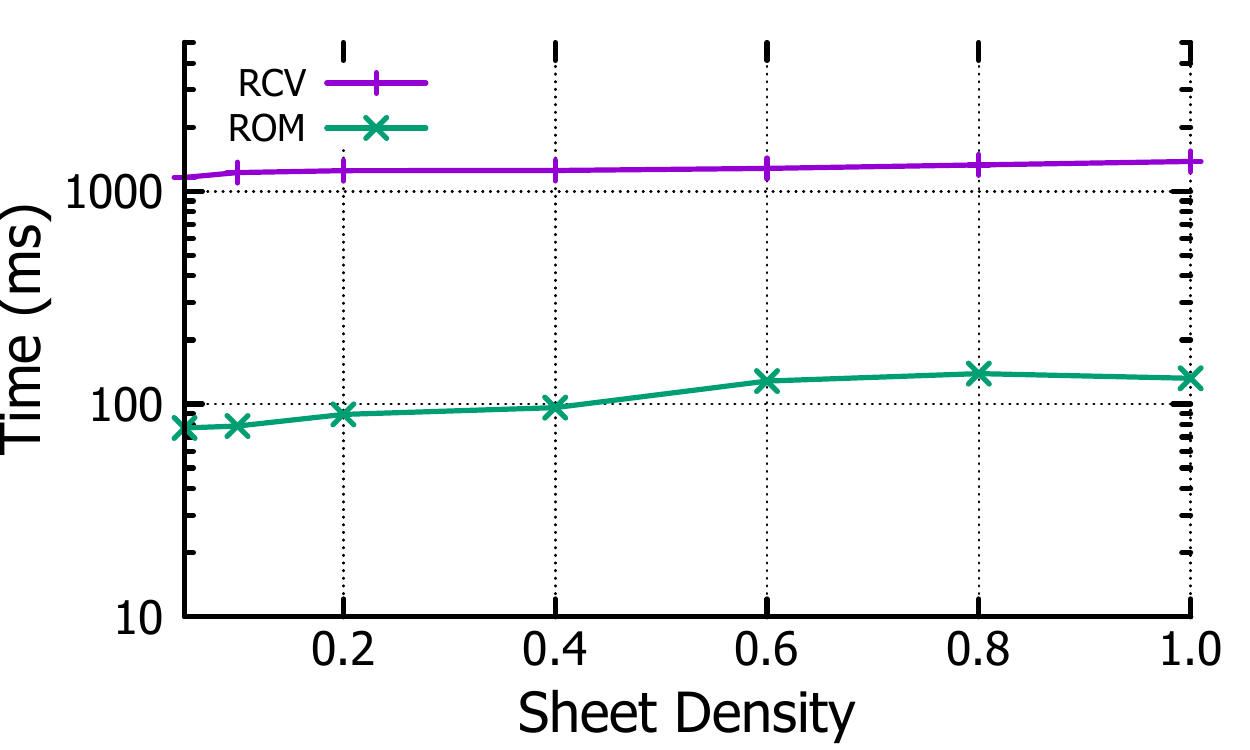}}
  \hspace{5pt}
  \subfloat{}{\includegraphics[width=0.3\textwidth,clip]{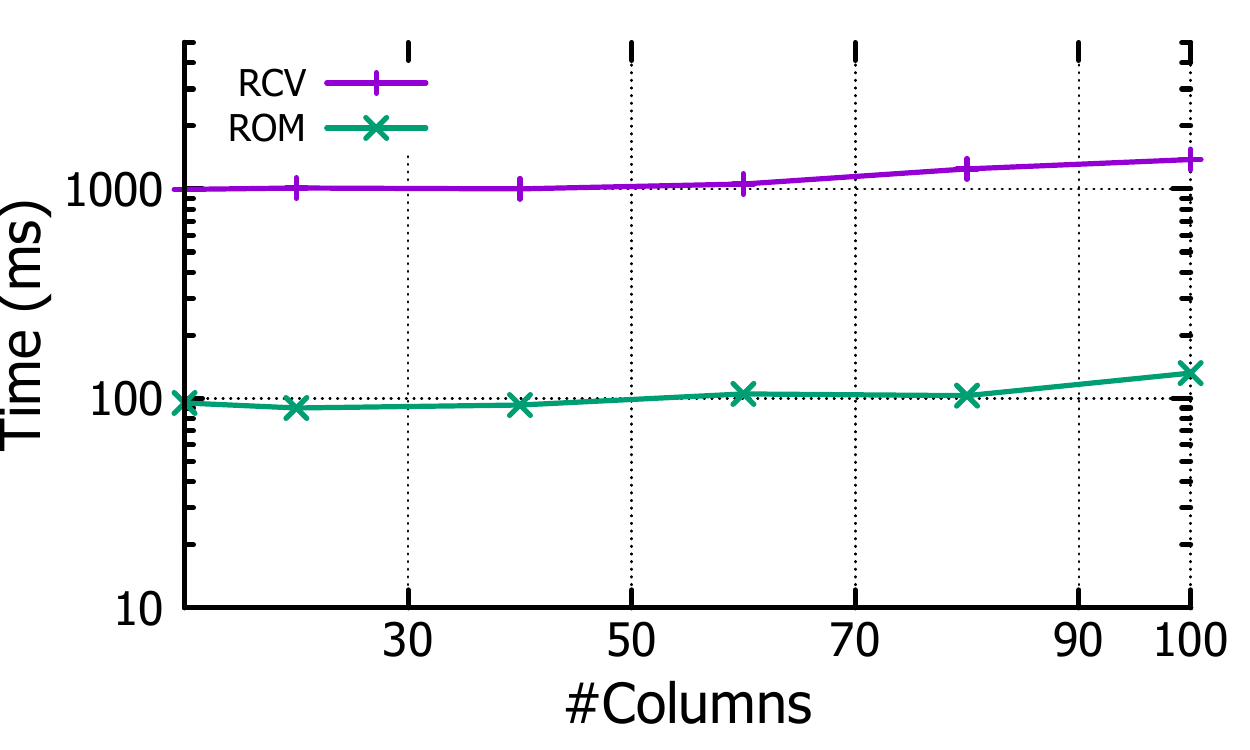}}
  \hspace{5pt}
  \subfloat{}{\includegraphics[width=0.3\textwidth,clip]{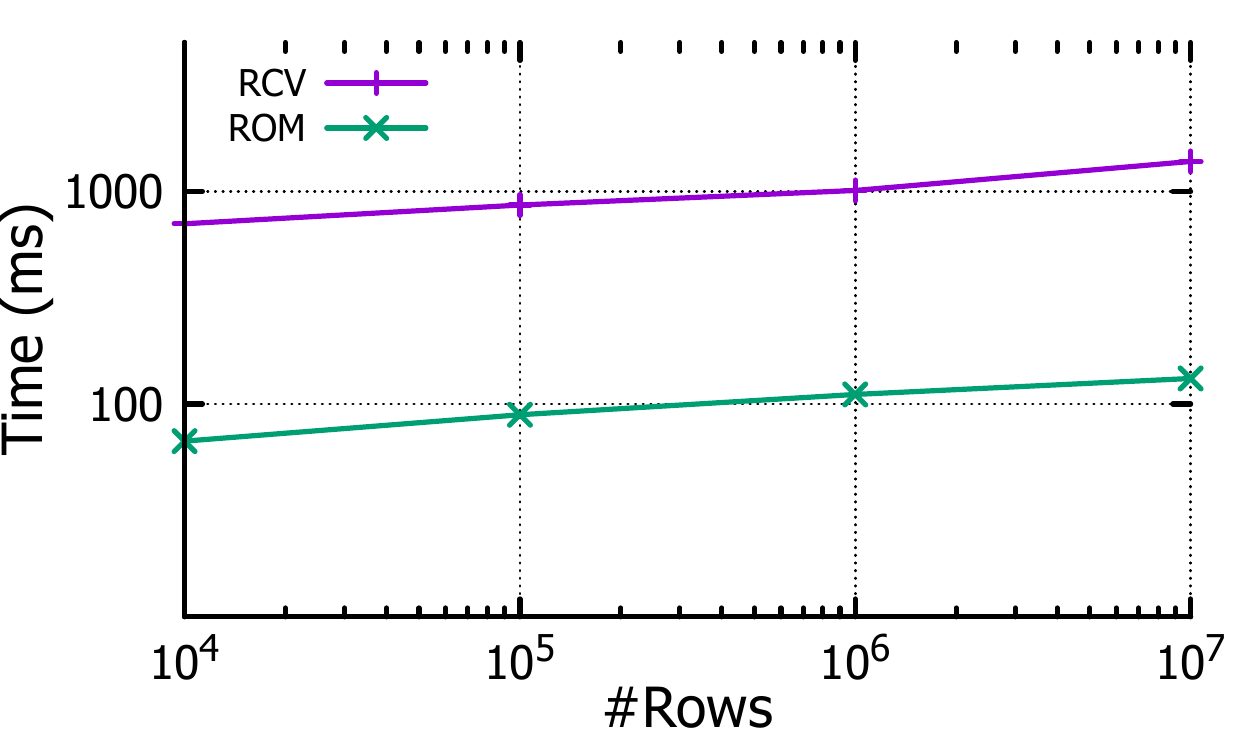}}
  \caption{Update range performance vs  (a) Sheet Density (b) Column Count (c) Row Count}
  \label{fig:dm_update_performance}
  \vspace{-12pt}
\end{figure*}

\begin{figure*}[t]
  \centering
  \subfloat{}{\includegraphics[width=0.3\textwidth,clip]{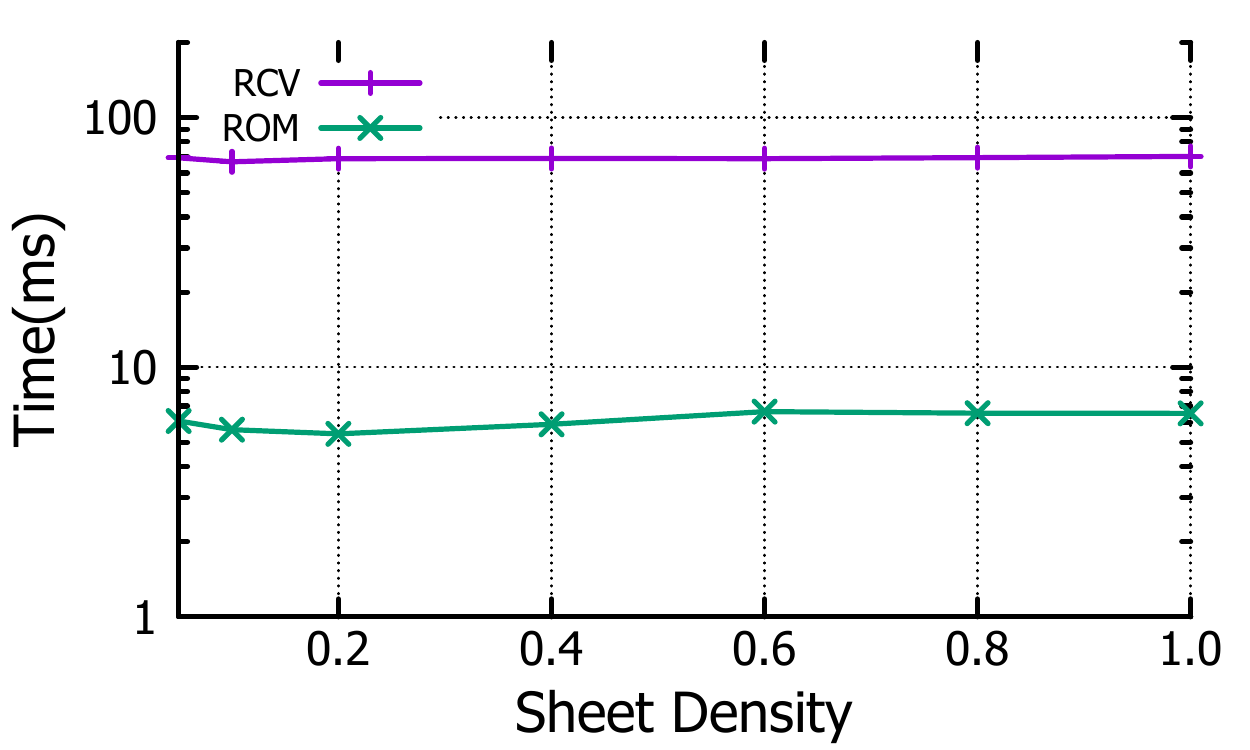}}
  \hspace{5pt}
  \subfloat{}{\includegraphics[width=0.3\textwidth,clip]{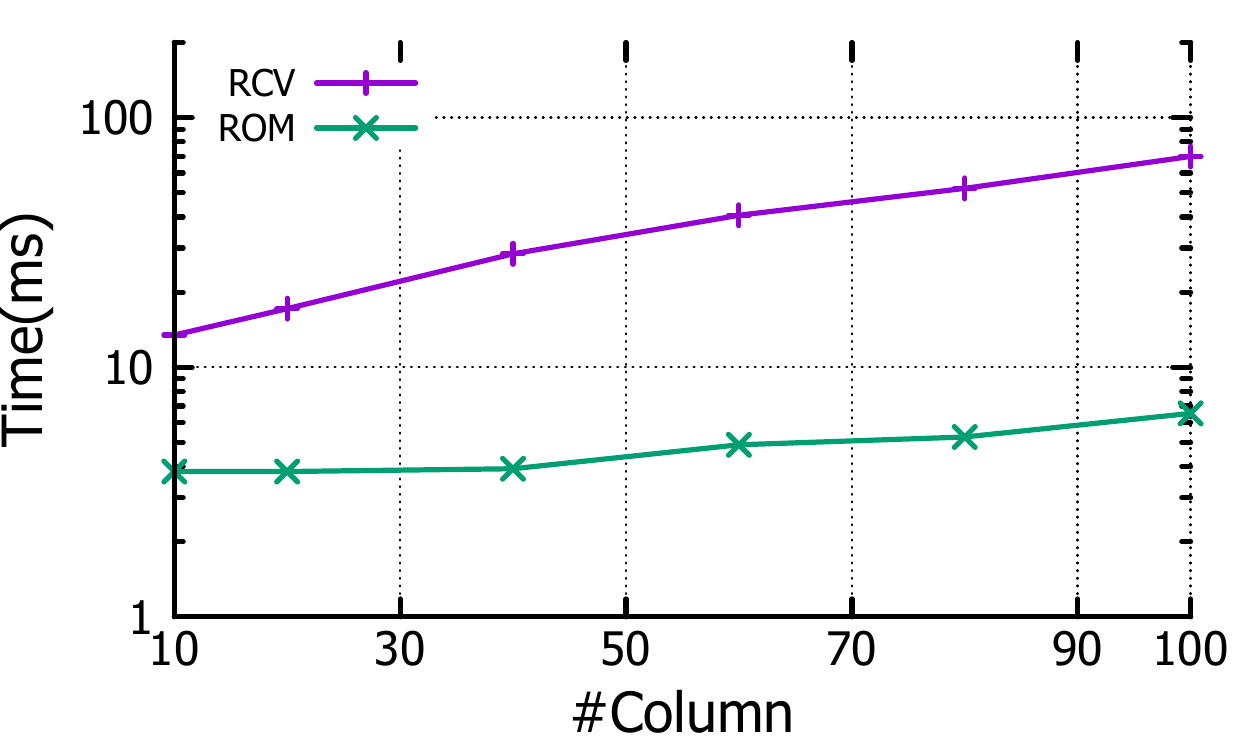}}
  \hspace{5pt}
  \subfloat{}{\includegraphics[width=0.3\textwidth,clip]{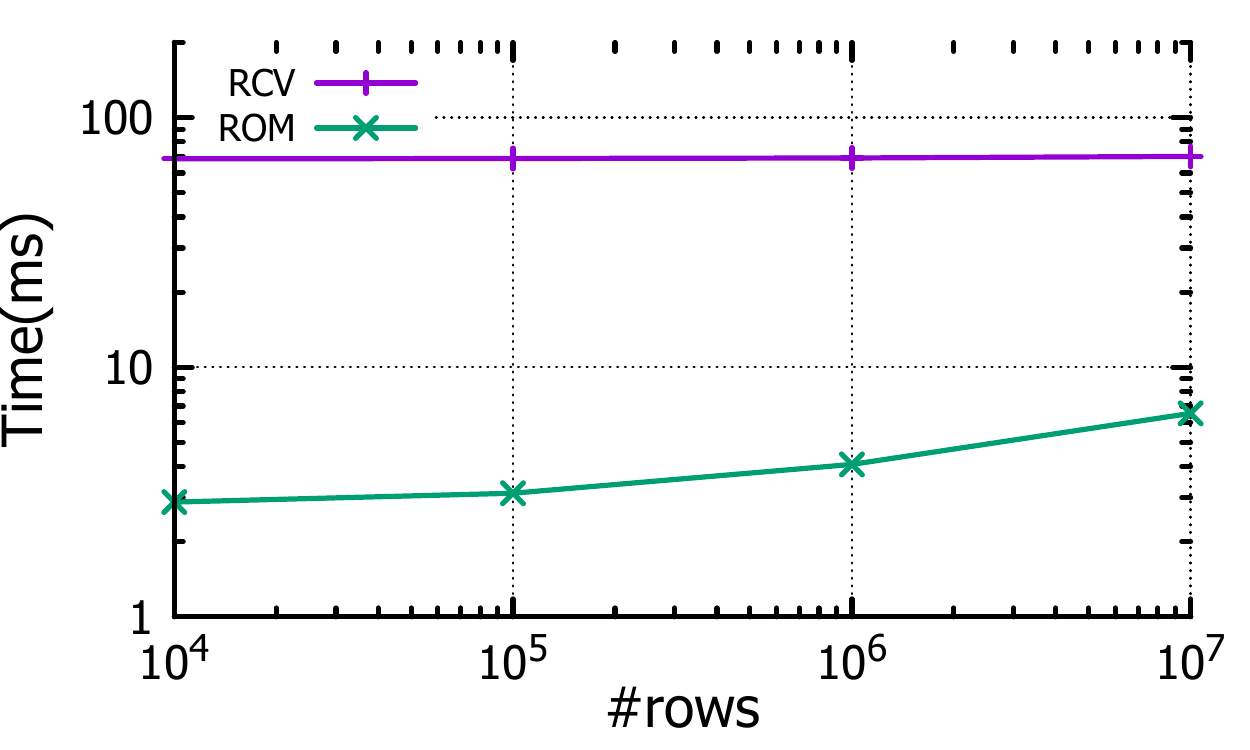}}
  \caption{Insert row performance vs (a) Sheet Density (b) Column Count (c) Row Count}
  \label{fig:dm_insert_row_performance}
  \vspace{-12pt}
\end{figure*}

\begin{figure*}[t]
  \centering
  \subfloat{}{\includegraphics[width=0.3\textwidth,clip]{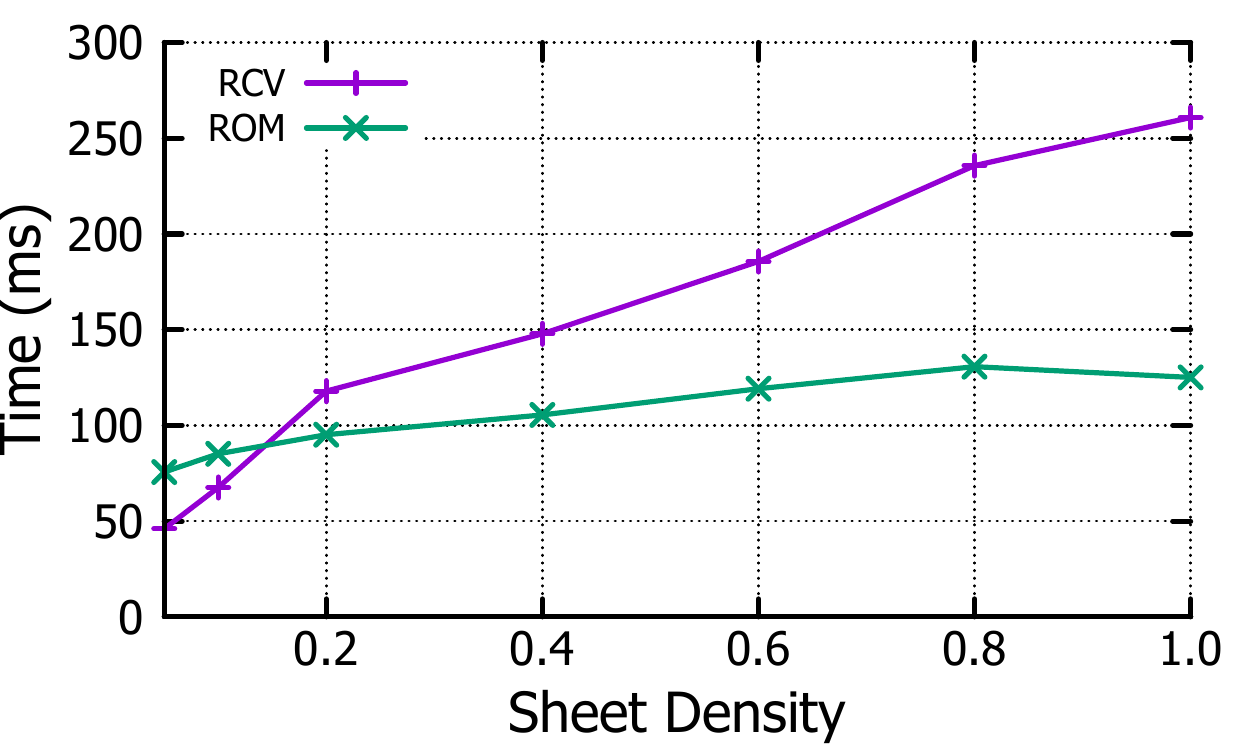}}
  \hspace{5pt}
  \subfloat{}{\includegraphics[width=0.3\textwidth,clip]{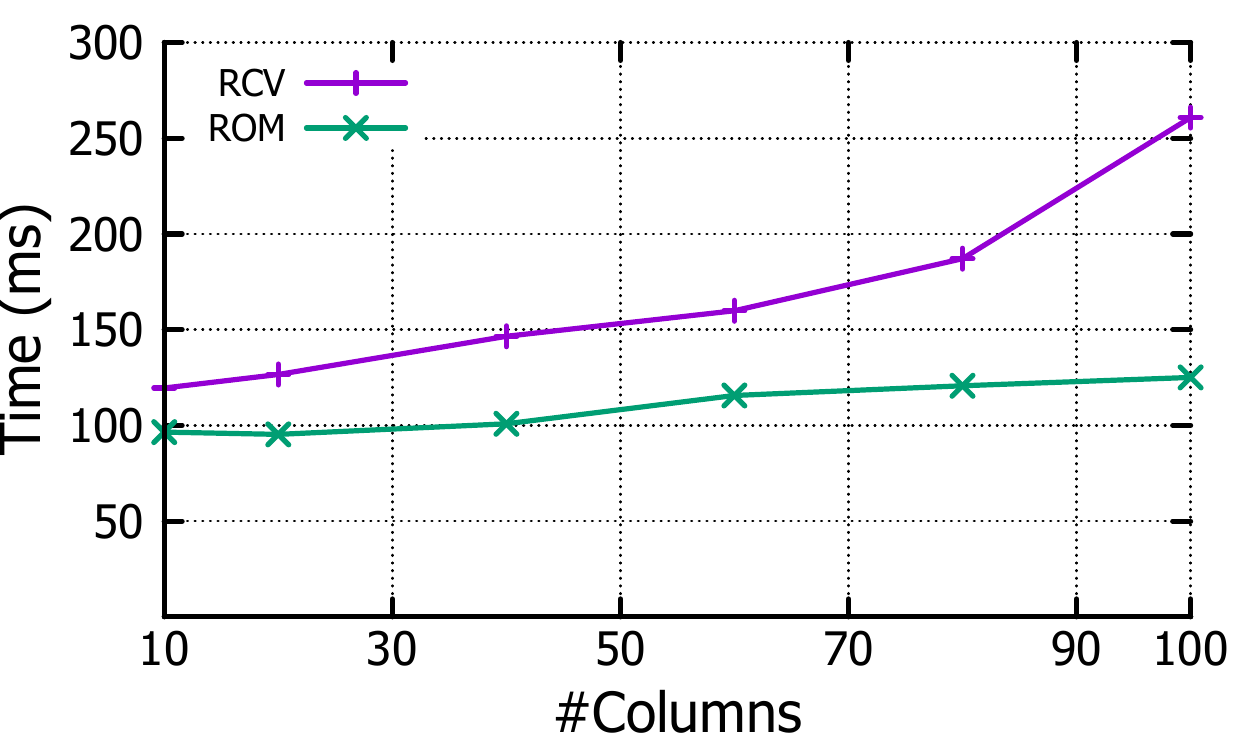}}
  \hspace{5pt}
  \subfloat{}{\includegraphics[width=0.3\textwidth,clip]{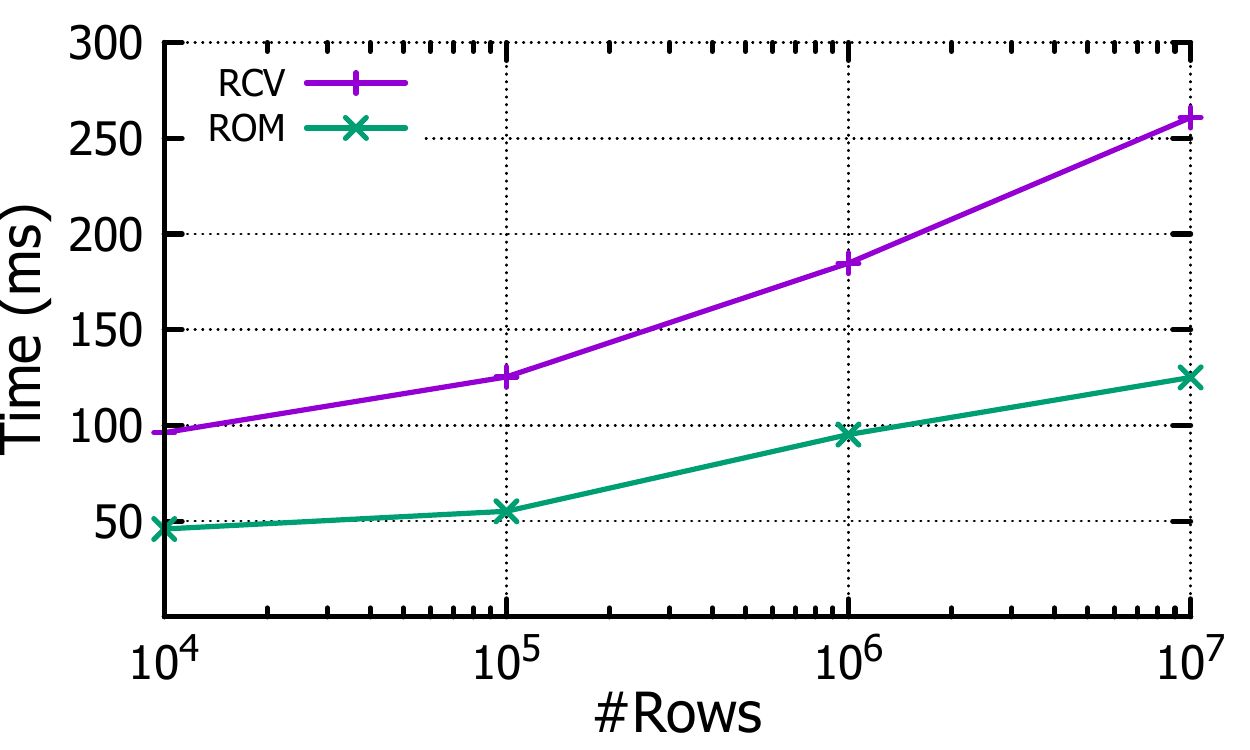}}
  \caption{Select performance vs --- (a) Sheet Density (b) Column Count (c) Row Count}
  \label{fig:dm_performance}
  \vspace{-12pt}
\end{figure*}

\section{Additional Experiments}\label{sec:app-exp}
In this section, we present additional evaluation of the storage engine of \system.

\subsection{Presentational Awareness and Access}
\begin{figure}[h]
	\centering
		\includegraphics[width=0.45\textwidth,clip]{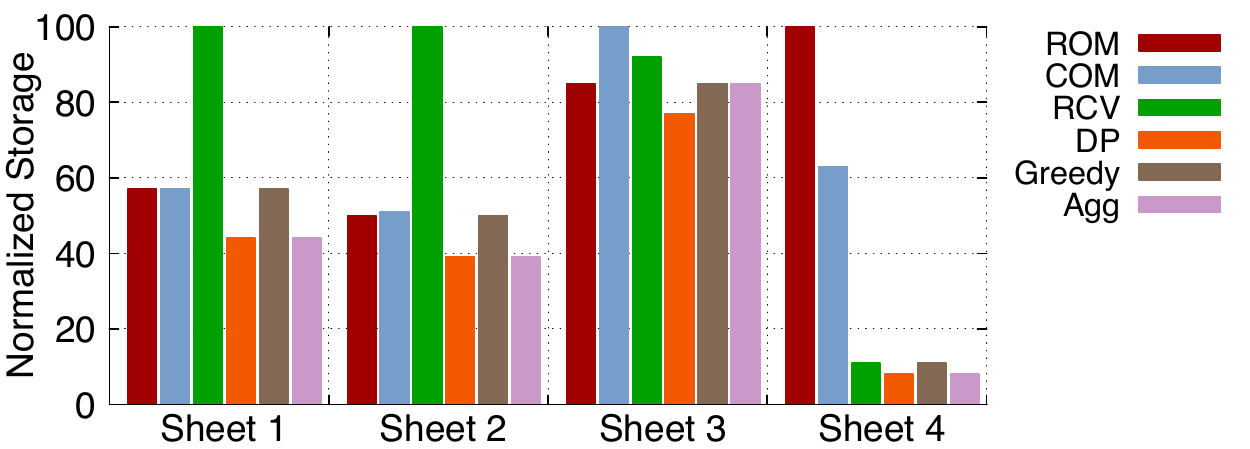}
	\caption{Storage comparison for sample spreadsheets.}
	\label{fig:storage_cost_samples}
\vspace{-5pt}
\end{figure}
\subsubsection{Drill-Down}\label{sec:app-drill-down}
We now drill deeper into the storage optimization algorithms 
and understand their behavior with respect to the characteristics
of spreadsheets. 
We selected four sample sheets from our dataset 
to represent variations in terms of data density and 
layout of data, which is either horizontal 
for most part or vertical for most part. 
For these sheets, we contrast their storage requirements 
for the different data models.
We plot the results in Figure~\ref{fig:storage_cost_samples}, 
where we depict the 
normalized storage across sheets. 
For each sheet we have normalized the data model 
that performs the worst
to $100$, and scaled the others accordingly.

The four spreadsheets show the variation 
among the different models in terms of storage requirements.
Sheets 1 and 2 have substantial storage savings for ROM and COM when compared with RCV since they are relatively high density.  
On the other hand, Sheet 4 has has substantial storage savings for RCV as compared to ROM and COM due to its relatively low density.   
For Sheet 3 (4), ROM's (COM's) storage requirement is less then that of COM (ROM). 
This is due to the distribution of the cells, which span for the most part in the vertical direction for Sheet 3 and in the horizontal direction for Sheet 4.
Except for Sheet 3, for all other sheets, the solution provided by Agg is close to DP.   
For Sheet 3, the optimization algorithms are not able to perform much better in terms of cost saving than the primitive data models. 
This is due to the fact that the sheet has both dense and sparse regions.

\begin{figure}[t]
	\centering
	\subfloat{}{\includegraphics[width=0.24\textwidth,clip]{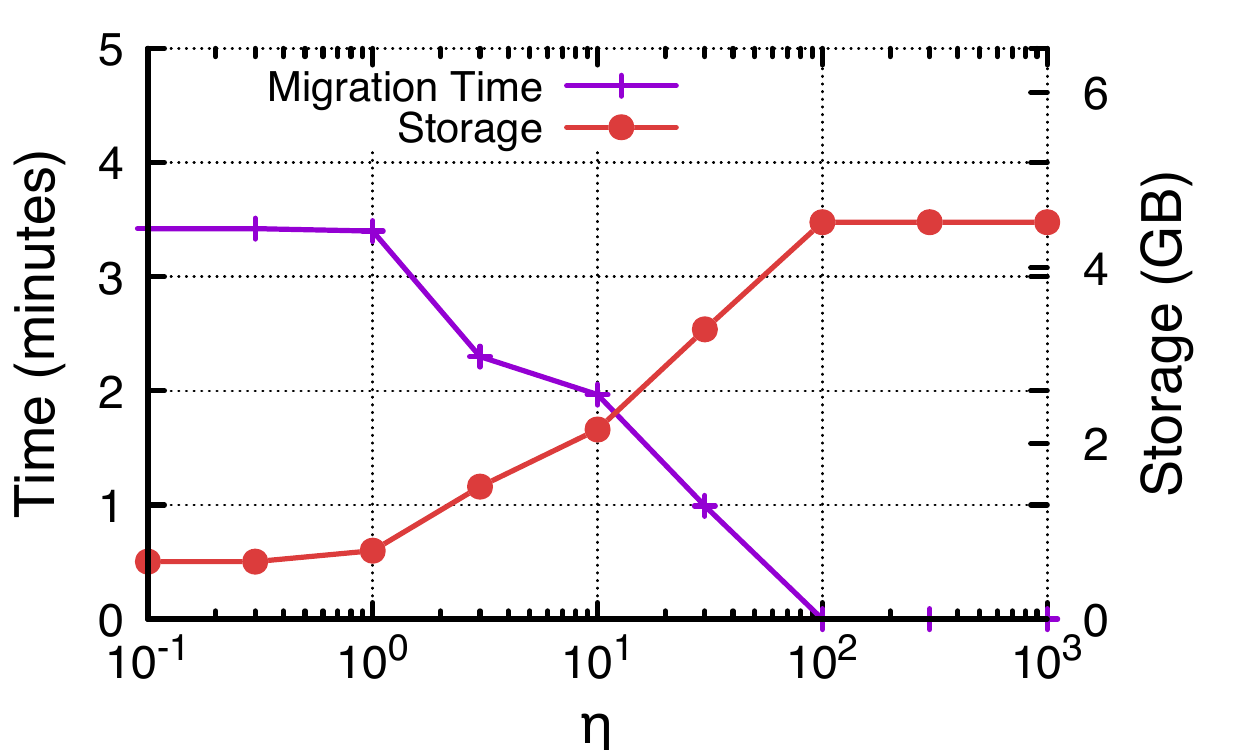}}
	\hspace{-7pt}
	\subfloat{}{\includegraphics[width=0.24\textwidth,clip]{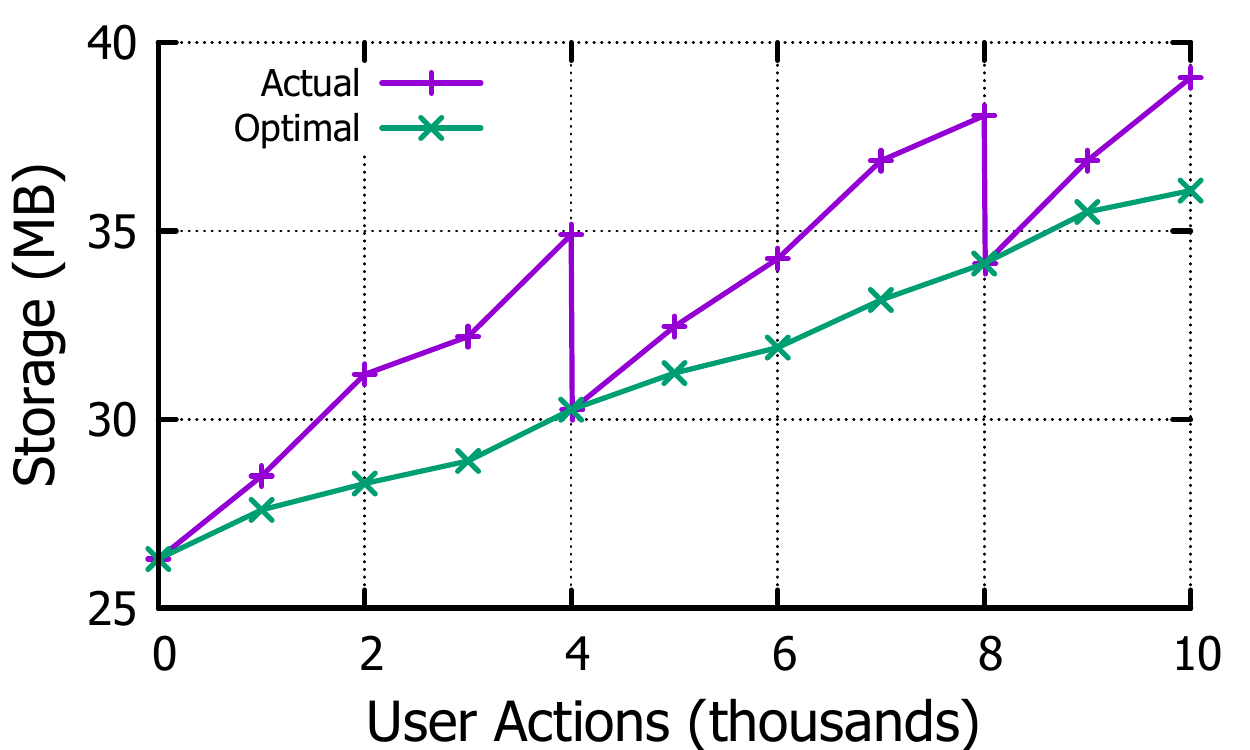}}	
	\caption{
	Incremental Hybrid Decomposition:
	(a) Trade-off with respect to $\eta$.
	(b) User operations vs. Storage.}
	\label{fig:inc_hybrid_decomp_others}
	\vspace{-12pt}
\end{figure}

\subsubsection{Incremental Maintenance} \label{sec:app-inc-hybrid-exp}
We now evaluate whether our representation schemes can be maintained
efficiently in the face of edits.
Note that in practice there will be periods where the \system is
idle, and so we can run the hybrid optimization algorithms then, but it is still 
valuable to ensure that the choice of data model is aware
of the existing layout. 
To illustrate our incremental decomposition approach from Section~\ref{sec:incr-decomp},
For this, we consider a synthetic spreadsheet 
as described in part e. of Section~\ref{sec:exp_impact_hybrid}.
We store the spreadsheet using the Agg-based hybrid data model.
In the absence of user operation traces, we develop a generative model for update operations. We consider the following four operations. 
\begin{paraenum}
\item Change the value of an existing cell.
\item Add a new cell at an arbitrary location. 
\item Add a new row.
\item Add a new column.
\end{paraenum}
Motivated from our user study, we consider that the above four operations are performed with probabilities $0.6$, $0.2$, $0.1999$, and $0.0001$ respectively. 
We fix the value of $\eta$ (the trade-off factor between migration and storage) 
to $1.0$ and run incremental maintenance with Agg
after each batch of $1000$ user updates are performed. 
We plot the storage requirement against the number of user updates in Figure~\ref{fig:inc_hybrid_decomp_others}(b). The \emph{actual} line in the graph indicates the storage requirement, which has a sawtooth like behavior. 
The drop in the graph correspond to the points where the incremental maintenance
algorithm chose 
a new decomposition and migrated to it: thus, there was no migration performed at 
batch 1, 2, 3, but there was one at batch 4. 
We also plot the storage for the non-incremental variant of Agg, 
which we obtain by running incremental decomposition and setting $\eta$ to $0$.
Overall, we find that a policy of this form (with $\eta=1$) only performs
migration when the structure within the spreadsheet has substantially changed. 

To study the impact of $\eta$, we consider one such point where
the spreadsheet has diverged from its original Agg-based data model. 
We run the Agg variant of 
incremental maintenance algorithm on varying $\eta$.
We plot $\eta$'s impact of the time taken to migrate 
and the storage requirement of the final decomposition in 
Figure~\ref{fig:inc_hybrid_decomp_others}(a).
Here, as we increase the value of $\eta$ we observe that the  migration time decreases and the storage requirement increases.
At lower values of $\eta$, the algorithm gives preference to finding the optimal solution while ignoring the migration cost. We can observe this from the low storage cost, and the high time required to migrate the data in to the new decomposition.
When $\eta>100$, the optimization aims at minimizing the migration cost at the expense of sacrificing the optimality of storage. 
\tr{Here, we observe a zero migration time, as the algorithm returns the original decomposition, and has the worst storage requirement. }

\subsection{Presentational Access with Updates}
\subsubsection{Varying Parameters}\label{sec:app-vary-parameters}
We now perform an evaluation of presentational access
with updates on varying various parameters of the
synthetic spreadsheets. 
For this evaluation, we focus on the two primitive data models \ie  
ROM and RCV, with the spreadsheet being represented
as a single table in these data models.
Since we use synthetic datasets where cells are ``filled in''
with a certain probability, we did not involve
hybrid data models, since they would (in this artificial context)
typically end up preferring the ROM data model.
These primitive data models are augmented
with hierarchical positional mapping.
We consider the performance on varying
several parameters of these datasets:
the density (\ie the number of filled in cells),
the number of rows, and the number of columns.
The default values of these parameters are $1$, $10^7$ and $100$ respectively.
We repeat each operation $500$ times and report the averages.

In Figure~\ref{fig:dm_performance}, we depict the charts
corresponding to average time to perform a random select
operation on a region of 1000 rows and 20 columns. 
This is, for example, the operation that would correspond 
to a user scrolling to a certain position on our spreadsheet.
As can be seen in Figure~\ref{fig:dm_performance}(a), 
ROM starts dominating RCV beyond a certain density,
at which point it makes more sense to store the data in
as tuples that span rows instead of incurring the penalty
of creating a tuple for every cell. 
Nevertheless, the best of these two models
takes less than 150ms across sheets of varying densities.
In Figure~\ref{fig:dm_performance}(b)(c),
since the spreadsheet is very dense (density=1),
ROM takes less time than RCV.
Overall, in all cases, even on spreadsheets with 100 columns and $10^7$ rows
and a density of 1, the average time to select a region
is well within 500ms.

Figures~\ref{fig:dm_update_performance} and~\ref{fig:dm_insert_row_performance}
depict the corresponding charts for updating
a region of 100 rows and 20 columns, and inserting one row of 100 columns
for the primitive data models.
In Figure~\ref{fig:dm_update_performance}, we find that the update time 
taken for RCV is a lot higher than the time for inserts or selects.
This is because in this benchmark, \system assumes
that the entire region update happens at once, and fires
$100 \times 20=2000$ update queries one at a time.
In practice, users may only update a small number of cells at a time;
and further, we may be able to batch these queries or issue them
in parallel to further save time.
In Figure~\ref{fig:dm_insert_row_performance}, we find that like 
in Figure~\ref{fig:dm_update_performance}, the time taken for updates on ROM is
faster than RCV since it only needs to issue one query, while RCV needs to issue 
multiple queries.
However, in this case, since the number of queries issued is small, 
the response time is always within $100$ms.

Overall, for both RCV and ROM, for inserting a row, the time is well
below 500ms for all of the charts;
for updates of a large region, 
while ROM is still highly interactive, RCV ends up taking longer
since 1000s of queries need to be issued to the database.
In practice, users won't update such a large region at a time, and
we can batch these queries.

\tr{

\section{Primitive Data Models: Optimality Argument }
\label{app:primitive_datamodels_optimality}

Earlier in Section \ref{sec:primitive_datamodels}, we developed three primitive data models, that represent reasonable extremes if we are to represent and store a spreadsheet in a single table in a database system. Here, we argue that these models are optimal choices with respect to cascading updates, among a large class of data models that we shall describe now.

\stitle{Requirements.}
We require each primitive data model in our class to have the following characteristics:
\begin{paraenum}
  \item {\em The data model should correspond to storing a rectangular region in the spreadsheet.} This constraint naturally stems from the way we perceive tables in a two-dimensional interface, in the sense that tables are rectangular, and our data models are stored as rectangular tables on disk.
  \item {\em The tuples in each table should correspond to a uniform geometric structure, and be contiguous in the sheet.} The first part of the constraint arises because we store our tables in a relational database, necessitating all tuples to have the same number of attributes. Additionally, we want our tuples to correspond to contiguous regions in the spreadsheet, \ie they should not have any ``holes'' in them.
\end{paraenum}

\stitle{Rectangular and Non-Rectangular Data Models.}
The data models which satisfy the aforementioned requirements fall into the following two classes:
\begin{paraenum}
  \item {\em Rectangular.} In rectangular data models, each tuple corresponds to a rectangle in the sheet. Clearly, they are uniform geometric units, and are contiguous. A typical example is provided in Figure~\ref{fig:data-models-examples}(a).
  \item {\em Non-rectangular.} Non-rectangular data models are essentially data-models where each tuple does not correspond to a rectangle. For instance, each tuple can either be diagonal with a fixed length, or have a ``zig-zag'' shape. A typical example where each tuple has zig-zag shape is provided in Figure~\ref{fig:data-models-examples}(b).
\end{paraenum}



\begin{figure}
  \centering
  \subfloat{}{\includegraphics[width=0.45\linewidth]{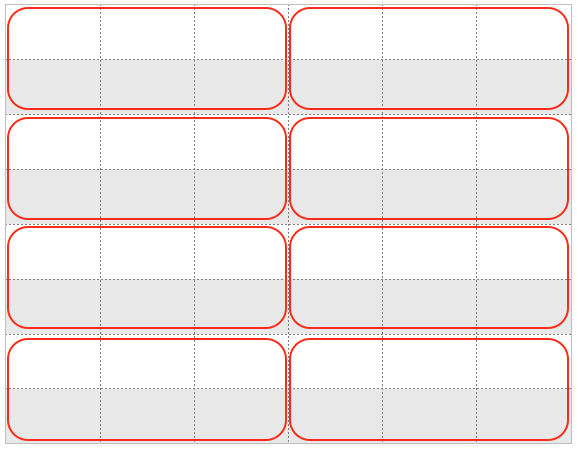}}
  \hspace{5pt}
  \subfloat{}{\includegraphics[width=0.45\linewidth]{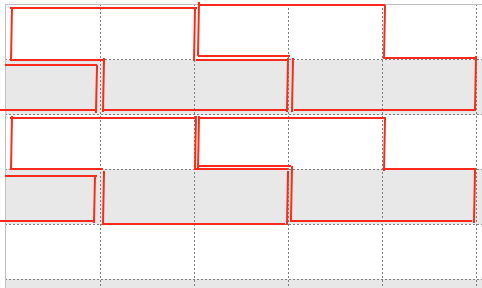}}
  \hspace{5pt}
  \caption{Data Model: (a) Rectangular (b) Non-rectangular}
  \label{fig:data-models-examples}
  \vspace{-12pt}
\end{figure}

\stitle{Updates as Optimality Criterion.}
We now discuss our optimality criterion. Since we consider a single table, storage is not a concern since every data model has to store all of the data in a table.
Furthermore, with any vanilla index, \eg \mbox{B+ tree}, access can be supported in all models in a similar manner, and likewise for single cells updates.
Hence, we focus on updates on the sheet, and how they correspond to reorganizations in backend. Specifically, we focus on row/column inserts/deletes since changing values of existing data in the sheet would result in the same time complexity across all data models.

As we shall soon describe, row/column inserts/deletes can greatly influence the performance of our data models.

\begin{theorem}[Optimality] Our primitive data models, coupled with our hierarchical positional mapping schemes, are the {\em only} models which do not result in cascading updates from the class of data models discussed above.
\end{theorem}

\begin{IEEEproof}
Consider any data model which can be rectangular or otherwise. We know all tuples are uniform in shape, and are contiguous in the sheet. Let say the tuple spans $p$ row and $q$ columns.

There are two possibilities: these tuples are either stored in row major form in the table or in column major form. If we use the former, then a row insert would result in data from $p$ rows to be {\em shifted} in the worst case. Equivalently, if the data is stored in column major form, then a column insert would result in data from $q$ columns to be shifted in the worst case.
Therefore, cascading updates can be avoided only when one among $p$ and $q$ equals $1$. There are three cases now:
\begin{enumerate}
  \item $p=1, q\neq 1$: This corresponds to ROM.
  \item $p\neq 1, q=1$: This corresponds to COM.
  \item $p=1, q=1$: This corresponds to RCV.
\end{enumerate}
\noindent
This completes our proof.\end{IEEEproof}
}
}

\end{document}